\documentclass[10.5pt,onecolumn]{article}
\usepackage{color}
\usepackage{amsmath}
\usepackage{amssymb}
\usepackage{amsfonts}
\usepackage{geometry}
\usepackage{setspace}
\usepackage{amsmath,bm}
\usepackage{graphicx}
\usepackage[section]{placeins}
\usepackage{subfigure}
\usepackage[numbers,sort&compress]{natbib}

\author{Shuwei Zhou$^{3,4}$, Timon Rabczuk$^{1,2*}$, Xiaoying Zhuang$^{4,5}$}
\title {Phase field modeling of quasi-static and dynamic crack propagation: COMSOL implementation and case studies}
\begin{document}

\bibliographystyle{unsrtnat}
\setcitestyle{numbers,square,aysep={},yysep={,}}
\date{}
\maketitle

\spacing {1.2}
\noindent
1 Division of Computational Mechanics, Ton Duc Thang University, Ho Chi Minh City, Vietnam \\
2 Faculty of Civil Engineering, Ton Duc Thang University, Ho Chi Minh City, Vietnam \\
3 Institute of Structural Mechanics, Bauhaus-University Weimar, Weimar 99423, Germany\\
4 Department of Geotechnical Engineering, College of Civil Engineering, Tongji University, Shanghai 200092, P.R. China\\
5 Institute of Continuum Mechanics, Leibniz University Hannover, Hannover 30167, Germany.

* Corresponding author: timon.rabczuk@tdt.edu.vn \\

\begin{abstract}
\noindent The phase-field model (PFM) represents the crack geometry in a diffusive way without introducing sharp discontinuities. This feature enables PFM to effectively model crack propagation compared with numerical methods based on discrete crack model, especially for complex crack patterns. Due to the involvement of ``phased field'', phase-field method can be essentially treated a multifield problem even for pure mechanical problem. Therefore, it is supposed that the implementation of PFM based on a software developer that especially supports the solution of multifield problems should be more effective, simpler and more efficient than PFM implemented on a general finite element software. In this work, the authors aim to devise a simple and efficient implementation of phase-field model for the modelling of quasi-static and dynamic fracture in the general purpose commercial software developer, COMSOL Multiphysics. Notably only the tensile stress induced crack is accounted for crack evolution by using the decomposition of elastic strain energy. The width of the diffusive crack is controlled by a length-scale parameter. Equations that govern body motion and phase-field evolution are written into different modules in COMSOL, which are then coupled to a whole system to be solved. A staggered scheme is adopted to solve the coupled system and each module is solved sequentially during one time step. A number of 2D and 3D examples are tested to investigate the performance of the present implementation. Our simulations show good agreement with previous works, indicating the feasibility and validity of the COMSOL implementation of PFM.
\end{abstract}


\section {Introduction}
Fracture induced failure has obtained extensive concern in engineering designs because of the potential serious risks for structures and machines being used \citep{anderson2005fracture}. The research on crack initiation and propagation in solids has therefore become very important \citep{liu2016abaqus}. Particularly, when experiments are difficult, or even impossible to perform for studying certain type of crack propagation, researchers have to employ numerical approaches to predict complicated crack paths \citep{klinsmann2015assessment} such as those in multiple scales \citep{budarapu2017multiscale,budarapu2017concurrently,yang2015meshless,budarapu2014efficient,
budarapu2014adaptive}. Consequently, a great number of numerical methods have been proposed to deal with crack problems in recent years. 

Most of these methods have to describe complex crack geometry in the discrete setting, such as the discrete crack models \citep{ingraffea1985numerical}, the extended finite element method (XFEM) \citep{moes2002extended,chen2012extended}, generalized finite-elements method (GFEM) \citep{fries2010extended}, and the phantom-node method \citep{chau2012phantom,rabczuk2008new}. These methods all enrich the displacement field with discontinuities. Particularly, the discrete crack model \citep{ingraffea1985numerical} introduces new boundaries for the freshly created crack surfaces by an adaptive reconstruction of the mesh. XFEM \citep{moes2002extended} enriches the cracked elements by adding a set of discontinuous shape functions to the standard parts of FEM. Another common option to model cracks is the so-called cohesive elements \citep{zhou2004dynamic,nguyen2001cohesive,rabczuk2008geometrically} that allow displacement jumps on element boundaries and cracks are therefore restricted to penetrate along the corresponding element edges. In addition, the element-erosion methods \citep{belytschko1987three,johnson1987eroding,liu2014regularized} also succeeds in dealing with the fracture surfaces by setting the stresses of the elements, which meet the fracture criterion, as zero. However, the element-erosion methods have the disadvantage that they cannot simulate crack branching correctly \citep{song2008comparative}. Therefore, the complicated and special treatments for complex crack topologies have made these numerical approaches not so easy to implement and apply in practical engineering.

A recently emerged and developed approach, the phase-field method (PFM) \citep{miehe2010phase,miehe2010thermodynamically,borden2012phase,hesch2014thermodynamically,rabczuk2013computational}, has attracted a lot of attention because of its relatively easier numerical implementation for fracture. The phase-field models utilize a scalar field (so-called phase-field) to represent the discrete cracks. The phase-field does not describe the crack as a physical discontinuity and just smoothly transits the intact material to the thoroughly broken one. The shape and propagation of the crack depend on the evolution equations of the phase-field. Thus, implementation of the phase-field does not require additional work to track the fracture surfaces algorithmically \citep{borden2012phase}. This results in that the phase-field methods have a large advantage over the discrete fracture models for modeling multiple and crack branching and merging in materials with arbitrary 2D and 3D geometries. 

The phase-field models for quasi-static brittle crack started from \citet{bourdin2008variational} and improved by \citet{miehe2010phase,miehe2010thermodynamically}. All these models are regarded as extension of the classical Griffith fracture theory and then extended to dynamic problems by \citet{borden2012phase}. In addition, Landau-Ginzburg type evolution equations \citep{karma2001phase} instead of the Griffith type have also been proposed and developed for the phase-field description of dynamic fracture. The progress in the phase-field models for quasi-static and dynamic crack problems has made PFM successfully applied  in different problems, such as cohesive fractures \citep{verhoosel2013phase}, ductile fractures \citep{ulmer2013phase,badnava2017phase}, large strain problems \citep{hesch2014thermodynamically}, hydraulic fracturing \citep{lee2016pressure},  thermo-elastic problems \citep{miehe2015phase,badnava2018h}, electrochemical problems \citep{miehe2010phase2}, thin shell \citep{amiri2014phase}, and stressed grain growth in polycrystalline metals \citep{jamshidian2014phase1,jamshidian2014phase2,jamshidian2016multiscale}. These attempts imply that the application of the phase-field methods is quite beyond purely mechanical problems. This naturally requires a much easier implementation approach for the phase-field models. Otherwise, extensive application of the phase-field models will be restricted, especially in multi-physical problems.

Due to the smooth characteristics of the phase-field, the phase-field method can be implemented in any existing standard finite element to model complex crack patterns as shown in \citep{miehe2010phase,miehe2010thermodynamically}. Therefore, to reduce the efforts in implementation, it is desirable to implement phase-field method to an extensively used FEM code or commercial software. In fact, \citet{msekh2015abaqus} and \citep{liu2016abaqus} have implemented the phase-field method for brittle cracks in Abaqus. However, the phase-field modeling itself is essentially a multi-field problem even in the case of pure mechanical problem \citep{miehe2010phase,miehe2010thermodynamically}. From the authors' experience, it is laborious and time consuming to implement a multifield problem in Abaqus. Therefore, a general purpose programme developer that especially supports the programming of multifield problem such as COMSOL has the potential to become a better solution than Abaqus.

In this paper, the possibility of simple and fast implementation of phase-field method is exploited for fracture modelling in a multifield programme developer, namely COMSOL Multiphysics. The phase-field modeling in COMSOL can be easily extended to problems that have more coupled fields by just adding suitable modules and coupling terms. It will be quite easy for readers to use this first-step implementation and augment it by other physical phenomena to solve multiphysics problems involving crack propagation. For example, the phase field implementation in COMSOL can be extended and applied to hydraulic fracturing, or compressed air energy storage \citep{zhou2015analytical,zhou2017numerical}, which involves fluid pressure field, temperature, and cyclic effects \citep{xia2015strength, zhou2015damage, zhou2017statistical, zhou2018analyicaltheory}. In this work, one phase-field model presented by \citet{miehe2010phase,miehe2010thermodynamically} for a quasi-static crack problem and another one presented by \citet{borden2012phase} for dynamic problems are implemented in COMSOL in a staggered scheme. The elastic strain energy density is decomposed into two individual parts resulting from compression and tension, respectively. Thus, the fractures only due to tension can be obtained. In COMSOL, we use an implicit time integration scheme to enable the simulation. We also calculate some 2D and 3D benchmarks for quasi-static and dynamic crack propagation to show the feasibility of our approach for modeling fracture.

The paper is organized as follows. We begin with a short introduction of the phase-field model for brittle fractures based on the variational approach in Section 2. Subsequently, we present the numerical implementation of the phase-field model in COMSOL in Section 3. In Section 4, we examine some 2D and 3D numerical examples for cracks under quasi-static and dynamic loading. Finally, we end with conclusions regarding our findings in Section 5. 
\section {Phase-field model for fracture}
\subsection {Theory of brittle fracture}

\begin{figure}[htbp]
	\centering
	\includegraphics[width = 10cm]{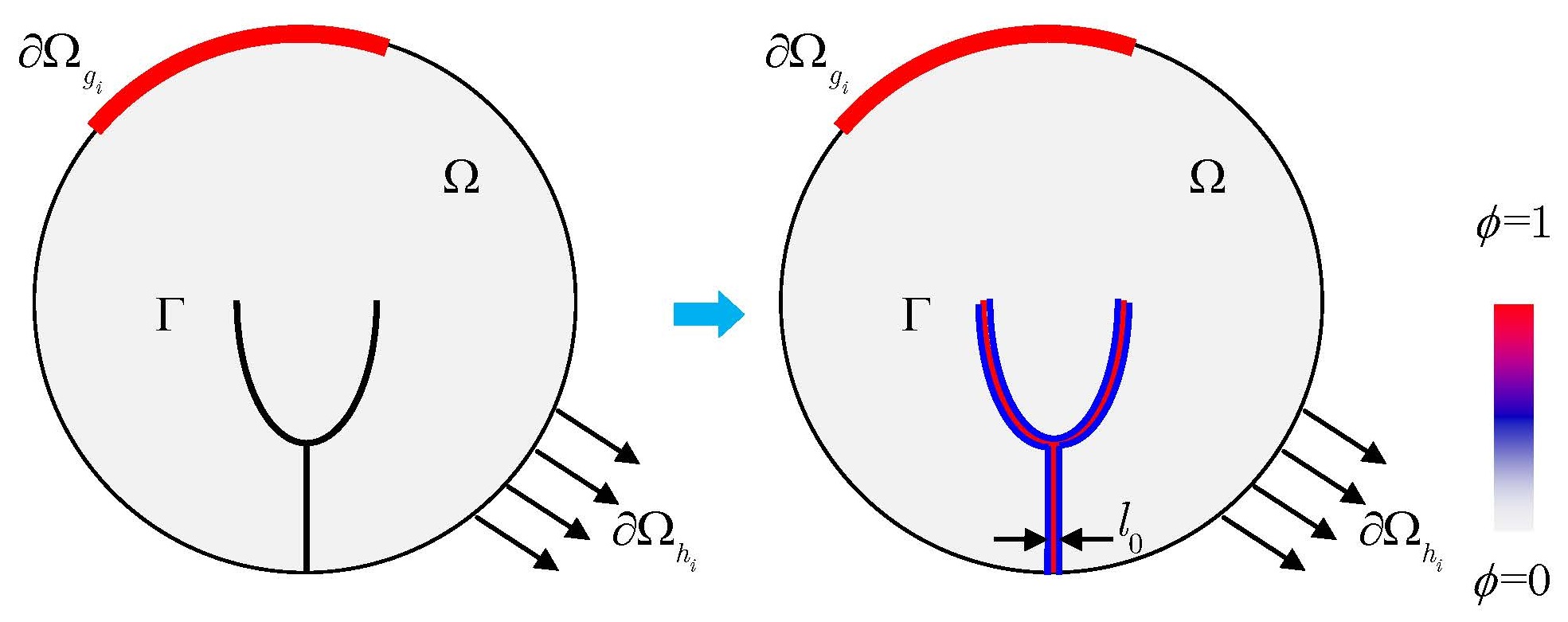}
	\caption{Phase-field approximation of the crack surface}
	\label{Figure1}
	\end{figure}

Let us consider an arbitrary body $\Omega\subset \mathbb R^d$ ($d\in \{1,2,3\} $) as shown in Fig. \ref {Figure1}. The body $\Omega $ has an external boundary $\partial \Omega$ and internal discontinuity boundary $\Gamma $. The displacement of body $\Omega $ at time $t$ is denoted by $\bm u(\bm x,t)\subset \mathbb R^d$ where $\bm x $ is the position vector. The displacement field satisfies the time-dependent Dirichlet boundary conditions, $u_i(\bm x,t)=g_i(\bm x,t)$, on $\partial \Omega_{g_i} \in \Omega$, and also the time-dependent Neumann conditions on $\partial \Omega_{h_i} \in \Omega$. We also consider a body force $\bm b(\bm x,t)\subset \mathbb R^d$ acted on the body $\Omega$  and a traction $\bm f(\bm x,t)$ on the boundary $\partial \Omega_{h_i}$.

A variational approach for fracture problems according to Griffith's theory has been proposed in \citep{francfort1998revisiting}. It states that the required energy to create a fracture surface per unit area is equal to the critical fracture energy density $G_c$, which is also commonly referred to as the critical energy release rate. The total potential energy $\Psi_{opt}(\bm u,\Gamma)$ can be expressed in terms of the elastic energy $\psi_{\varepsilon}(\bm \varepsilon)$, fracture energy and energy due to external forces:

	\begin{equation}
	\Psi_{opt}(\bm u,\Gamma) = \int_{\Omega}\psi_{\varepsilon}(\bm \varepsilon) \mathrm{d}{\Omega}+\int_{\Gamma}G_c \mathrm{d}S-\int_{\Omega} \bm b\cdot{\bm u}\mathrm{d}{\Omega} - \int_{\partial\Omega_{h_i}} \bm f\cdot{\bm u}\mathrm{d}S
	\end{equation}

\noindent with the linear strain tensor $\bm\varepsilon = \bm\varepsilon(\bm u)$ given by

	\begin{equation}
	\varepsilon_{ij}=\frac 1 2 \left(\frac{\partial u_i}{\partial x_j}+\frac{\partial u_j}{\partial x_i}\right)
	\end{equation}
	
Isotropic linear elasticity is assumed and the elastic energy density $\psi_{\varepsilon}(\bm \varepsilon)$ is given by \citep{miehe2010phase}

	\begin{equation}
	\psi_{\varepsilon}(\bm \varepsilon) = \frac{1}{2}\lambda\varepsilon_{ii}\varepsilon_{jj}+\mu\varepsilon_{ij}\varepsilon_{ij}
	\end{equation}

\noindent where $\lambda$ and $\mu$ are Lam\'e constants.

In addition, the variational approach \citep{francfort1998revisiting} states that initiation, propagation and branching of the crack $\Gamma(\bm x,t)$ at the time $t\in(0,T)$ for a point $x\in\Omega$ occur when the potential reaches  the minimum value and the irreversible condition $\Gamma(\bm x,s)\in\Gamma(\bm x,t)(s<t)$ is satisfied. The irreversible condition means that a crack cannot be recovered to the uncracked state after its formation.

\subsection{Phase filed approximation for fracture energy}

The variational approach for brittle fracture was successfully implemented by \citet{miehe2010phase,miehe2010thermodynamically}, \citet{borden2012phase} and \citet{bourdin2008variational} with the introduction of a scalar phase filed. In this paper, we define a phase-field $\phi(\bm x,t)\in[0,1]$ to approximate the fracture surface, $\Gamma$ (see also Fig. 1). $\phi(\bm x,t)\in[0,1]$ represents naturally a diffusive crack topology. $\phi=1$ represents the crack and $\phi=0$ means that the body is uncracked. Thus, the crack surface density per unit volume of the solid is given by \citep{miehe2010phase},

	\begin{equation}
	\gamma(\phi,\bigtriangledown\phi)=\frac{\phi^2}{2l_0}+\frac{l_0}2\frac{\partial\phi}{\partial x_i}\frac{\partial\phi}{\partial x_i}
	\end{equation}

\noindent where $l_0$ is a parameter that controls the transition region of the phase-field from 0 to 1. We call  $l_0$ the length scale parameter that reflects the shape of a crack. The crack region will have a larger width with a larger $l_0$ and vice versa, see Fig. 1. 

Applying Eq. (4), the fracture energy is approximated by

	\begin{equation}
	\int_{\Gamma}G_c \mathrm{d}S=\int_{\Omega}G_c\left[\frac{\phi^2}{2l_0}+\frac{l_0}2\frac{\partial\phi}{\partial x_i}\frac{\partial\phi}{\partial x_i}\right]\mathrm{d}{\Omega}
	\end{equation}

The crack surface energy is transformed from the elastic energy as shown in \citep{francfort1998revisiting}, indicating that the elastic energy drives the evolution of the phase-field. In order to ensure that the crack is only driven by tensile load, it is important to decompose the elastic energy into tensile and compressive parts \citep{bourdin2008variational}. Here, the decomposition approach in \citet{miehe2010phase} is adopted to ensure the evolution of the phase-field will only be induced by the tensile part of the elastic energy density while compressive stress will not contribute to the propagation of crack. Therefore, the strain tensor $\bm\varepsilon$ is decomposed as follows

	\begin{equation}
	  \left\{
	   \begin{aligned}
	\bm\varepsilon_+=\sum_{a=1}^d \langle\varepsilon_a\rangle_+\bm n_a\otimes\bm n_a \\ \bm\varepsilon_-=\sum_{a=1}^d \langle\varepsilon_a\rangle_-\bm n_a\otimes\bm n_a
	   \end{aligned}\right .
	\end{equation}

\noindent where $\bm\varepsilon_+$  and $\bm\varepsilon_-$  are the tensile and compressive strain tensors, respectively. $\varepsilon_a$ is the principal strain and $\bm n_a$ is its direction vector. The operators $\langle\centerdot\rangle_+$  and $\langle\centerdot\rangle_-$  are defined as \cite{miehe2010phase}: $\langle\centerdot\rangle_+=(\centerdot+|\centerdot|)/2$, $\langle\centerdot\rangle_-=(\centerdot-|\centerdot|)/2$.

By applying the decomposed strain tensor, the elastic energy density is represented as follows:
	
	\begin{equation}
	  \left\{
	   \begin{aligned}
	\psi_{\varepsilon}^+(\bm \varepsilon) = \frac{\lambda}{2}\langle tr(\bm\varepsilon)\rangle_+^2+\mu tr \left(\bm\varepsilon_+^2\right) 
	\\ \psi_{\varepsilon}^-(\bm \varepsilon) = \frac{\lambda}{2}\langle tr(\bm\varepsilon)\rangle_-^2+\mu tr \left(\bm\varepsilon_-^2\right) 
	   \end{aligned}\right.
	\end{equation}

It is assumed that only the tensile part of the elastic energy density is affected by the phase-field and then the following equation is used to model the stiffness reduction \citep{borden2012phase}:

	\begin{equation}
	\psi_{\varepsilon}(\bm\varepsilon)=\left[(1-k)(1-\phi)^2+k\right]\psi_{\varepsilon}^+(\bm \varepsilon)+\psi_{\varepsilon}^-(\bm \varepsilon)
	\end{equation}

\noindent where $k$ is a model parameter that prevents the positive part of the elastic energy density from disappearing and the numerical singularity when phase-field $\phi$  tends to 1. In addition, it is required that $k>0$  and $k\ll1$.

\subsection{Governing equations}

We consider also the kinetic energy of body $\Omega$:

	\begin{equation}
	\psi_{kin}(\bm{\dot u})=\frac 1 2\int_{\Omega}\rho\dot u_i\dot u_i \mathrm{d}{\Omega}
	\end{equation}

\noindent with  $\bm{\dot u}=\frac {\partial {\bm u}}{\partial t}$ and $\rho$ being the density of body $\Omega$.

The total Lagrange energy functional can be expressed by the sum of the phase-field approximation for the fracture energy (5), the elastic energy (8), the kinetic energy (9) and the external potential energy by the external loads:

	\begin{multline}
	L=\frac 1 2\int_{\Omega}\rho\dot u_i\dot u_i \mathrm{d}{\Omega}-\int_{\Omega}\left\{\left[(1-k)(1-\phi)^2+k\right]\psi_{\varepsilon}^+(\bm \varepsilon)+\psi_{\varepsilon}^-(\bm \varepsilon)\right\}\mathrm{d}{\Omega}-\int_{\Omega}G_c\left[\frac{\phi^2}{2l_0}+\frac{l_0}2\frac{\partial\phi}{\partial x_i}\frac{\partial\phi}{\partial x_i}\right]\mathrm{d}{\Omega}+\int_{\Omega} b_iu_i\mathrm{d}{\Omega}+\\ \int_{\partial\Omega_{h_i}} f_iu_i\mathrm{d}S
	\end{multline}

The variation of the functional $L$ can be derived and its first variation should be zero, which leads to the following governing equations:

	\begin{equation}
	  \left\{
	   \begin{aligned}
	\frac {\partial {\sigma_{ij}}}{\partial x_j}+b_i=\rho{\ddot u_i}
	\\ \left[\frac{2l_0(1-k)\psi_{\varepsilon}^+}{G_c}+1\right]\phi-l_0^2\frac{\partial^2 \phi}{\partial {x_i^2}}=\frac{2l_0(1-k)\psi_{\varepsilon}^+}{G_c}
	   \end{aligned}\right.
	\end{equation}

\noindent where $\ddot u_i=\frac {\partial^2 u}{\partial t^2}$  and  $\sigma_{ij}$ is the component of Cauchy stress tensor given by

	\begin{equation}
	\sigma_{ij}=\left [(1-k)(1-\phi)^2+k \right]\frac {\partial{\psi_\varepsilon^+}}{\partial {\varepsilon_{ij}}}+\frac {\partial{\psi_\varepsilon^-}}{\partial {\varepsilon_{ij}}} \ ,
	\end{equation}

and it can be rewritten as
	
	\begin{equation}
	\bm \sigma=\left [(1-k)(1-\phi)^2+k \right]\left[\lambda \langle tr(\bm\varepsilon)\rangle_+ \bm I+ 2\mu \bm\varepsilon_+ \right]+\lambda \langle tr(\bm\varepsilon)\rangle_- \bm I+ 2\mu \bm\varepsilon_-
	\end{equation}

\noindent where $\bm I$ is a unit tensor $\in \mathbb R^{d\times d}$.

In order to ensure a monotonically increasing phase-field, the irreversibility condition is needed during compression or unloading. Thus, we introduce a strain-history field  $H(\bm x,t)$ \citep{miehe2010phase,miehe2010thermodynamically} defined by

	\begin{equation}
	H(\bm x,t) = \max \limits_{s\in[0,t]}\psi_\varepsilon^+\left(\bm\varepsilon(\bm x,s)\right)
	\end{equation}

Replacing $\psi_\varepsilon^+$  by  $H(\bm x,t)$  in Eq. (11), the strong form is obtained by

	\begin{equation}
	  \left\{
	   \begin{aligned}
	\frac {\partial {\sigma_{ij}}}{\partial x_j}+b_i=\rho{\ddot u_i}
	\\ \left[\frac{2l_0(1-k)H}{G_c}+1\right]\phi-l_0^2\frac{\partial^2 \phi}{\partial {x_i^2}}=\frac{2l_0(1-k)H}{G_c}
	   \end{aligned}\right.
	\label{Strong form}
	\end{equation}

In addition, the zero first variation of the functional $L$ also achieves the Neumann conditions,

	\begin{equation}
	  \left\{
	   \begin{aligned}
	&\sigma_{ij}m_j=f_i \hspace{2cm} \mathrm{on}\hspace{0.5cm} \partial\Omega_{h_i}
	\\ &\frac{\partial \phi}{\partial x_i} m_i = 0 \hspace{2cm} \mathrm{on}\hspace{0.5cm} \partial\Omega
	\end{aligned}\right.
	\end{equation}

\noindent with $m_j$  the component of the outward-pointing normal vector of the boundary. 

For dynamic problems, the following initial conditions must be fulfilled,

	\begin{equation}
	  \left\{
	   \begin{aligned}
	&\bm u(\bm x,0)=\bm u_0(\bm x)\hspace{2cm} &\bm x\in\Omega
	\\ &\bm v(\bm x,0)=\bm v_0(\bm x)\hspace{2cm} &\bm x\in\Omega
	\\&\phi(\bm x,0)=\phi_0(\bm x)\hspace{2cm} &\bm x\in\Omega
	\end{aligned}\right. \ .
	\end{equation}

Here, the initial phase-field  $\phi_0(\bm x)$ can be used to model pre-existing cracks or geometrical features by setting it locally equal to 1 \citep{borden2012phase}.

\subsection{Estimation of $l_0$}\label{Estimation of $l_0$}

Selecting the length scale parameter $l_0$ remains an open topic in phase field models for fracture. An analytical solution for the critical tensile stress $\sigma_{cr}$ that a 1D bar can sustain under tension has been derived by \citet{borden2012phase}:
	\begin{equation}
	\sigma_{cr}=\frac 9 {16} \sqrt{\frac{EG_c}{3l_0}}
	\label{critical stress}
	\end{equation}

\noindent where $E$ is the Young's modulus and $G_c$ the critical energy release rate. There is an apparent singularity when $l_0$ tends to zero, i.e. in case of a sharp crack which leads to a phyiscally meaningless infinite tensile strength. However, when all other parameters except $l_0$ are known, Eq. (\ref{critical stress}) can be solved for $l_0$ yielding
	\begin{equation}
	l_0=\frac {27EG_c} {256\sigma_{cr}^2}
	\label{l0}
	\end{equation}

In Eq. \eqref{l0}, the critical energy release rate $G_c$, Young's modulus $E$, and critical stress $\sigma_{cr}$ can be estimated through some regular tests. Though the extension of this approach into higher-order dimensions and complex mixed-mode fracture is difficult, it gives at least some estimate how to choose $l_0$. It should be noted here again that no external fracture criterion is needed in the phase field method. The crack path is obtained by the evolution equation of phase field.

\section {Implementation method in COMSOL}
\subsection{Overall framework}
The phase-field modeling in this work is naturally a two-field problem ($\bm u$ and $\phi$). The phase-field model is implemented into the software COMSOL Multiphysics. We establish three main modules namely, Solid Mechanics Module, History-strain Module and phase-field Module. These modules are used to solve the three fields, $\bm u$, $H$ and $\phi$, respectively. These modules are all written in strong forms and solved based on standard finite element discretization in space domain and finite difference discretization in time domain. We also establish a preset Storage Module to calculate and store the internal field variables during each time step, such as the principal strains and their corresponding direction vectors.

\subsection{Module setup}
The Solid Mechanics Module is set up based on a linear elastic material library. The boundary and initial conditions in the Solid Mechanics Module are implemented as shown in Section 2. However, the elasticity matrix during each time step requires modification. The elasticity matrix is calculated on basis of the elasticity tensor of four order $\bm D$:

	\begin{equation}
	\bm D = \frac {\partial \bm\sigma}{\partial \bm\varepsilon}=\lambda\left\{ \left[(1-k)(1-\phi)^2+k \right]H_\varepsilon(tr(\bm\varepsilon))+H_\varepsilon(-tr(\bm\varepsilon))\right\}\bm J +2\mu\left\{\left[(1-k)(1-\phi)^2+k \right]\frac{\partial \bm\varepsilon_+}{\partial \bm\varepsilon}+\frac{\partial \bm\varepsilon_-}{\partial \bm\varepsilon}\right\}
	\end{equation}

\noindent where $H_\varepsilon \langle x \rangle$  is a Heaviside function with $H_\varepsilon \langle x \rangle=1$  for  $x>0$ and $H_\varepsilon \langle x \rangle=0$  for $x\leq 0$  and $J_{ijkl}=\delta_{ij}\delta_{kl}$  with $\delta_{ij}$ and $\delta_{kl}$ are Kronecker deltas. Finally, the elasticity matrix $\bm D_e$ is rewritten as following:

	\begin{equation}
	\bm D_e= \left [
	\begin{array}{cccccc}
	D_{1111} & D_{1122} & D_{1133} & D_{1112} & D_{1123} & D_{1113}\\
	D_{2211} & D_{2222} & D_{2233} & D_{2212} & D_{2223} & D_{2213}\\
	D_{3311} & D_{3322} & D_{3333} & D_{3312} & D_{3323} & D_{3313}\\
	D_{1211} & D_{1222} & D_{1233} & D_{1212} & D_{1223} & D_{1213}\\
	D_{2311} & D_{2322} & D_{2333} & D_{2312} & D_{2323} & D_{2313}\\
	D_{1311} & D_{1322} & D_{1333} & D_{1312} & D_{1323} & D_{1313}
	\end{array}
	\right ]
	\end{equation}

\noindent with $D_{ijkl}=\bar D_{ijkl}+\tilde D_{ijkl}$.

The component  $\bar D_{ijkl}$ is expressed as

	\begin{equation}
	\bar D_{ijkl}=\lambda\left\{ \left[(1-k)(1-\phi)^2+k \right]H_\varepsilon(tr(\bm\varepsilon))+H_\varepsilon(-tr(\bm\varepsilon))\right\} \delta_{ij}\delta_{kl}
	\end{equation}

According to the algorithm for fourth-order isotropic tensor \citep{miehe1998comparison}, the component  $\tilde D_{ijkl}$ is calculated based on the following:

	\begin{equation}
	\tilde D_{ijkl}=2\mu\left\{ \left[(1-k)(1-\phi)^2+k \right]P_{ijkl}^++P_{ijkl}^-\right\}
	\label{Dijkl}
	\end{equation}

\noindent with

	\begin{equation}
	P_{ijkl}^+ = \sum_{a=1}^3\sum_{b=1}^3 H_\varepsilon(\varepsilon_a)\delta_{ab}n_{ai}n_{aj}n_{bk}n_{bl}+\sum_{a=1}^3\sum_{b\neq a}^3 \frac 1 2 \frac {\langle \varepsilon_a\rangle_+ - \langle \varepsilon_b\rangle_+}{\varepsilon_a-\varepsilon_b}n_{ai}n_{bj}(n_{ak}n_{bl}+n_{bk}n_{al})
		\label{Dijkl+}
	\end{equation}

\noindent and

	\begin{equation}
	P_{ijkl}^- = \sum_{a=1}^3\sum_{b=1}^3 H_\varepsilon(-\varepsilon_a)\delta_{ab}n_{ai}n_{aj}n_{bk}n_{bl}+\sum_{a=1}^3\sum_{b\neq a}^3 \frac 1 2 \frac {\langle \varepsilon_a\rangle_- - \langle \varepsilon_b\rangle_-}{\varepsilon_a-\varepsilon_b}n_{ai}n_{bj}(n_{ak}n_{bl}+n_{bk}n_{al})
		\label{Dijkl-}
	\end{equation}

\noindent in which $n_{ai}$ denotes the $i$-th component of vector $\bm n_a$.

It can be seen from Eqs. \eqref{Dijkl+} and \eqref{Dijkl-} that Eq. \eqref{Dijkl} cannot be evaluated if $\varepsilon_a=\varepsilon_b$. We therefore adopt a ``perturbation'' technology for the principal strains \citep{miehe1993computation} and make a change for better application in COMSOL:

	\begin{equation}
	  \left\{
	   \begin{aligned}
	&\varepsilon_1 = \varepsilon_1(1+\delta)\hspace{0.5cm} &\mathrm{if}\hspace{0.1cm}\varepsilon_1 = \varepsilon_2
	\\ &\varepsilon_3 = \varepsilon_3(1-\delta)\hspace{0.5cm} &\mathrm{if}\hspace{0.1cm}\varepsilon_2 = \varepsilon_3
	\end{aligned}\right.
	\end{equation}

\noindent with the perturbation $\delta=1\times 10^{-9}$ for this paper. The second principal strain remains unchanged. 

The Solid Mechanics Module has an inertial term in the governing equation as shown in Eq. (15). Thus, the governing equation is automatically suitable for a dynamic crack problem. For a quasi-static problem, the inertia term vanishes.

The phase-field Module is established by revising a module governed by the Helmholtz equation. The governing equation in Eq. (15), boundary condition in Eq. (16) and initial condition (17) are implemented in this module. The History-strain Module is set up based on the Distributed ODEs and DAEs Interfaces in COMSOL, which provide the possibility to solve distributed ODEs and DAEs in domains. The history strain field $H(\bm x,t)$ is obtained by solving the following equation:

	\begin{equation}
	\dot H =  \left\{
	   \begin{aligned}
	&\dot\psi_\varepsilon^+,\hspace{0.4cm} &\psi_\varepsilon^+>0 \hspace{0.1cm}\mathrm{and} \hspace{0.1cm} H=\psi_\varepsilon^+
	\\&0, &\mathrm{else}
	\end{aligned}\right.
	\end{equation}

Initial conditions are also required for the History-strain Module. Commonly $H_0(\bm x)=0$ unless pre-existing cracks are modelled as the induced ones by the following expression \citep{borden2012phase}:

	\begin{equation}
	H_0(\bm x) =  \left\{
	   \begin{aligned}
	&\frac {BG_c}{2l_0}\left (1-\frac {2d(\bm x,l)}{l_0}\right ),\hspace{0.1cm}&d(\bm x, l)\leq \frac {l_0} 2
	\\&0,\hspace{0.5cm} & d(\bm x, l)> \frac {l_0} 2
	\end{aligned}\right.
	\end{equation}

\noindent where B is a scalar that controls the magnitude of the induced history field.

Letting  $d = 0$ and substituting $H_0$ into the second equation of \eqref{Strong form}, one will get:

	\begin{equation}
	B=\frac\phi {(1-k)(1-\phi)}
	\end{equation}

$B$  will become quite large if $\phi$ is close to 1, the value of the phase-field for the initial crack. Here, we chose $B=1\times 10^6$  for the simulation in this paper.

\subsection{Finite element method and discretization}\label{Finite element method and discretization}

In COMSOL, the finite element method is used with the weak form of the governing equations given  by
	\begin{equation}
	\int_{\Omega}\left(-\rho \bm{\ddot u} \cdot \delta \bm u -\bm\sigma:\delta \bm {\varepsilon}\right) \mathrm{d}\Omega +\int_{\Omega}\bm b \cdot \delta \bm u  \mathrm{d}\Omega +\int_{\Omega_{h_i}}\bm f \cdot \delta \bm u  \mathrm{d}S=0
	\label{weak form 1}
	\end{equation}

\noindent and
	\begin{equation}
	\int_{\Omega}-2(1-k)H(1-\phi)\delta\phi\mathrm{d}\Omega+\int_{\Omega}G_c\left(l_0\nabla\phi\cdot\nabla\delta\phi+\frac{1}{l_0}\phi\delta\phi\right)\mathrm{d}\Omega=0
	\label{weak form 2}
	\end{equation}

The standard vector-matrix notation is used with $\bm u_i$ and $\phi_i$ being the nodal values of the displacement and phase field. Then, we let the discretization as 
	\begin{equation}
	\bm u = \bm N_u \bm d,\hspace{0.5cm} \phi = \bm N_{\phi} \hat{\bm\phi}
	\end{equation}

\noindent where $\bm d$ and $\hat{\bm\phi}$ are the vectors consisting of node values $\bm u_i$ and $\phi_i$. $\bm N_u$ and $\bm N_{\phi}$ are shape function matrices:
	\begin{equation}		
			\bm N_u = \left[ \begin{array}{ccccccc}
			N_{1}&0&0&\dots&N_{n}&0&0\\
			0&N_{1}&0&\dots&0&N_{n}&0\\
			0&0&N_{1}&\dots&0&0&N_{n}
			\end{array}\right], \hspace{0.5cm}
			\bm N_\phi = \left[ \begin{array}{cccc}
			N_{1}&N_{2}&\dots&N_{n}
			\end{array}\right]
	\end{equation}

\noindent where $n$ is the node number in one element and $N_i$ is the shape function at node $i$. Assuming that the test functions have the same discretization, we obtain
	\begin{equation}
	\delta \bm u = \bm N_u \delta \bm d,\hspace{0.5cm} \delta \phi = \bm N_{\phi} \delta \hat{\bm\phi}
	\end{equation}

\noindent where $\delta \bm d$ and $\delta \hat{\bm\phi}$ are the vectors consisting of node values of the test functions.

The gradients are thereby as follows
	\begin{equation}
	\bm \varepsilon =  \bm B_u \bm d,\hspace{0.5cm} \nabla\phi = \bm B_\phi \hat{\bm\phi}, \hspace{0.5cm}\bm \delta \varepsilon =  \bm B_u \delta \bm d,\hspace{0.5cm} \nabla\phi = \bm B_\phi \delta \hat{\bm\phi}
	\end{equation}

\noindent where $\bm B_u$ and $\bm B_\phi$ are the derivatives of the shape functions:
	\begin{equation}
	\bm B_u=\left[
		\begin{array}{ccccccc}
		N_{1,x}&0&0&\dots&N_{n,x}&0&0\\
		0&N_{1,y}&0&\dots&0&N_{n,y}&0\\
		0&0&N_{1,z}&\dots&0&0&N_{n,z}\\
		N_{1,y}&N_{1,x}&0&\dots&N_{n,y}&N_{n,x}&0\\
		0&N_{1,z}&N_{1,y}&\dots&0&N_{n,z}&N_{n,y}\\
		N_{1,z}&0&N_{1,x}&\dots&N_{n,z}&0&N_{n,x}
		\end{array}\right],\hspace{0.2cm}
		\bm B_\phi=\left[
		\begin{array}{cccc}
		N_{1,x}&N_{2,x}&\dots&N_{n,x}\\
		N_{1,y}&N_{2,y}&\dots&N_{n,y}\\
		N_{1,z}&N_{2,z}&\dots&N_{n,z}
		\end{array}\right]
		\label{BiBu}
	\end{equation}

The equations of weak form \eqref{weak form 1} and \eqref{weak form 2} are then written as
	\begin{equation}
	-(\delta\bm d)^\mathrm{T} \left[\int_{\Omega}\rho\bm N_u^\mathrm{T}\bm N_u \mathrm{d}\Omega \ddot{\bm d}+\int_{\Omega} \bm B_u^\mathrm{T} \bm D_e \bm B_u \mathrm{d}\Omega \bm d \right]+ 
	(\delta\bm d)^\mathrm{T} \left[\int_{\Omega}\bm N_u^\mathrm{T}\bm b \mathrm{d}\Omega+\int_{\Omega_{h_i}} \bm N_u^\mathrm{T} \bm f\mathrm{d}S \right]=0
	\label{discrete equation 1}
	\end{equation}
	\begin{equation}
	-(\delta\hat{\bm \phi})^{\mathrm{T}} \int_{\Omega}\left\{\bm B_\phi^{\mathrm{T}} G_c l_0 \bm B_\phi +\bm N_\phi^{\mathrm{T}} \left [ \frac{G_c}{l_0} + 2(1-k)H \right ]  \bm N_\phi \right \} \mathrm{d}\Omega \hat{\bm \phi}+ (\delta\hat{\bm \phi})^\mathrm{T} \int_{\Omega}2(1-k)H\bm N_\phi^{\mathrm{T}} \mathrm{d}\Omega = 0
	\label{discrete equation 2}      
	\end{equation}

For admissible arbitrary test functions, Eqs. \eqref{discrete equation 1} and \eqref{discrete equation 2} produces the discretized weak form as
	\begin{equation}
	-\underbrace{\int_{\Omega}\rho\bm N_u^\mathrm{T}\bm N \mathrm{d}\Omega \ddot{\bm d}}_{\bm F_u^{ine}=\bm M \ddot{\bm d}}-\underbrace{\int_{\Omega} \bm B_u^\mathrm{T} \bm D_e \bm B_u \mathrm{d}\Omega \bm d}_{\bm F_u^{int}=\bm K_u \bm d} + \underbrace{\int_{\Omega}\bm N_u^\mathrm{T}\bm b \mathrm{d}\Omega+\int_{\Omega_{h_i}} \bm N_u^\mathrm{T} \bm f\mathrm{d}S}_{\bm F_u^{ext}}=0
	\end{equation}
	\begin{equation}
	-\underbrace{ \int_{\Omega}\left\{\bm B_\phi^{\mathrm{T}} G_c l_0 \bm B_\phi +\bm N_\phi^{\mathrm{T}} \left [ \frac{G_c}{l_0} + 2(1-k)H \right ] \bm N_\phi \right \} \mathrm{d}\Omega \hat{\bm \phi}}_{\bm F_\phi^{int}=\bm K_\phi \hat{\bm \phi}}+ \underbrace{\int_{\Omega}2(1-k)H\bm N_\phi^{\mathrm{T}} \mathrm{d}\Omega}_{\bm F_\phi^{ext}} = 0 
	\end{equation}

\noindent where $\bm F_u^{ine}$, $\bm F_u^{int}$, and $\bm F_u^{ext}$ are the inertial, internal, and external forces for the displacement field and $\bm F_\phi^{int}$ and $\bm F_\phi^{ext}$ are the internal and external force terms of the phase field. Additionally, the mass and stiffness matrices follow
	\begin{equation}
	\left\{\begin{aligned}\bm M &= \int_{\Omega}\rho\bm N_u^\mathrm{T}\bm N \mathrm{d}\Omega\\
\bm K_u &= \int_{\Omega} \bm B_u^\mathrm{T} \bm D_e \bm B_u \mathrm{d}\Omega\\
	\bm K_\phi &= \int_{\Omega}\left\{\bm B_\phi^{\mathrm{T}} G_c l_0 \bm B_\phi +\bm N_\phi^{\mathrm{T}} \left [ \frac{G_c}{l_0} + 2(1-k)H \right ] \bm N_\phi \right \} \mathrm{d}\Omega
	\end{aligned}\right .
	\end{equation}

\subsection{Staggered method}

The relationship between all the modules established is shown in Fig. 2. The ``Storage Module'' stores the results obtained from the ``Solid Mechanics Module'', such as the magnitude of principal strains, direction of principal strain and elastic energy. The positive part of the elastic energy is then calculated and imported into the ``History-strain Module'' to solve and update the local history strain field. Then the updated history strain is used to solve the phase-field. The updated phase-field solution and the previously stored principal strains and their corresponding directions are used to modify the stiffness in the  ``Solid Mechanics Module'' and then update the solution for the mechanical field. Fig. 2 shows the coupling for the solution of each module. Based on this, we employ a staggered scheme to solve the coupled system of equations as indicated in Fig. 3. Thus, the Newton-Raphson approach is adopted to obtain the residual of the discrete equations $\bm R_{u}=\bm F_u^{ext} - \bm F_u^{ine} - \bm F_u^{int} = 0 $ and $ \bm R_{\phi}=\bm F_\phi^{ext} - \bm F_\phi^{int}=0$, respectively. 

For the staggered time integration scheme, the equations of displacement, history strain and phase-field are solved independently. To obtain unconditional stability for the calculation, we use the implicit Generalized-$\alpha$ method \citep{borden2012phase}. When the time comes to a new value $t_i$, a new guess for the three field variables ($\bm u_i^{j=0}$, $H_i^{j=0}$ and $\phi_i^{j=0}$) is made first based on the results that have been solved in previous time steps. That is, linear extrapolation of the previous solution is used to construct the initial guess for the nonlinear system of equations to be solved at the present time step. For the given time step $i$ and iteration step $j+1$, the displacement $\bm u_i^{j+1}$ is first solved based on one Newton-Raphson iteration by using the results ($\bm u_i^j$, $H_i^j$ and $\phi_i^j$) of previous iteration step $j$. Using the updated displacements $\bm u_i^{j+1}$, the equation concerning history strain is then solved based on another Newton-Raphson iteration. Subsequently, the phase-field $\phi_i^{j+1}$ is obtained by the updated $\bm u_i^{j+1}$, $H_i^{j+1}$ and also a Newton-Raphson iteration. We finally compare the total relative error between the solution in previous and present iteration steps. If the error is less than the tolerance $\varepsilon_t$, the calculation is finished for current time step and will switch to the next step. Otherwise, the calculation will go through another iteration process until the tolerance requirement is satisfied. We choose the tolerance $\varepsilon_t=1\times 10^{-6}$ for our simulation. Thus, we succeed in obtaining all the solutions in the whole time domain by the implicit staggered time integration scheme.

It should be noted here that the iteration is slow to converge and more iteration steps are required when the material starts to fracture. Therefore, Anderson acceleration, a nonlinear convergence acceleration method that uses information from previous Newton iterations, is used to accelerate convergence \citep{comsol2005comsol}. The dimension of iteration space field is chosen as more than 50 to control the number of iteration increments in our work. In addition, we take standard Lagrangian elements (see the examples in the following section) to discretize the space domain for the three physical fields. Finally, Fig. 4 gives the flow chart of our implementation of phase-field method for crack problems in COMSOL. Our original codes can be downloaded from "https://sourceforge.net/projects/phasefieldmodelingcomsol/".
	
	\begin{figure}[htbp]
	\centering
	\includegraphics[width = 12cm]{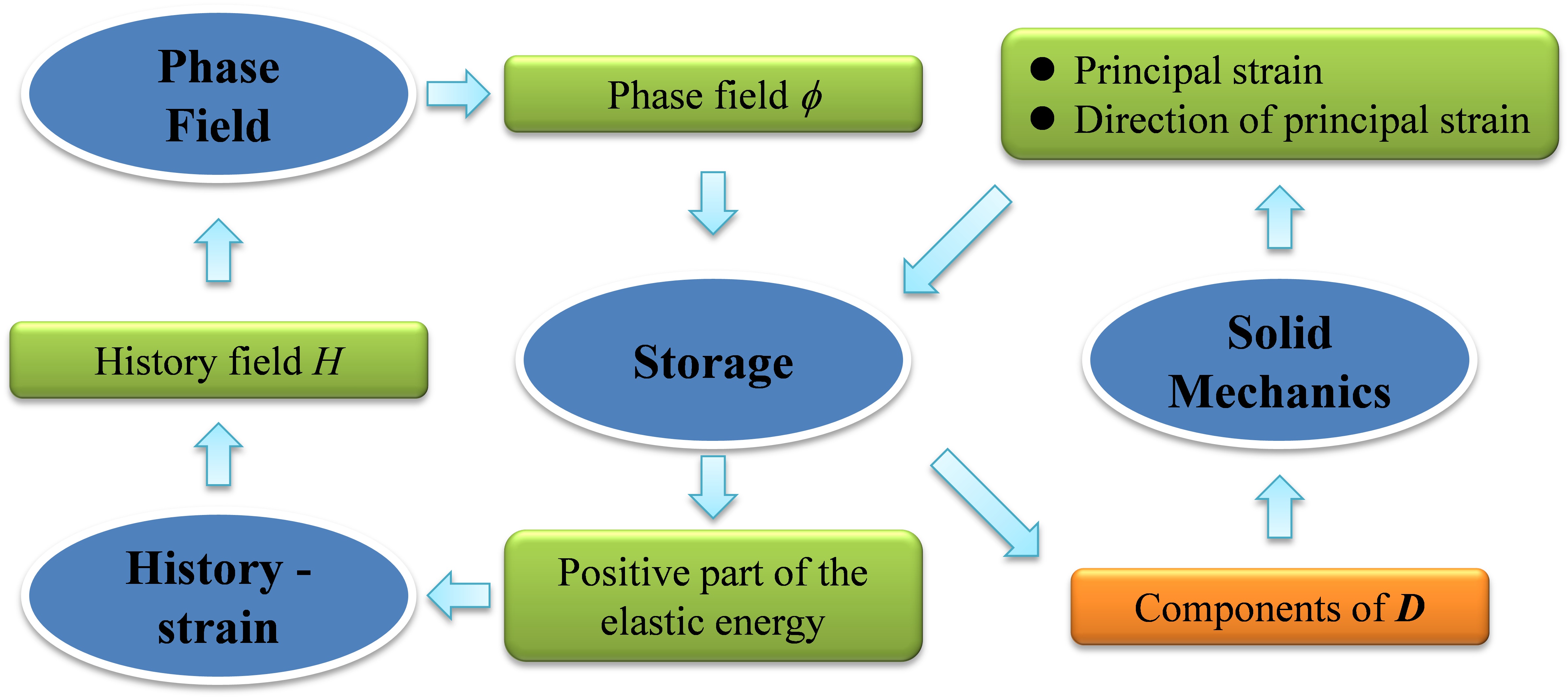}
	\caption{Relationship between all the modules established}
	\label{Figure2}
	\end{figure}

	\begin{figure}[htbp]
	\centering
	\includegraphics[width = 15cm]{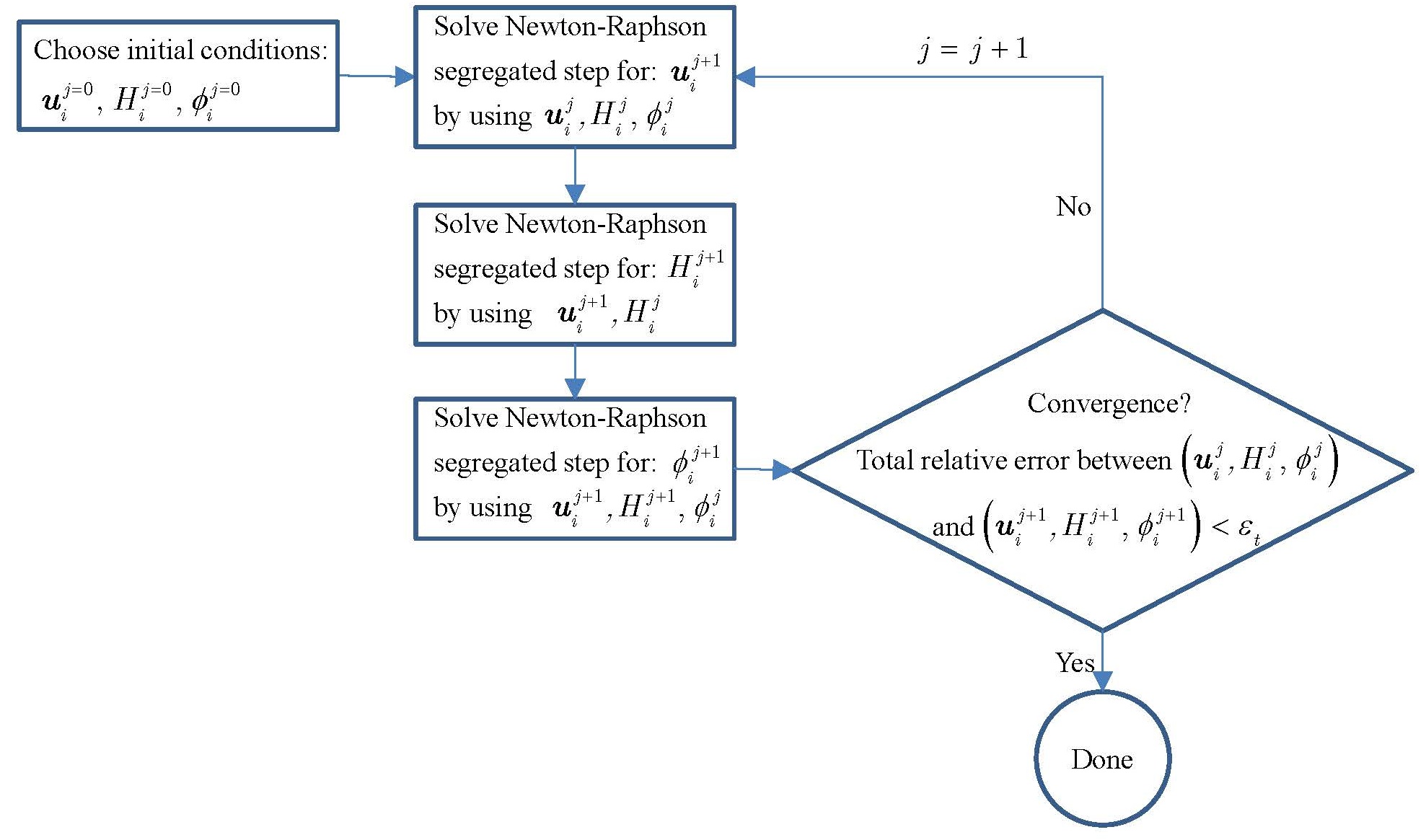}
	\caption{Staggered scheme for the coupled calculation in phase-field modeling}
	\label{Figure3}
	\end{figure}
	
	\begin{figure}[htbp]
	\centering
	\includegraphics[width = 10cm]{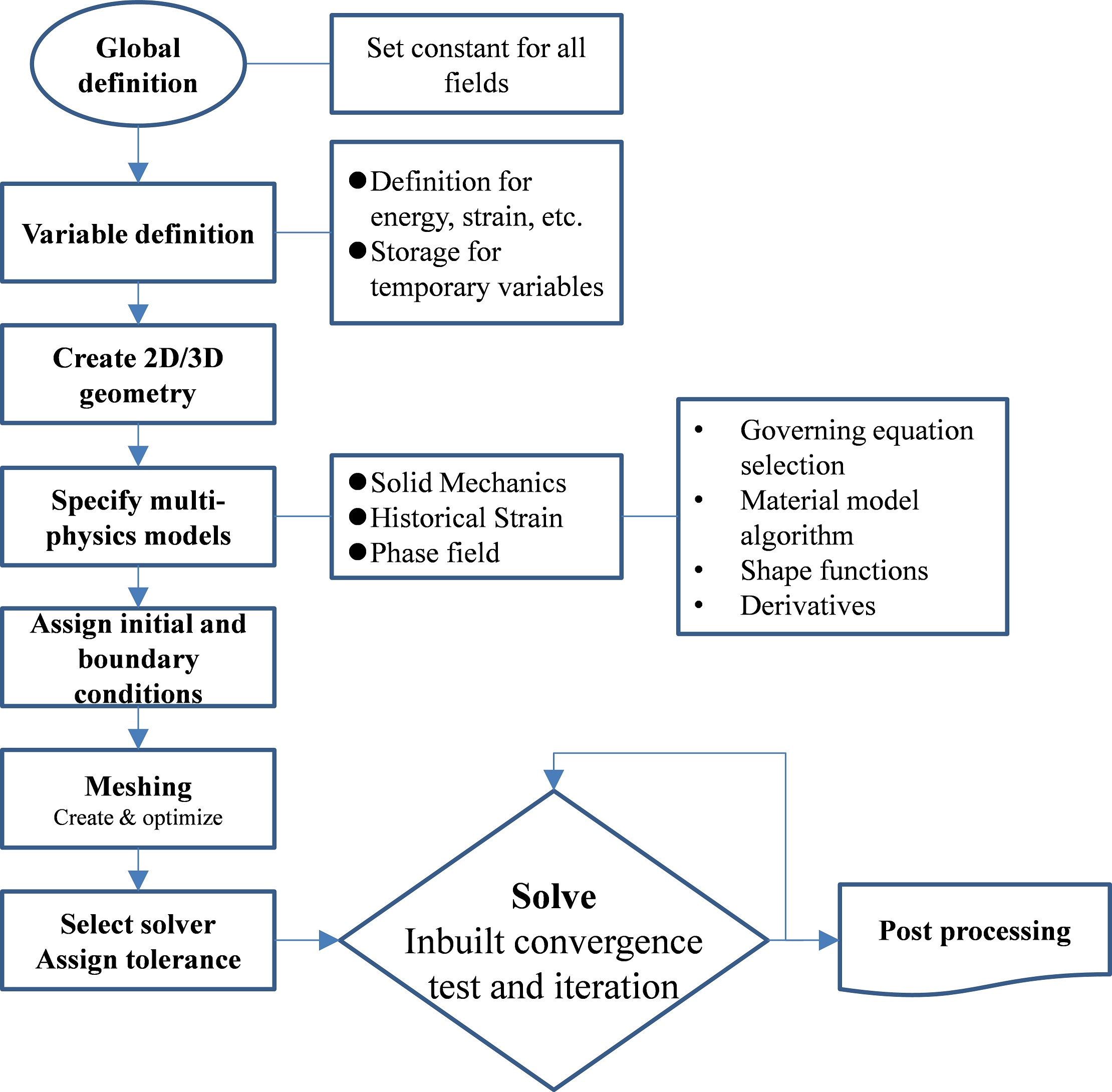}
	\caption{COMSOL implementation of phase-field method for crack problems}
	\label{Figure4}
	\end{figure}

\section {Numerical examples}
In this section, we present several quasi-static and dynamic benchmark problems testing the influence of the length-scale parameter $l_0$, the element size, load and time step sizes as well as the critical energy release rate $G_c$ on the numerical results.

\subsection {2D notched square plate subjected to tension}

\begin{figure}[htbp]
\centering
\includegraphics[width = 6cm]{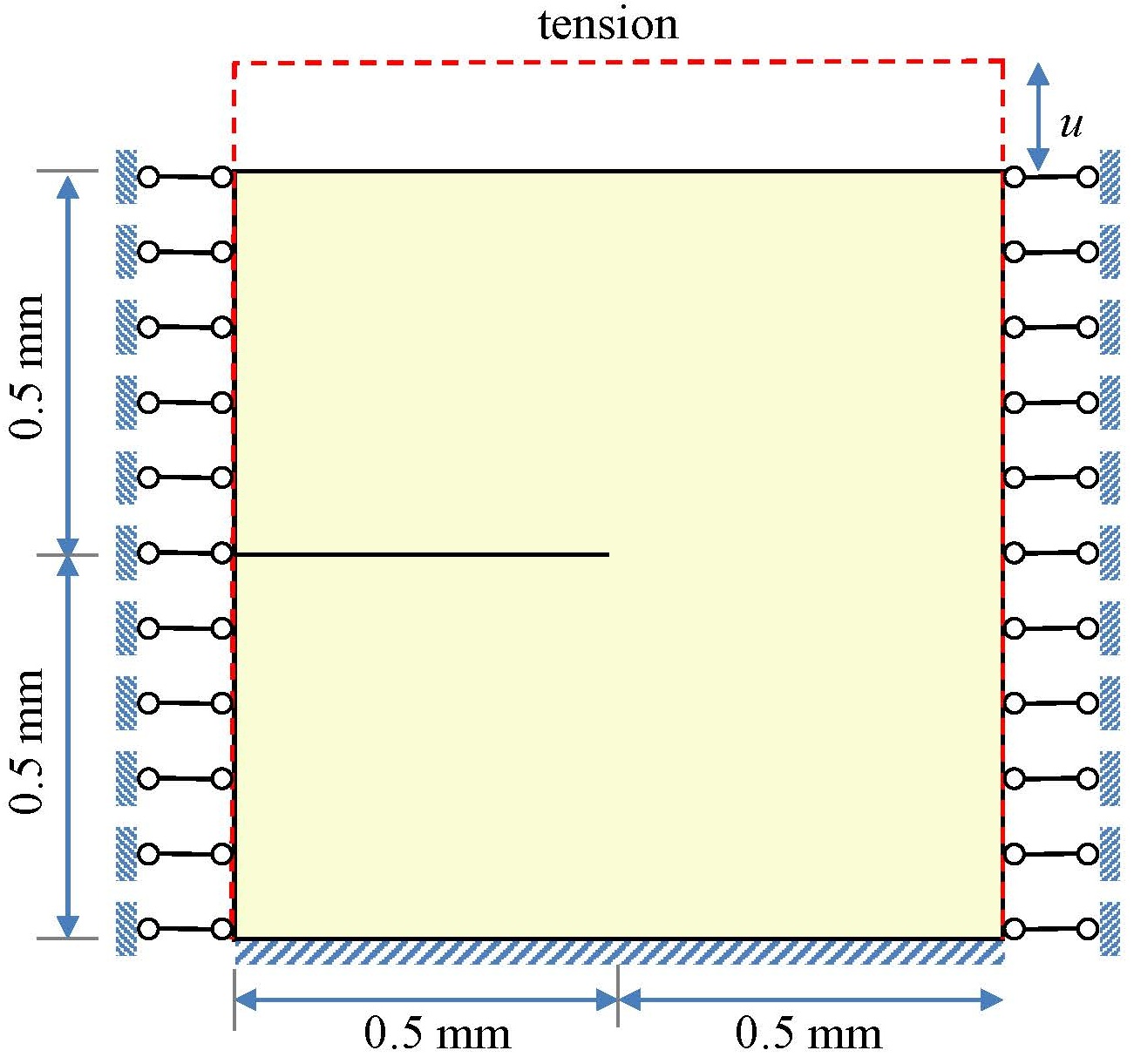}
\caption{Geometry and boundary condition of the single-edge-notched square plate subjected to tension}
\label{Geometry and boundary condition of the single-edge-notched square plate subjected to tension}
\end{figure}

Consider a square plate with an initial notch subjected to static tension loading. This benchmark test has been calculated and analyzed by Miehe et al. \citep{miehe2010phase,miehe2010thermodynamically} and \citet{hesch2014thermodynamically}. The geometry and loading condition are shown in Fig. \ref{Geometry and boundary condition of the single-edge-notched square plate subjected to tension}. A vertical displacement $u_y$ is applied on the upper boundary of the plate with $u_x=0$.  The material parameters are: $E$ = 210 GPa, $\nu$ = 0.3, and $G_c$ = 2700 J/m$^2$. We choose $k=1\times 10^{-9}$ to avoid a singular stiffness matrix. The length parameter $l_0=7.5\times 10^{-3}$ mm and $1.5\times10^ {-2}$ mm, respectively. Plane strain conditions are assumed. The domain is discretized with 64516 Q4 (4 node quadrilateral) elements (with bi-linear shape functions). The element size $h$ is around $3.96\times 10^ {-3}$ mm yielding $l_0=2h$ and $l_0=4h$ for $l_0=7.5\times10^ {-3}$ mm and $1.5\times10^ {-2}$ mm, respectively.

\begin{figure}[htbp]
\centering
\subfigure[]{\includegraphics[width = 5cm]{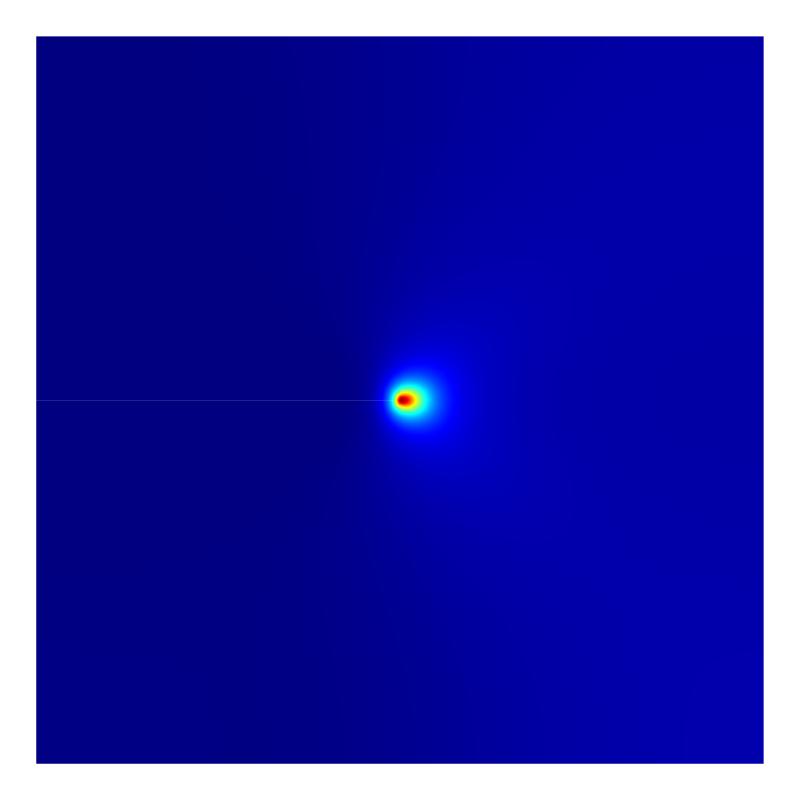}}
\subfigure[]{\includegraphics[width = 5cm]{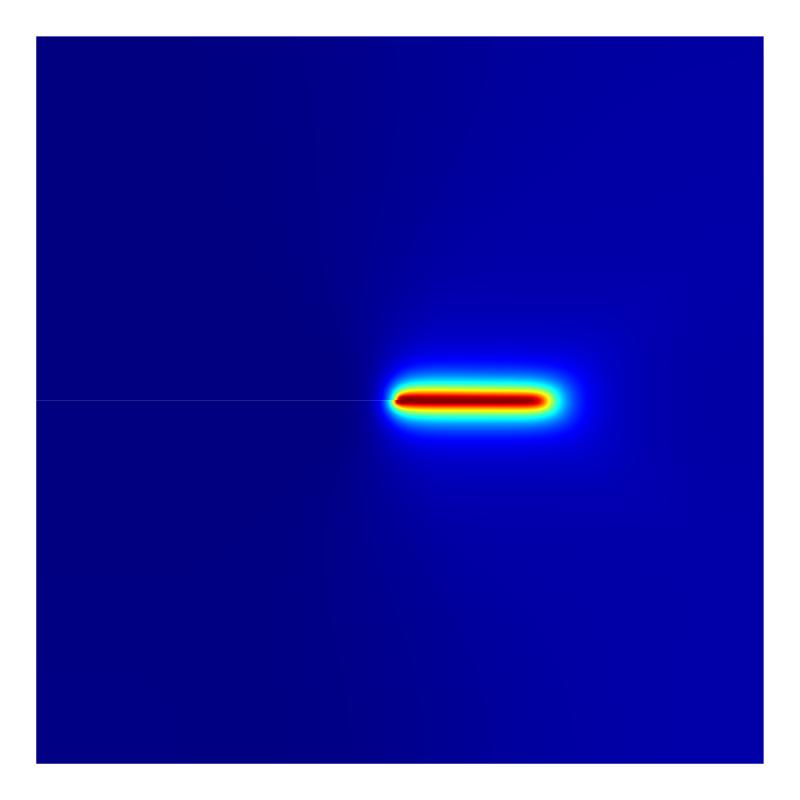}}
\subfigure[]{\includegraphics[width = 5cm]{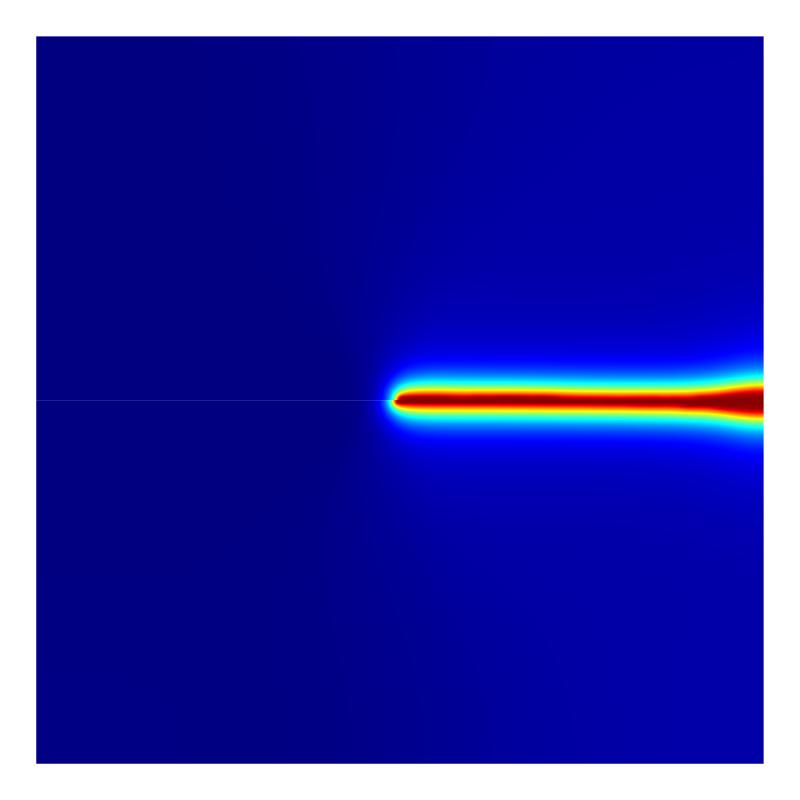}}

\subfigure[]{\includegraphics[width = 5cm]{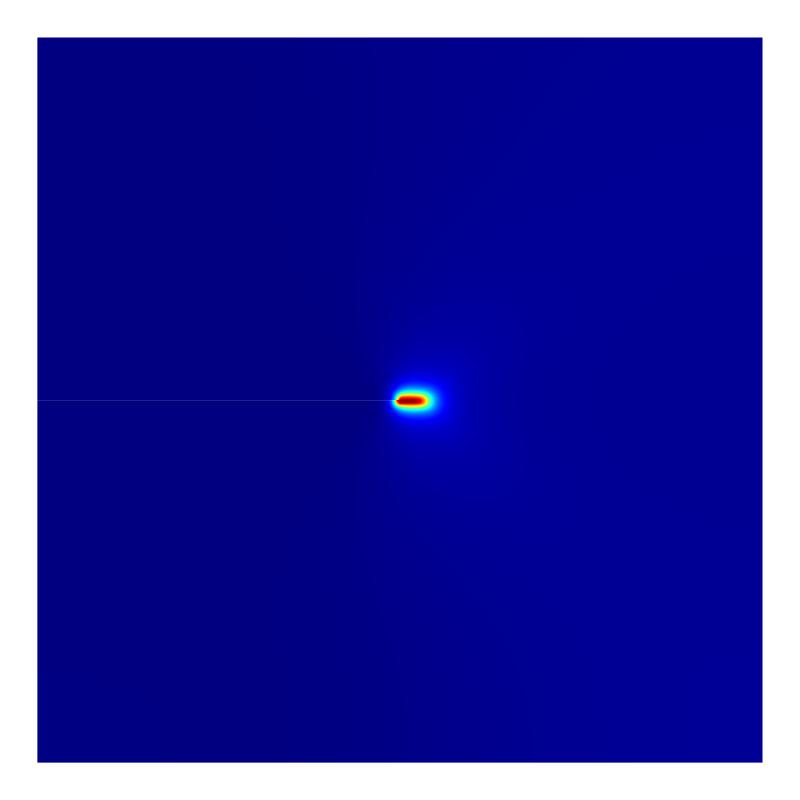}}
\subfigure[]{\includegraphics[width = 5cm]{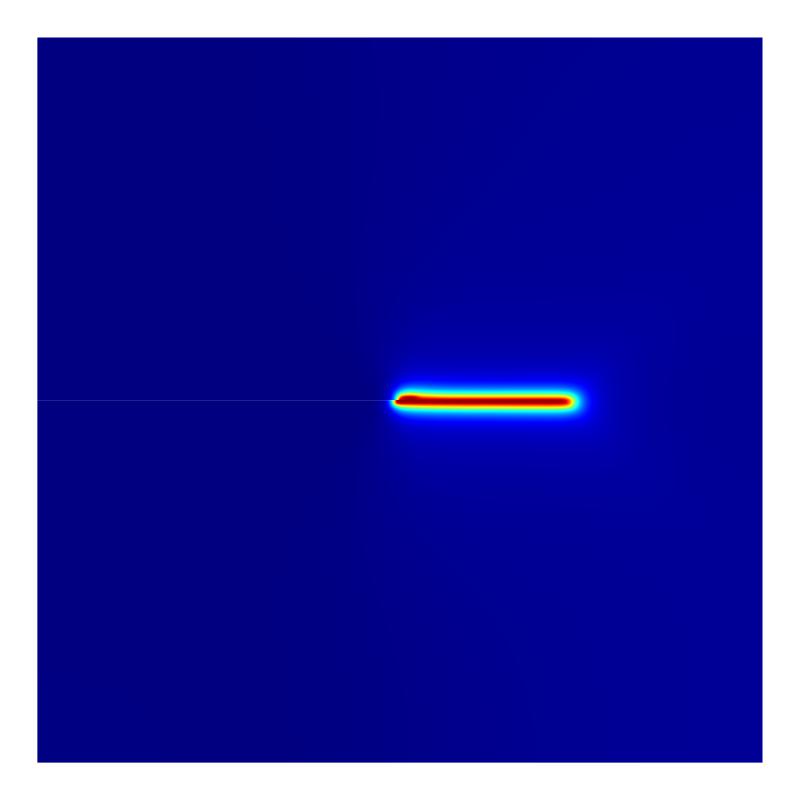}}
\subfigure[]{\includegraphics[width = 5cm]{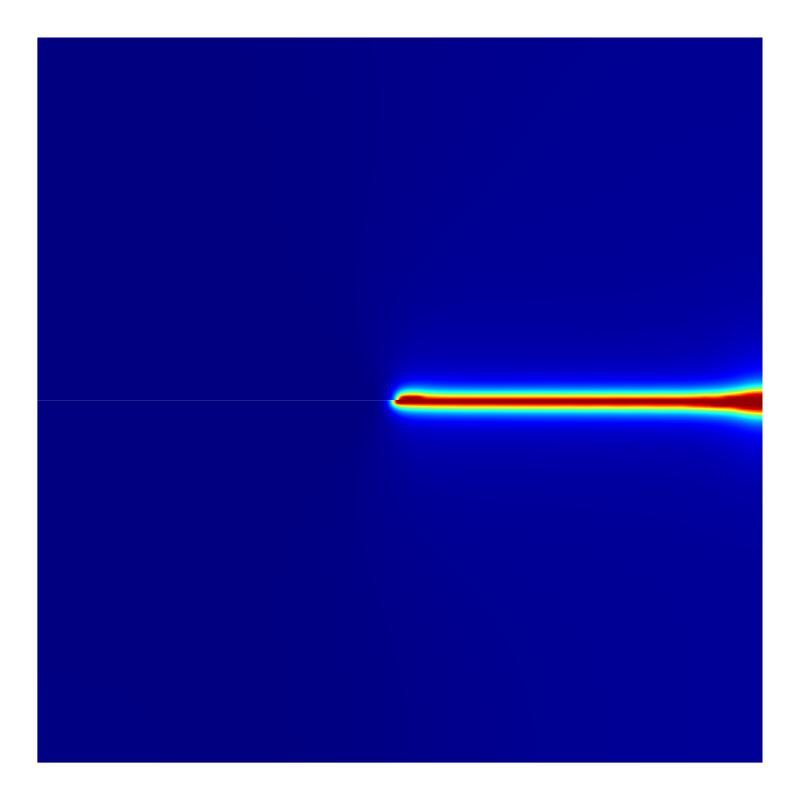}}

\caption{2D single-edge-notched square subjected to tension. Crack pattern at a displacement of (a) $u = 5.3\times10^{-3}$ mm, (b) $u = 5.5\times10^{-3}$ mm, (c) $u = 5.8\times10^{-3}$ mm for a length scale $l_0$ of $1.5\times10^{-2}$ mm and (d) $u = 5.55\times10^{-3}$ mm, (e) $u = 5.9\times10^{-3}$ mm, and (f) $u = 6.25\times10^{-3}$ mm for a length scale $l_0$ of $7.5\times10^{-3}$ mm.}
\label{2D single-edge-notched square subjected to tension Crack pattern}

\end{figure}

\begin{figure}[htbp]
	\centering
	\includegraphics[width = 8cm]{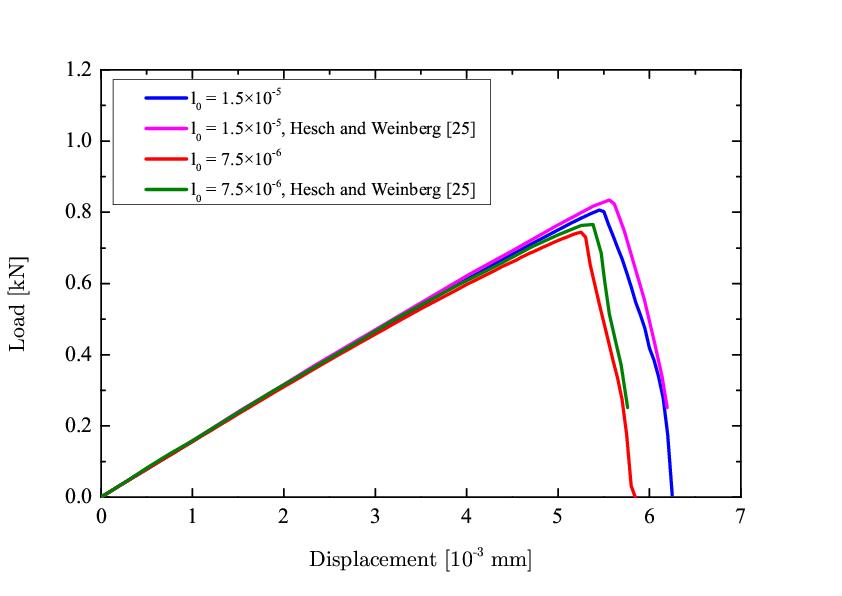}
	\caption{Load-displacement curves of the 2D single-edge-notched tension test}
	\label{Load-displacement curves of the 2D single-edge-notched tension test}
	\end{figure}

We apply an displacement increment of $\Delta u = 1\times10^{-5}$ mm for the first 450 time steps. Then, a displacement increment of  $\Delta u = 1\times10^{-6}$ mm is chosen for the remaining time steps. We obtain the crack patterns at different displacements for the two fixed length scale parameters $l_0$, as shown in Fig. \ref{2D single-edge-notched square subjected to tension Crack pattern}. As expected, the crack is less diffused for a smaller length scale parameter. The presented crack patterns are the same as those reported by \citet{miehe2010phase, miehe2010thermodynamically, hesch2014thermodynamically, liu2016abaqus}. The cracks propagate in horizontal direction. The load-displacement curves on the top boundary of the plate are shown in Fig. \ref{Load-displacement curves of the 2D single-edge-notched tension test} in comparison with the results by \citet{hesch2014thermodynamically}. The loads obtained by this work are in good agreement with those by \citet{hesch2014thermodynamically} with the increase in the vertical displacement. A minor difference exists due to the different algorithm used in both methods. For a total of 453,520 degrees of freedom, COMSOL required 4 h 20 min ($l_0 =1.5\times10^ {-2}$ mm) and 4 h 11 min ($l_0=7.5\times 10^{-3}$ mm) on two I5-6200U CPUs. 

\begin{figure}[htbp]
	\centering
	\subfigure[$l_0=1.5\times10^{-2}$ mm]{\includegraphics[width = 8cm]{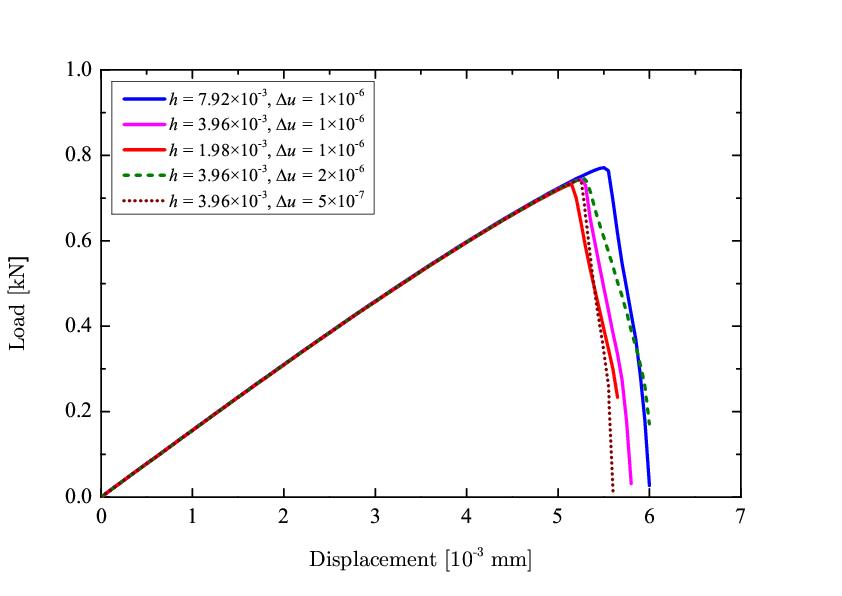}}
	\subfigure[$l_0=7.5\times10^{-3}$ mm]{\includegraphics[width = 8cm]{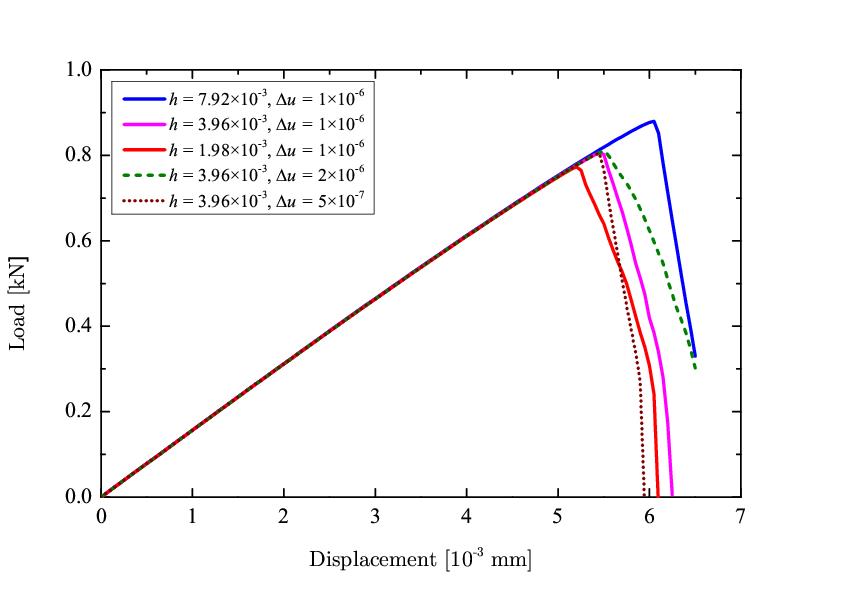}}
	\caption{2D single-edge-notched square subjected to tension. Influence of mesh size $h$ and displacement increment $\Delta u$ on the load-displacement curves.}
	\label{2D single-edge-notched square subjected to tension Influence of mesh size}
	\end{figure}

We also perform the simulation by changing the mesh size $h$ to $7.92\times10^{-3}$ and $1.98\times10^{-3}$ mm and the displacement increment $\Delta u$ to $2\times10^{-6}$ and $5\times10^{-7}$ mm in the remaining time steps. The results show that mesh size and displacement increment have no effect on the crack pattern. Figure \ref{2D single-edge-notched square subjected to tension Influence of mesh size} represents the load-displacement curves for different mesh sizes and displacement increments. A larger mesh size leads to a larger peak load. As \citet{miehe2010phase} have suggested a mesh size $h\le 0.5l_0$ to obtain a precise crack topology, the mesh size $h=7.92\times 10^{-3}$ mm achieves a much larger peak load than the other mesh sizes for $l_0=7.5\times10^{-3}$ mm. In addition, a much steeper post-peak stage can be seen for a smaller displacement increment.

\subsection{3D notched square plate subject to tension}

We now extend the benchmark problem in section 4.1 to 3D. The geometry of the plate in $x-y$ plane is the same as that in Section 4.1 with a thickness of 0.05 mm in $z$ direction. The material parameters are identical to the 2D example and we present results for a length scale parameter of $l_0 = 1.5\times10^{-2}$,  $1.25\times10^{-2}$,  $1.0\times10^{-2}$ and  $7.5\times10^{-3}$ mm. All boundaries of the plate are fixed in the normal direction except the top boundary, which is subjected to a displacement of $u$  in the $y$ direction and is fixed in the $x$  and $z$  direction ($u_x=0$ and $u_z=0$).

\begin{figure}[htbp]
	\centering
	\subfigure[]{\includegraphics[width = 5cm]{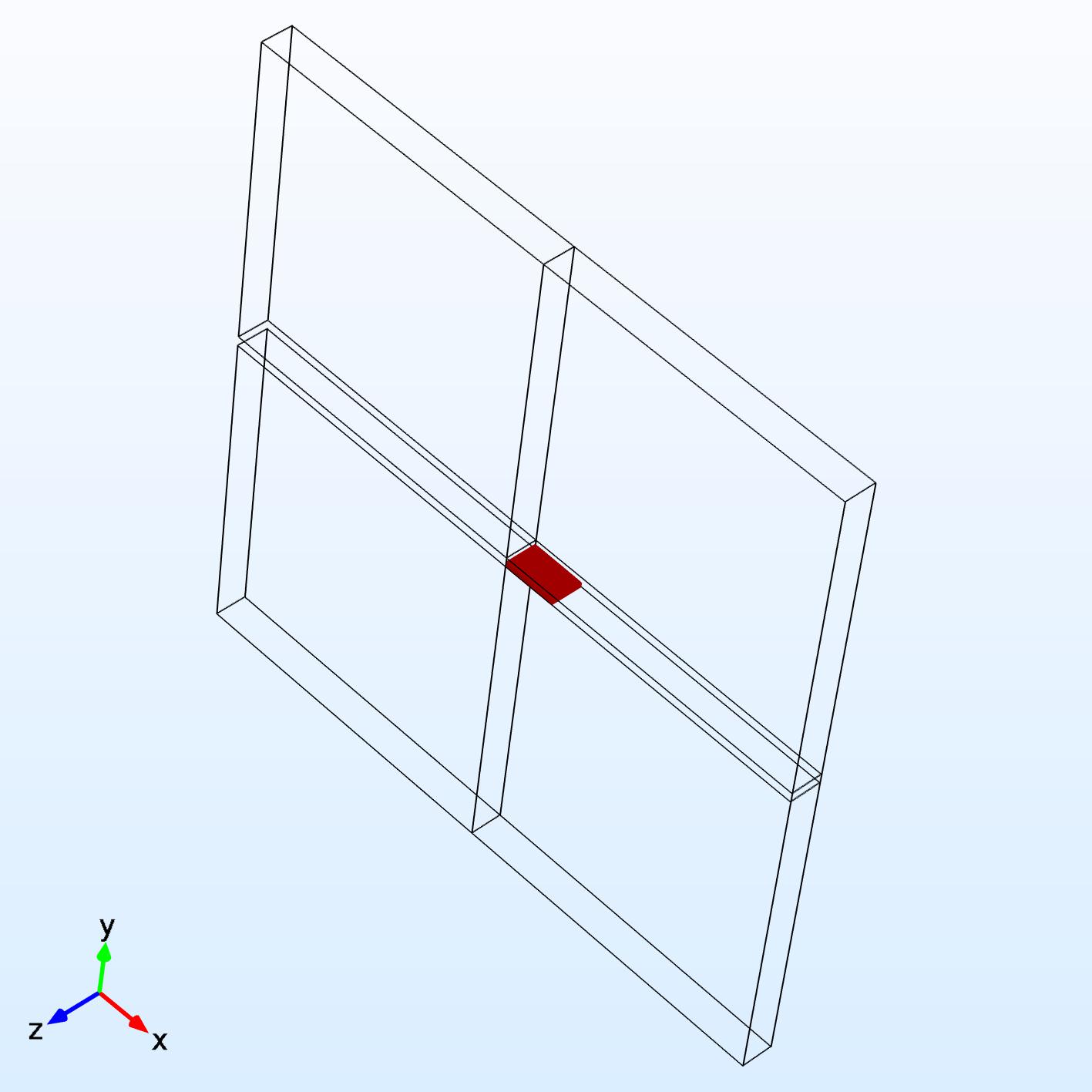}}
	\subfigure[]{\includegraphics[width = 5cm]{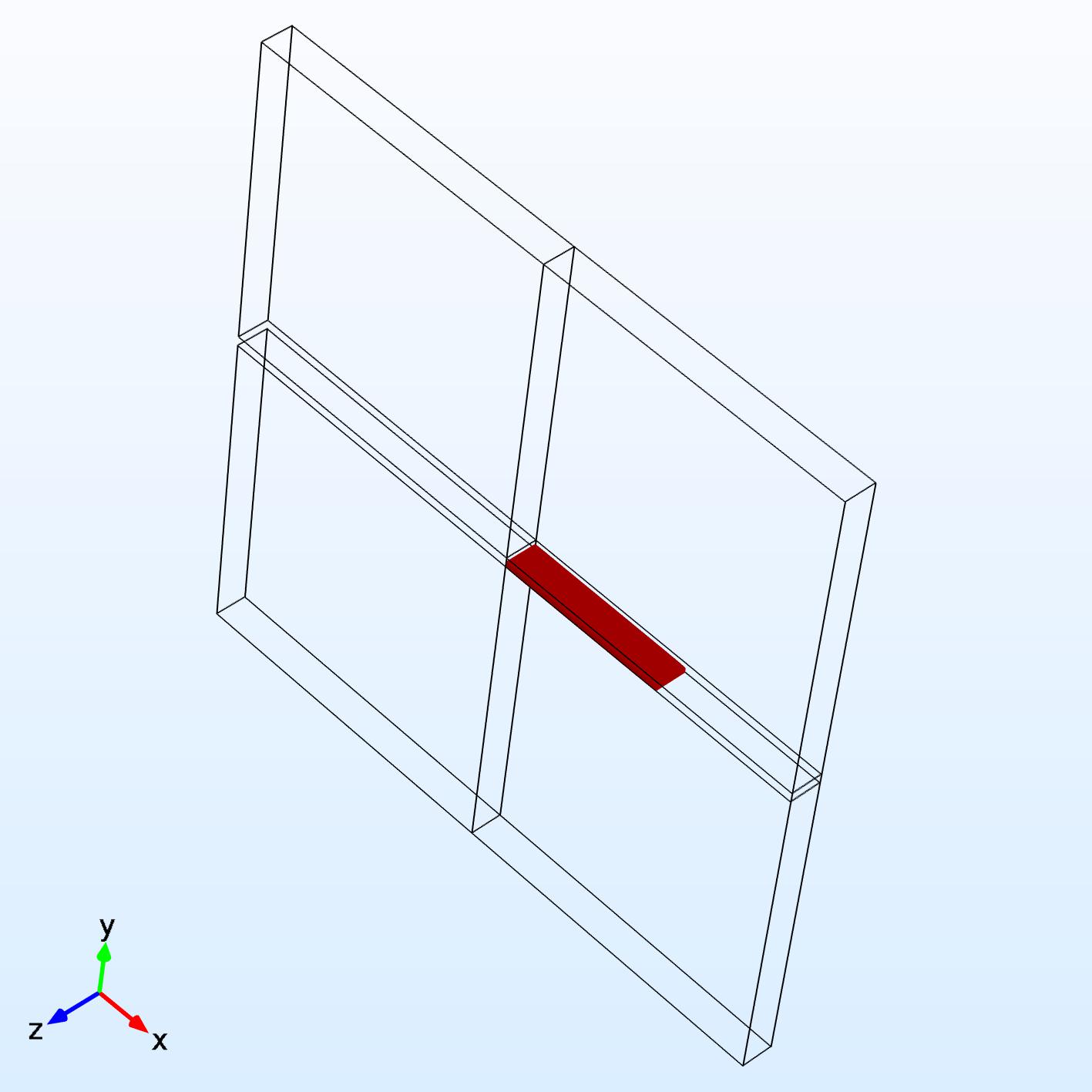}}
	\subfigure[]{\includegraphics[width = 5cm]{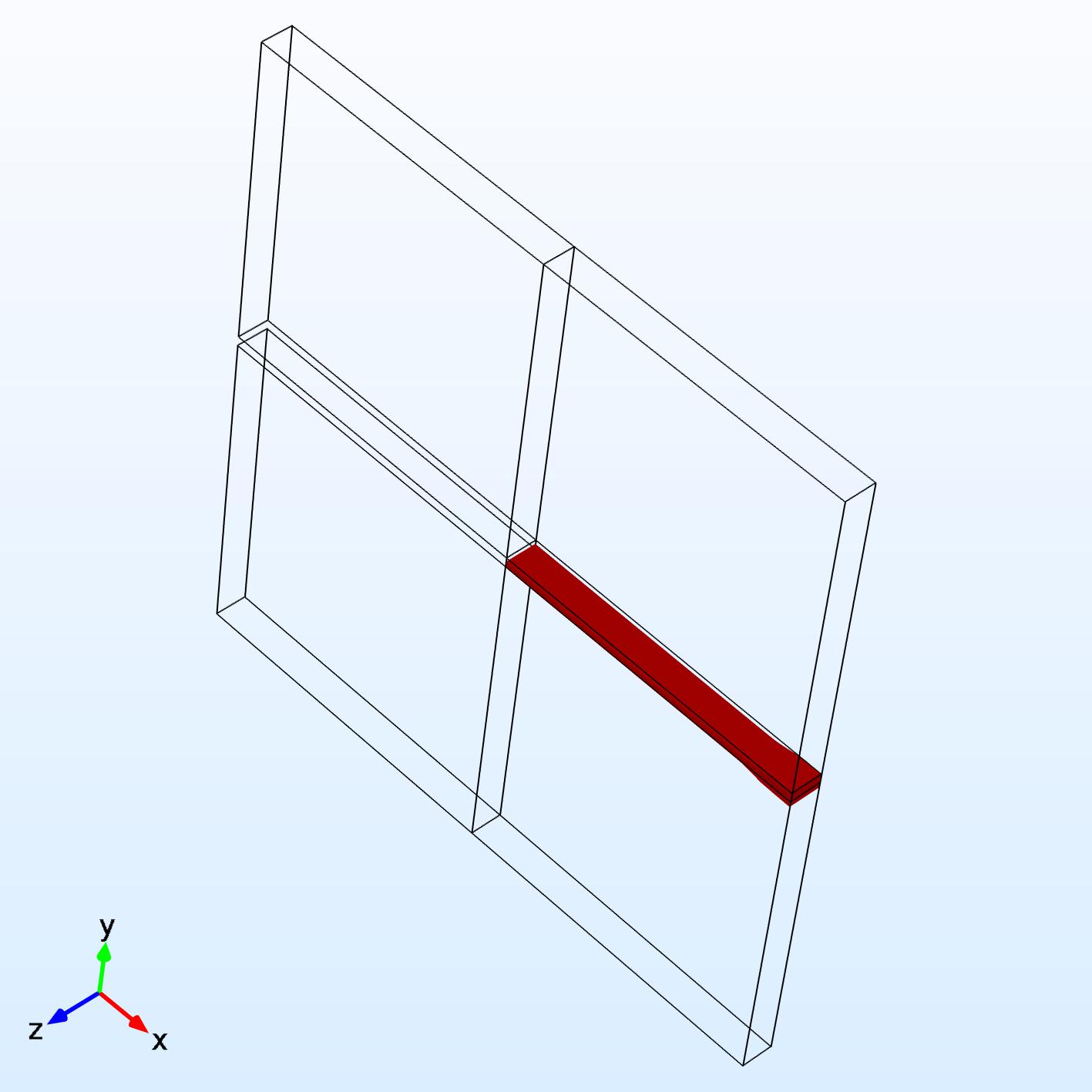}}
	
	\subfigure[]{\includegraphics[width = 5cm]{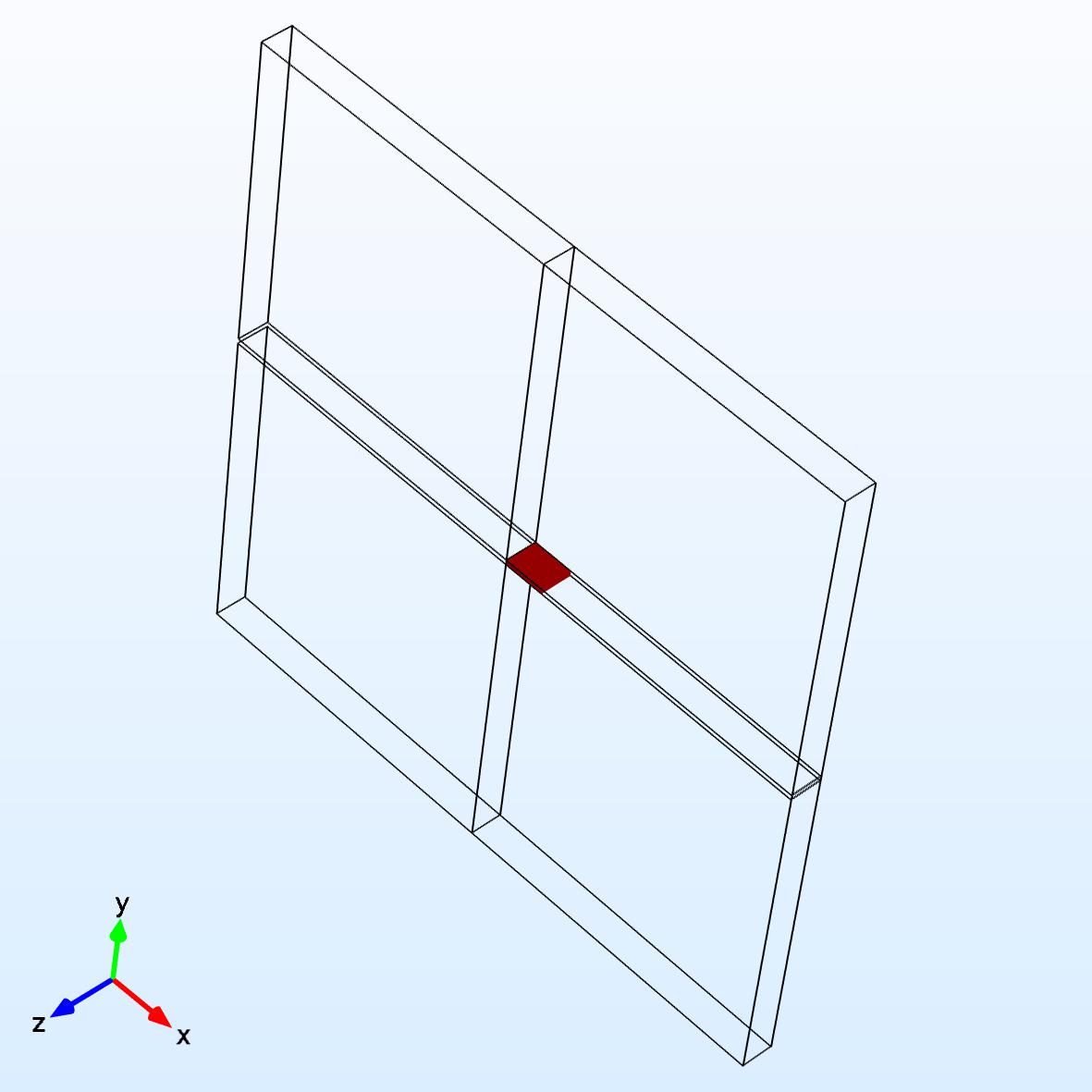}}
	\subfigure[]{\includegraphics[width = 5cm]{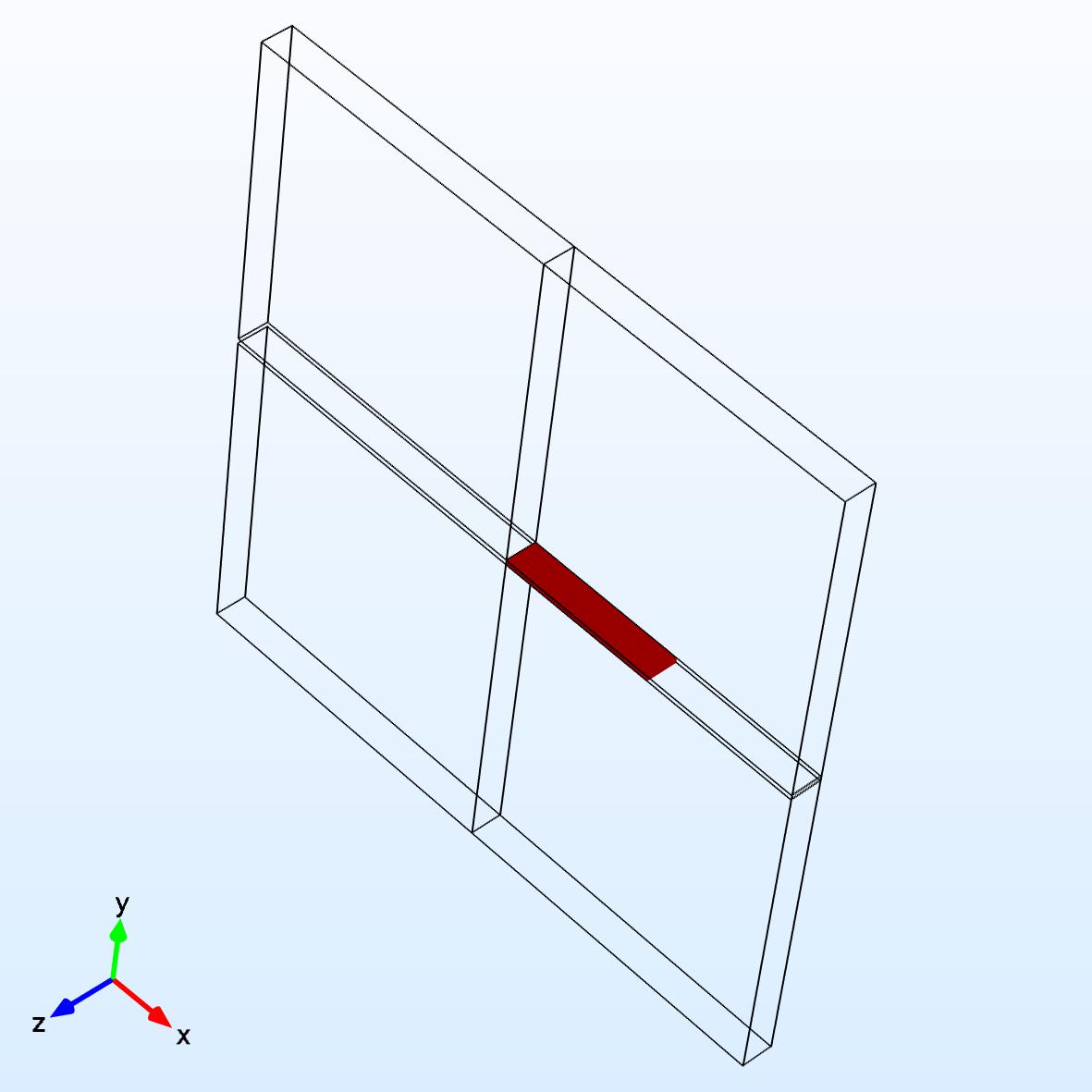}}
	\subfigure[]{\includegraphics[width = 5cm]{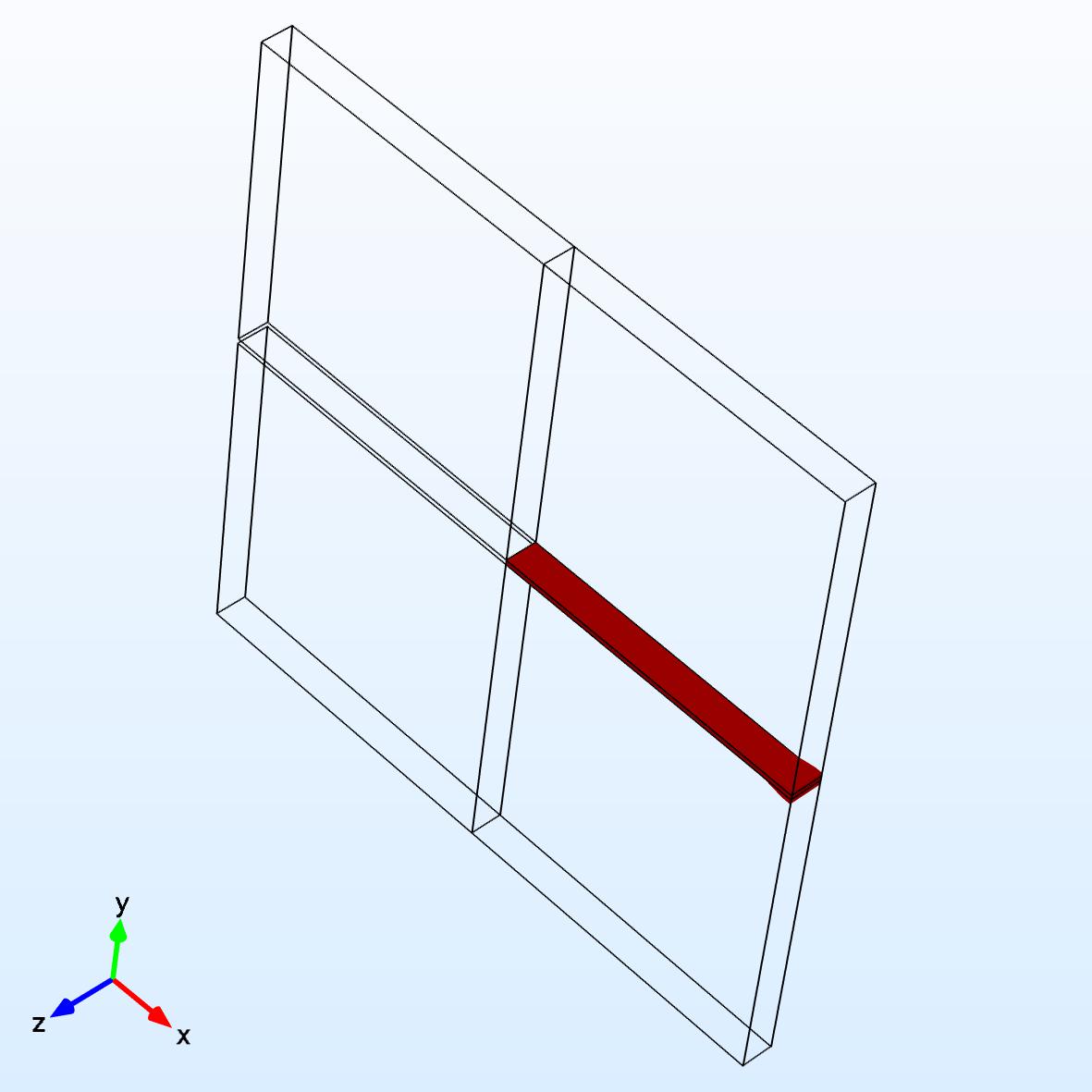}}
	\caption{3D single-edge-notched square plate subjected to tension. Crack pattern (display for $\phi>0.95$ at a displacement of (a) $u = 5.3\times10^{-3}$ mm, (b) $u = 5.6\times10^{-3}$ mm, (c) $u = 5.8\times10^{-3}$ mm for a length scale $l_0$ of $1.5\times10^{-2}$ mm and (d) $u = 5.7\times10^{-3}$ mm, (e) $u = 6.1\times10^{-3}$ mm, and (f) $u = 6.4\times10^{-3}$ mm for a length scale $l_0$ of $7.5\times10^{-3}$ mm.}
	\label{3D single-edge-notched square plate subjected to tension Crack pattern}
	
	\end{figure}

\begin{figure}[htbp]
	\centering
	\includegraphics[width = 8cm]{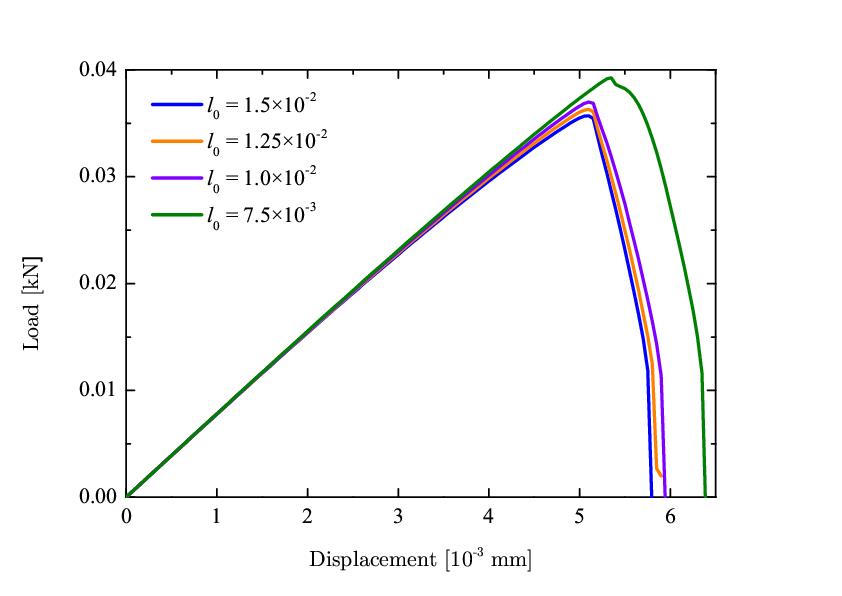}
	\caption{Load-displacement curves of the 3D single-edge-notched tension test}
	\label{Load-displacement curves of the 3D single-edge-notched tension test}
	\end{figure}

The plate is discretized with 8-node Lagrangian elements of the same size $h = 7.5\times10^{-3}$ mm without special refinement in the expected path for crack propagation. For the staggered scheme, we apply the displacement increment  $\Delta u = 1\times10^{-5}$ mm for the first 400 time steps and then adopt the displacement increment as $\Delta u  = 1\times10^{-6}$ mm for the remaining time steps. Our simulation show that different length scale parameters $l_0$ achieve the same crack pattern. Crack patterns for $\phi>0.95$  in the 3D simulation with $l_0  = 1.5\times10^{-2}$ and $7.5\times10^{-3}$ mm are shown in Fig.  \ref{3D single-edge-notched square plate subjected to tension Crack pattern}. The load-displacement curve for the top boundary of the plate is shown in Fig. \ref{Load-displacement curves of the 3D single-edge-notched tension test}. As observed, the crack patterns and load-displacement curve are quite similar to those of the 2D case.The peak load of the plate increases as the length scale parameter decreases. 

\subsection{2D notched square plate subjected to shear loading}

\begin{figure}[htbp]
	\centering
	\includegraphics[width = 8cm]{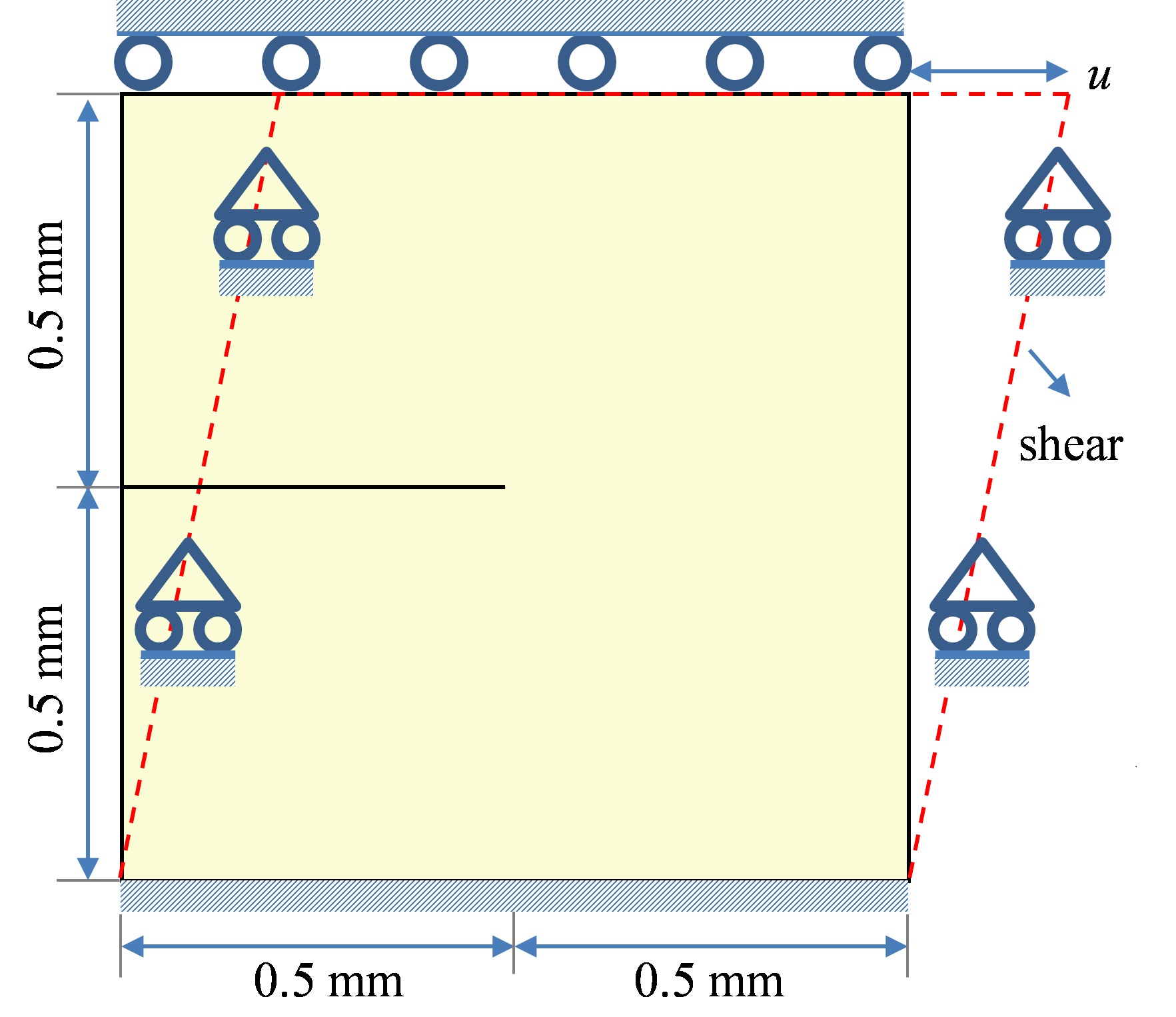}
	\caption{Geometry and boundary conditon of the single-edge-notched plate subjected to shear load}
	\label{Geometry and boundary conditon of the single-edge-notched plate subjected to shear load}
	\end{figure}

\begin{figure}[htbp]
	\centering
	\subfigure[]{\includegraphics[width = 5cm]{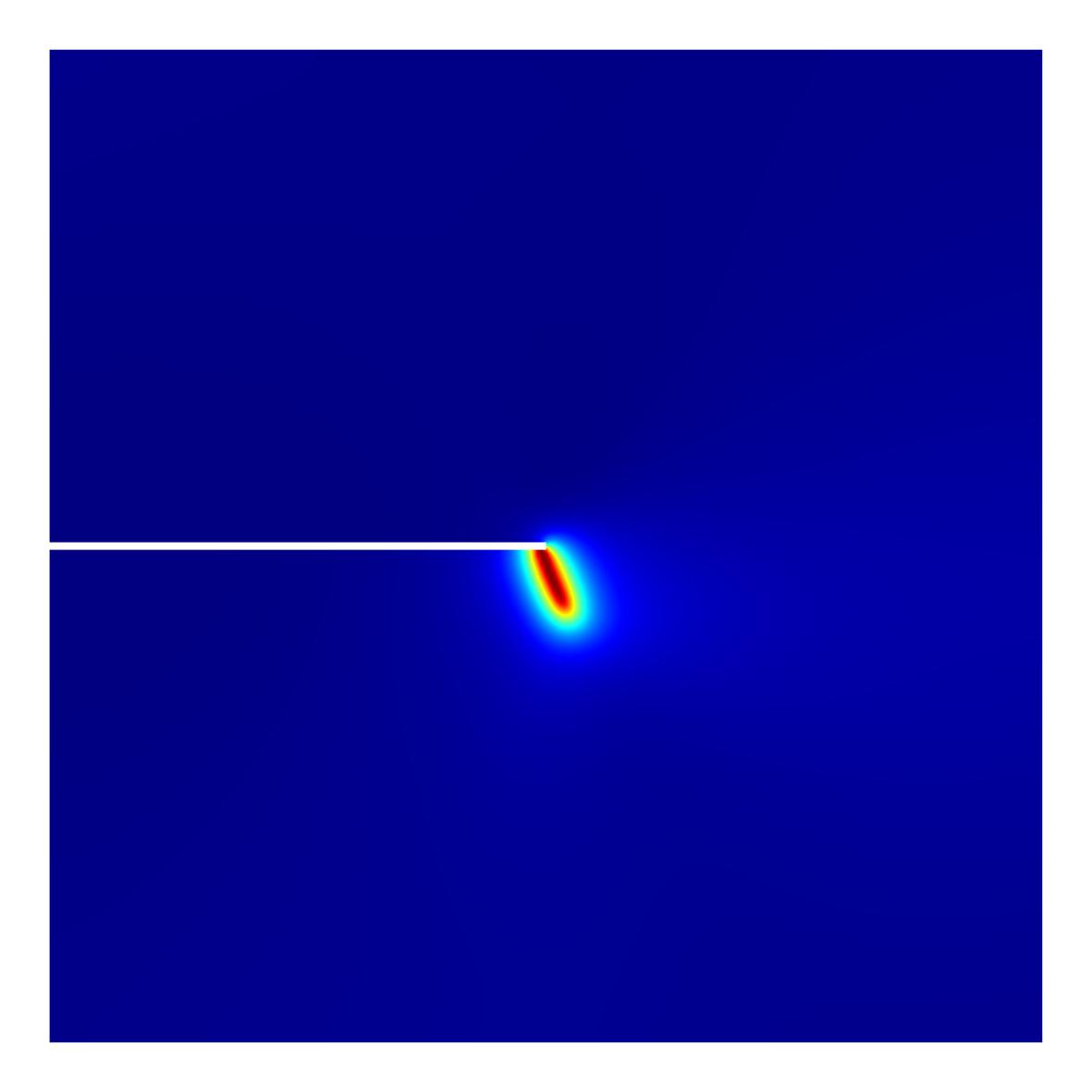}}
	\subfigure[]{\includegraphics[width = 5cm]{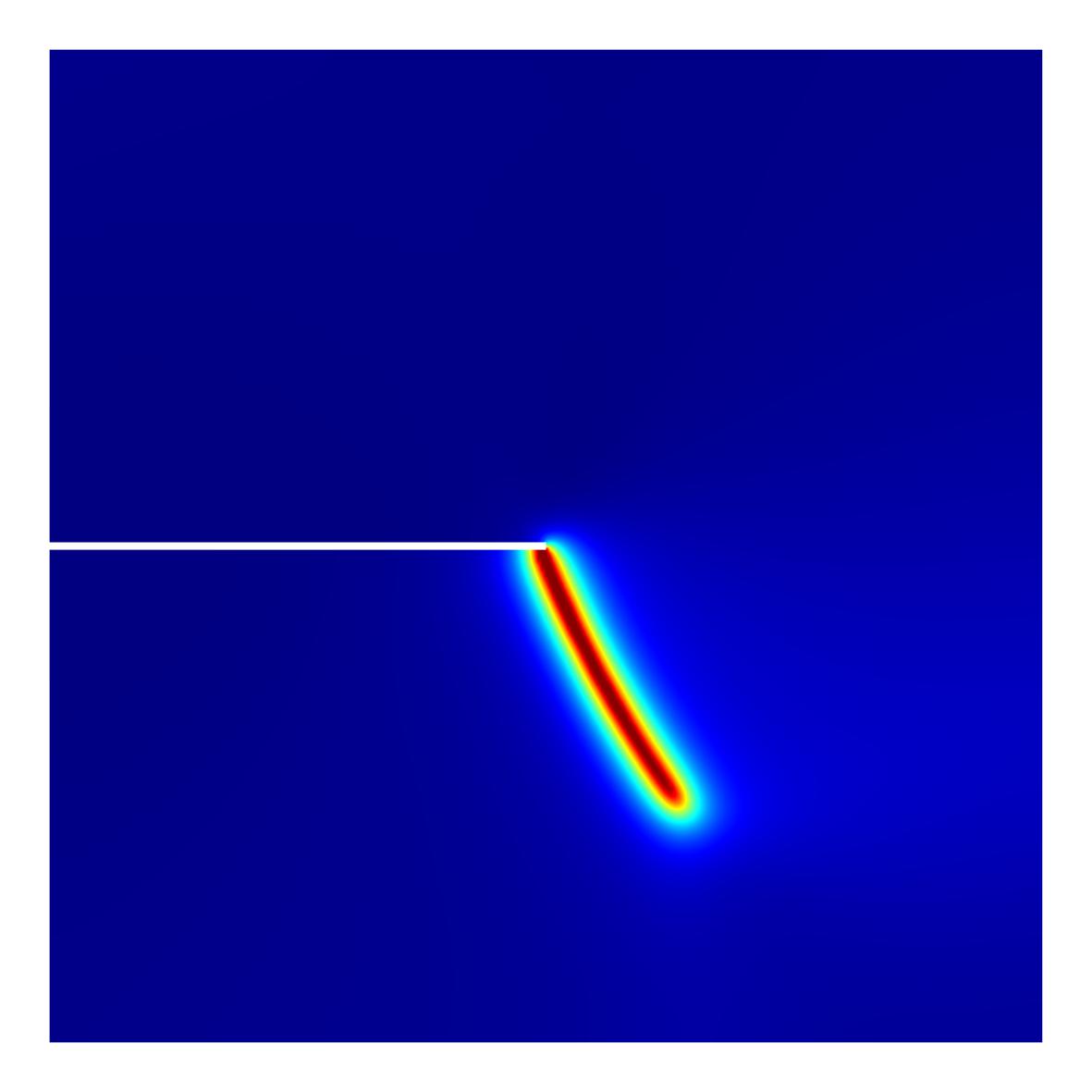}}
	\subfigure[]{\includegraphics[width = 5cm]{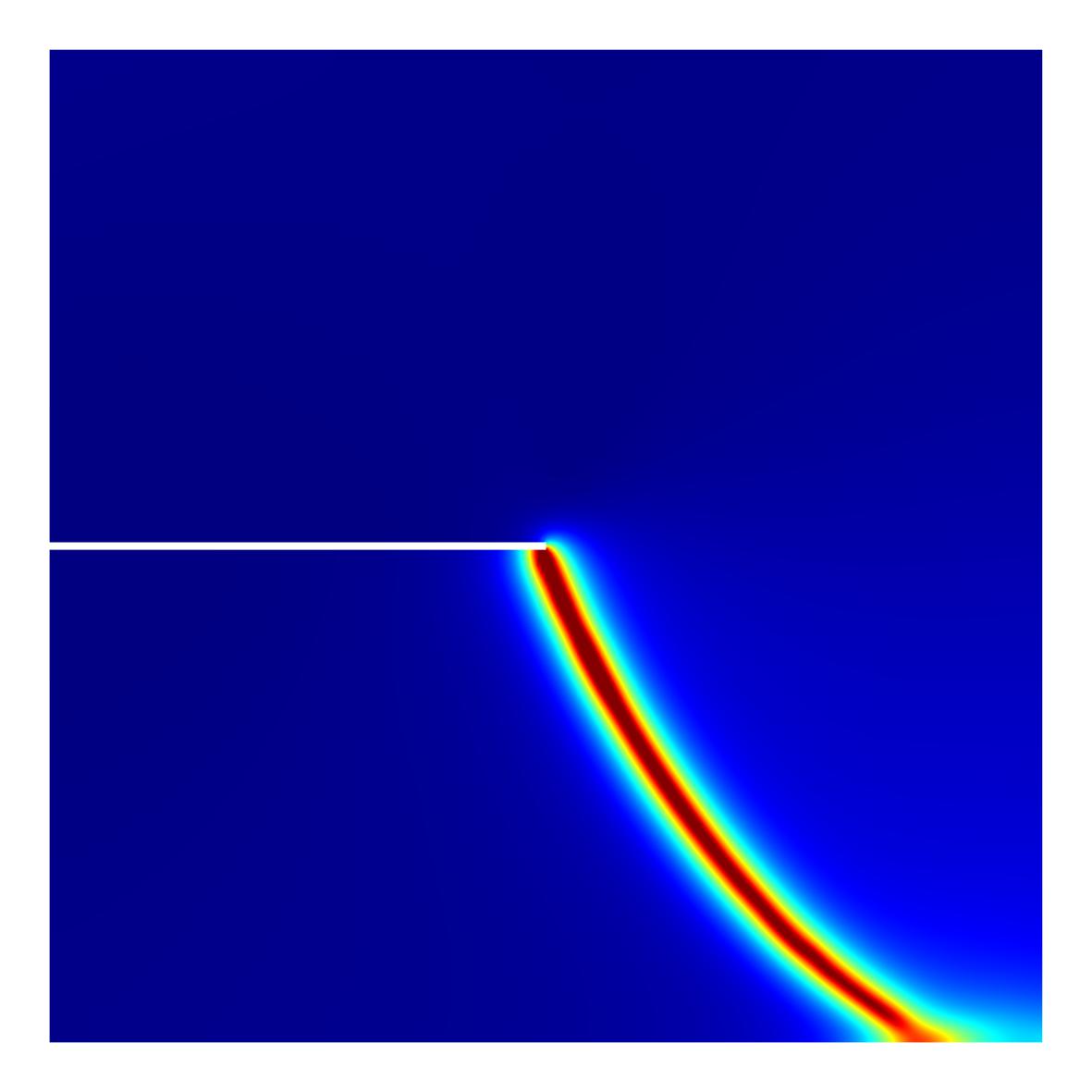}}
	
	\subfigure[]{\includegraphics[width = 5cm]{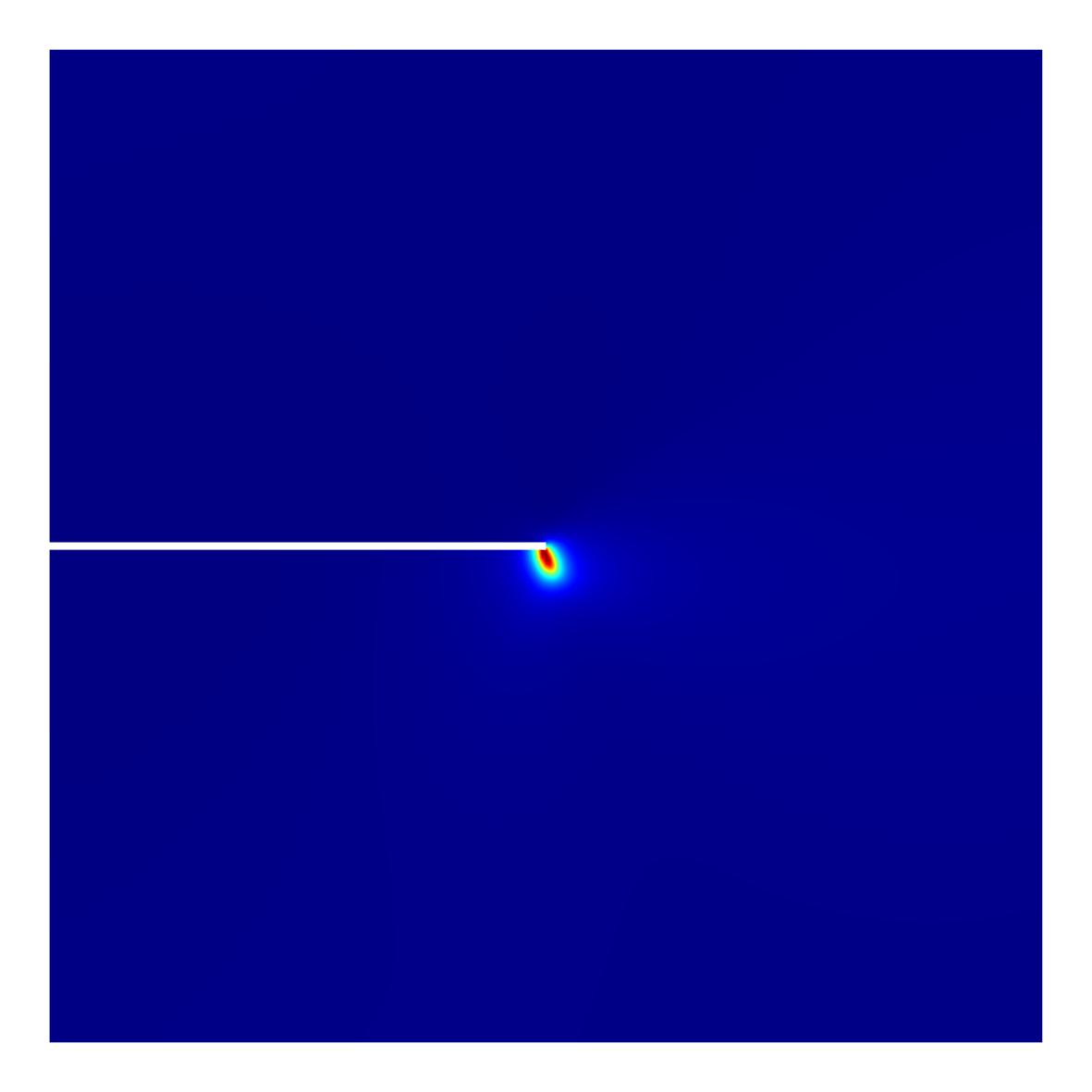}}
	\subfigure[]{\includegraphics[width = 5cm]{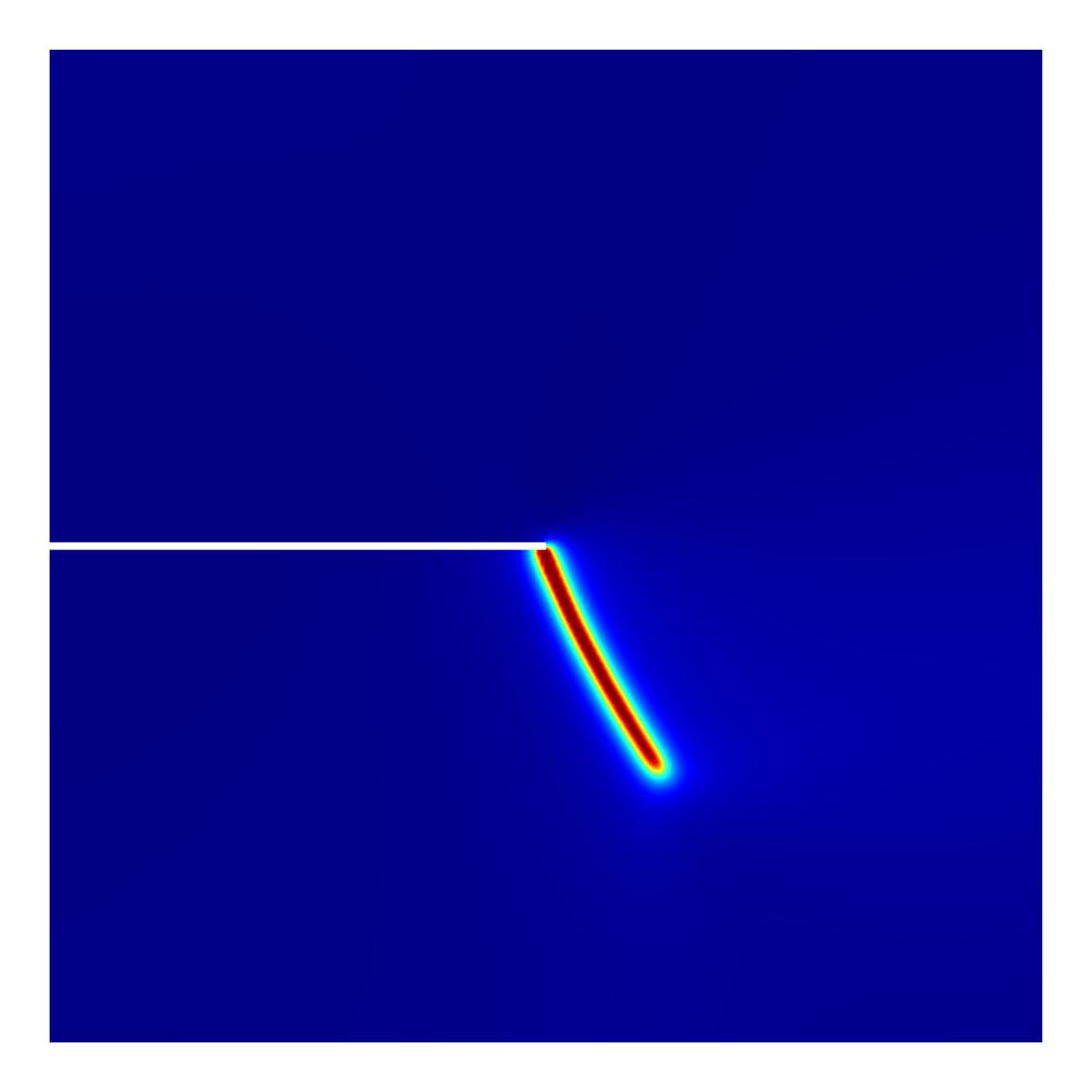}}
	\subfigure[]{\includegraphics[width = 5cm]{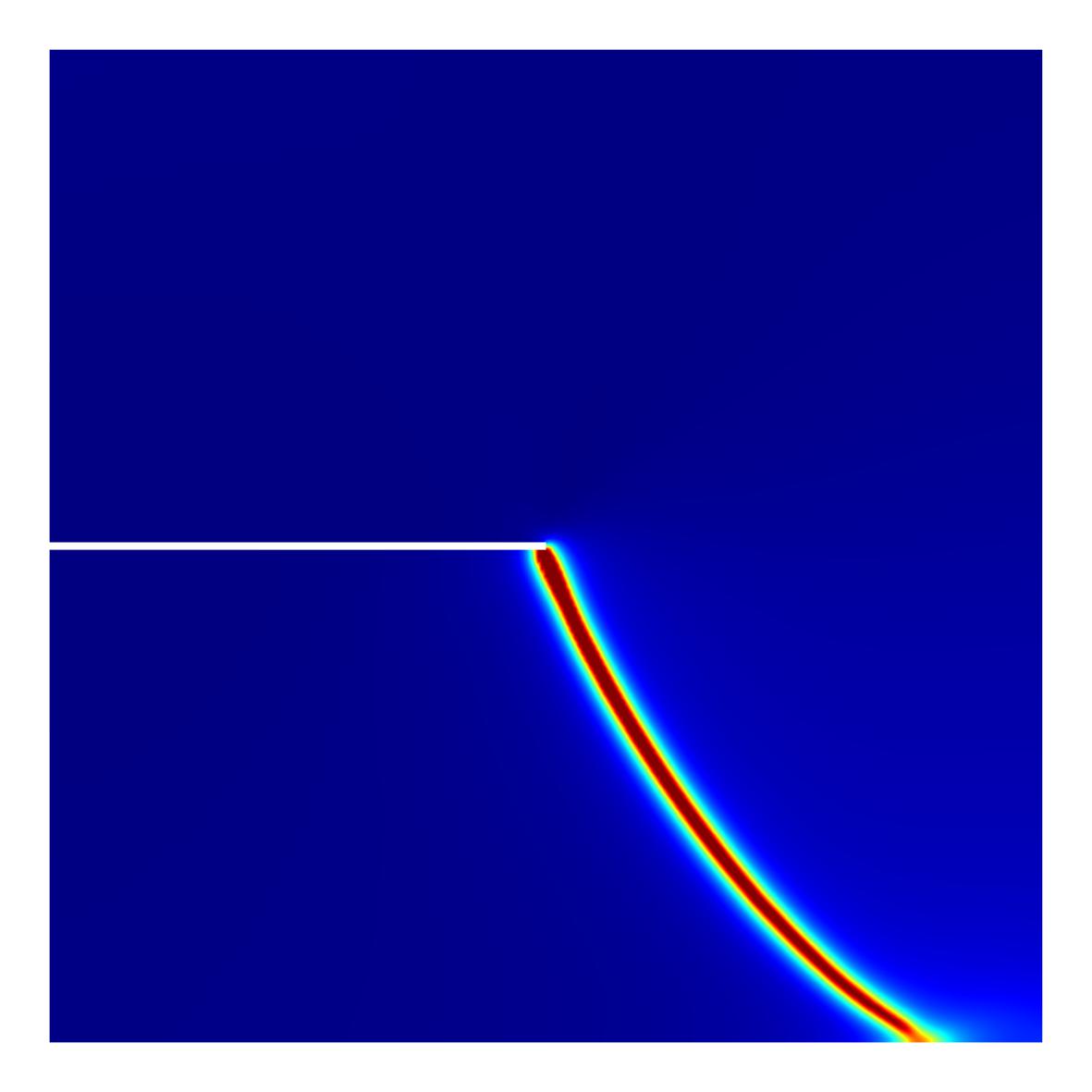}}
	\caption{2D Single-edge-notched square subjected to shear loading. Crack pattern at a displacement of (a) $u = 1.0\times10^{-2}$ mm, (b) $u = 1.2\times10^{-2}$ mm, (c) $u = 1.4\times10^{-2}$ mm for a length scale $l_0$ of $1.5\times10^{-2}$ mm and (d) $u = 1.0\times10^{-2}$ mm, (e) $u = 1.3\times10^{-2}$ mm, and (f) $u = 1.6\times10^{-2}$ mm for a length scale $l_0$ of $7.5\times10^{-3}$ mm.}
	\label{2D Single-edge-notched square subjected to shear loading Crack pattern}
	\end{figure}

We now test the 2D notched square plate from Section 4.1 under shear loading. The geometry and boundary conditions of the plate subjected to shear is depicted in Fig. \ref{Geometry and boundary conditon of the single-edge-notched plate subjected to shear load}. For the first 80 time steps, we take the displacement increment $\Delta u = 1\times10^{-4}$ mm, afterwards, $\Delta u = 1\times10^{-5}$ mm. We calculate the crack patterns under shear loading for two length scale parameters $l_0 = 1.5\times10^{-2}$ mm and $l_0 = 7.5\times10^{-3}$ mm as shown in Fig. \ref{2D Single-edge-notched square subjected to shear loading Crack pattern}. As expected, the crack propagates under the shear loading and the crack has wider crack width when $l_0 = 1.5\times10^{-2}$ mm. Figure \ref{2D Single-edge-notched square subjected to shear loading Crack pattern} shows a curved crack path under shear, which is the same as those reported by \citet{miehe2010phase, miehe2010thermodynamically, hesch2014thermodynamically, liu2016abaqus}.

\begin{figure}[htbp]
	\centering
	\includegraphics[width = 8cm]{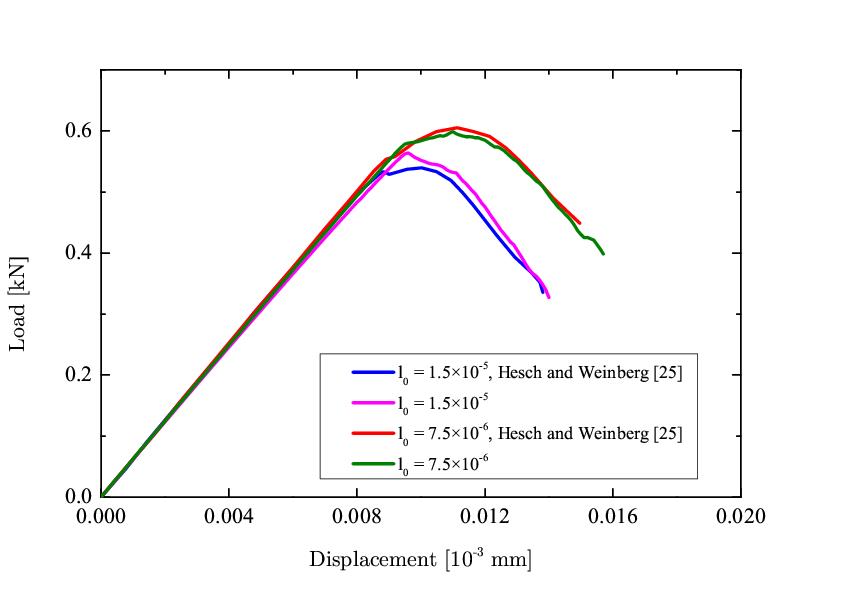}
	\caption{Load-displacement curves of the 2D single-edge-notched shear test}
	\label{Load-displacement curves of the 2D single-edge-notched shear test}
	\end{figure}

\begin{figure}[htbp]
	\centering
	\subfigure[$l_0=1.5\times10^{-2}$ mm]{\includegraphics[width = 8cm]{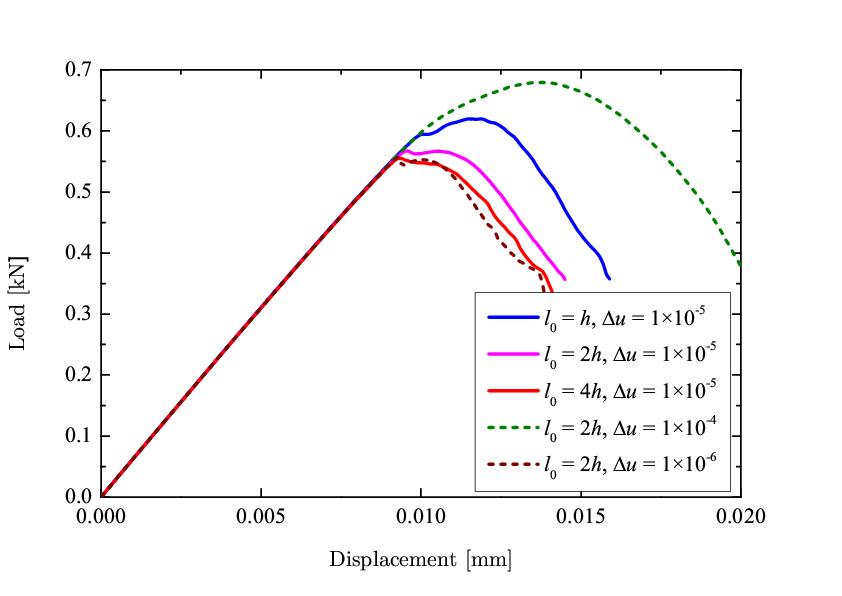}}
	\subfigure[$l_0=7.5\times10^{-3}$ mm]{\includegraphics[width = 8cm]{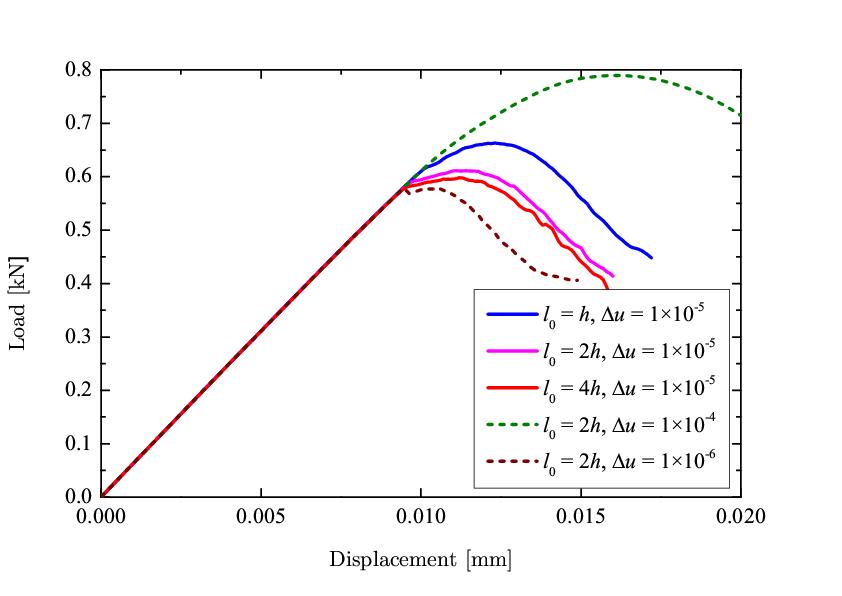}}
	\caption{2D single-edge-notched square subjected to shear. Influence of mesh size $h$ and displacement increment $\Delta u$ on the load-displacement curves.}
	\label{2D single-edge-notched square subjected to shear Influence of mesh size}
	\end{figure}

The load-displacement curves for the top edge of the plate are depicted in Fig. \ref{Load-displacement curves of the 2D single-edge-notched shear test} in comparison with the results by \citet{hesch2014thermodynamically}. A close observation is shown in Fig. \ref{Load-displacement curves of the 2D single-edge-notched shear test}. Thus, the loads obtained by this work and \citet{hesch2014thermodynamically} are exactly matching as the displacement increases, particularly for a smaller length scale $l_0 = 7.5\times10^{-3}$ mm. The excellent agreement in the crack pattern and load-displacement curves indicates the feasibility and practicability of the presented phase field modeling approach in COMSOL. We then show the influence of mesh size and step size on the load-displacement curves at a fixed $l_0$ in Fig. \ref{2D single-edge-notched square subjected to shear Influence of mesh size}. Larger mesh size and displacement increment achieve larger peak load.

\subsection{2D and 3D notched square plate subjected to tension and shear}

\begin{figure}[htbp]
	\centering
	\includegraphics[width = 8cm]{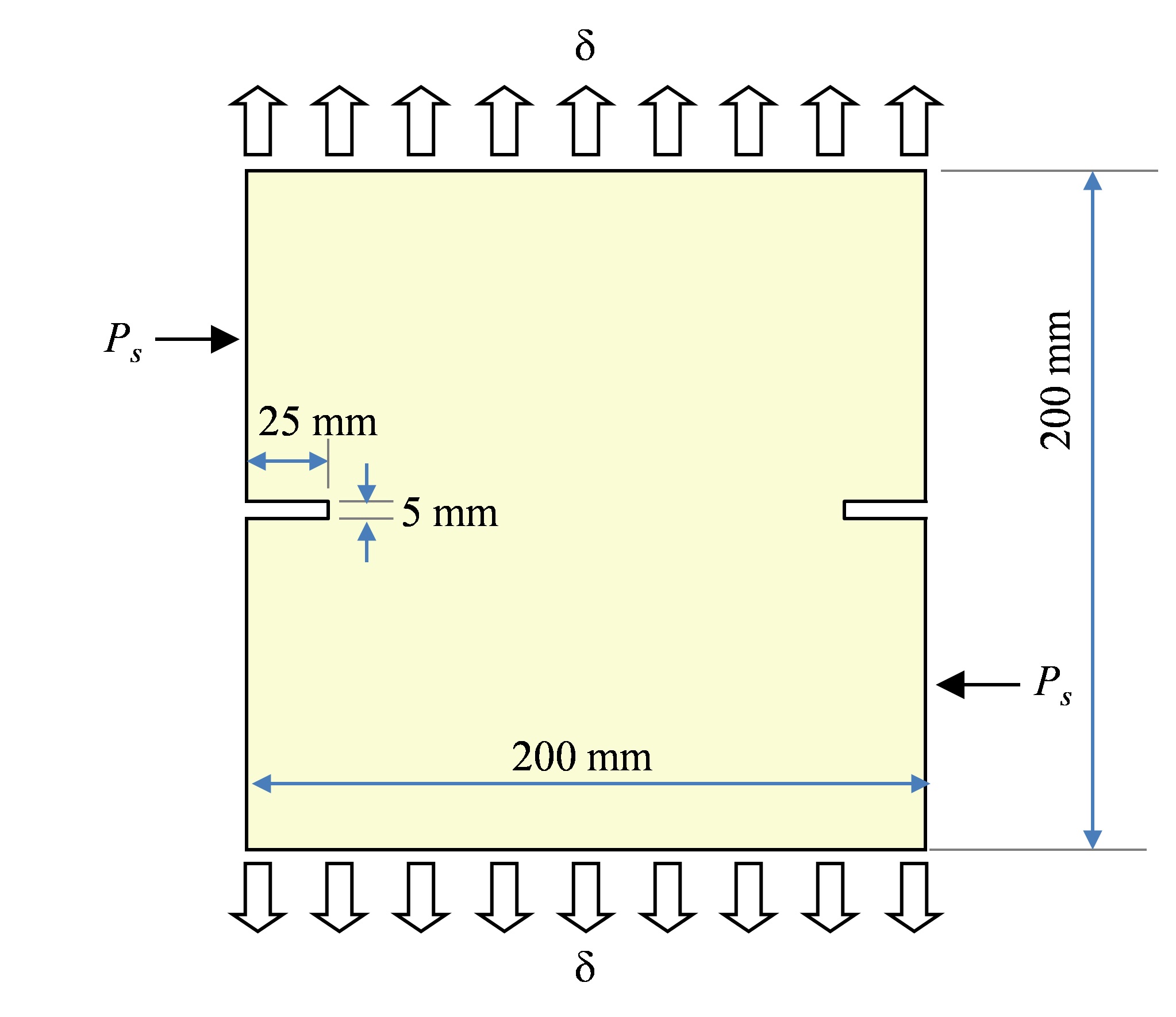}
	\caption{Geometry and boundary condition of the 2D and 3D notched plate subjected to tension and shear}
	\label{Geometry and boundary condition of the 2D and 3D notched plate subjected to tension and shear}
	\end{figure}

\begin{figure}[htbp]
	\centering
	\includegraphics[width = 6cm]{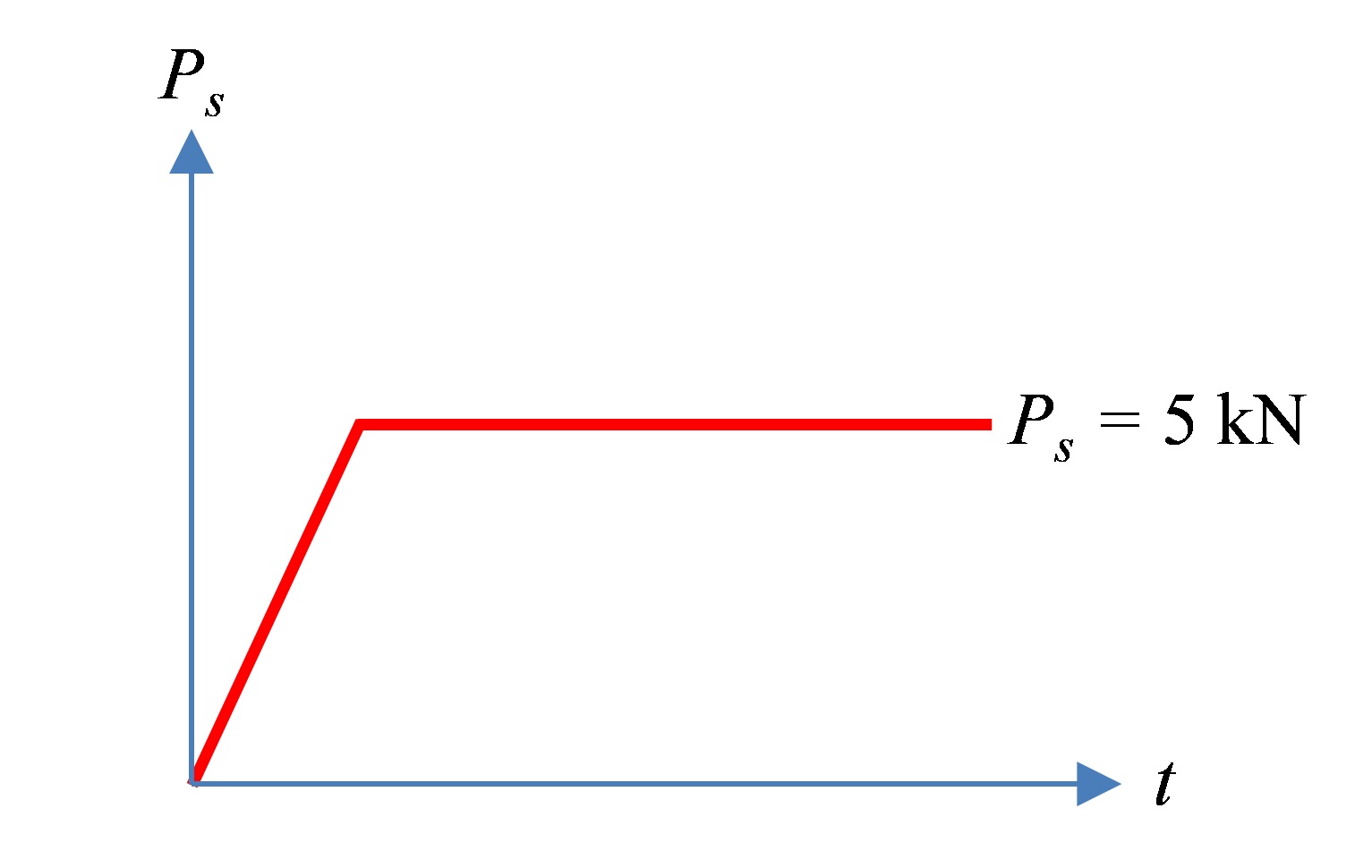}
	\caption{Load-time curve of the shear force}
	\label{Load-time curve of the shear force}
	\end{figure}

Nooru-Mohammed carried out experiments on double-edge-notched plates subjected to both tension and shear loading \citep{bobinski2011simulations}. Figure \ref{Geometry and boundary condition of the 2D and 3D notched plate subjected to tension and shear} shows the geometry and boundary conditions of their experiments. The length, height and thickness of the plate are 200, 200 and 50 mm, respectively. Two horizontal notches of 25 mm $\times$ 5 mm exist in the middle of the left and right edges of the plate. The shear force $P_s$  is applied as Fig. \ref{Load-time curve of the shear force}. The tensile load is zero with the increase in $P_s$ and the vertical displacement increment $\delta$ is then prescribed on the upper and lower boundaries when $P_s=5$ kN.

We conduct both 2D and 3D simulations of the Nooru-Mohammed experiment \citep{bobinski2011simulations}. In the simulations, these elastic constants are used: Young's modulus  $E$ = 32.8 GPa and Poisson's ratio $\nu  = 0.2$. We fix the length scale of  $l_0 = 2.5$ mm and the plate is discretized into uniform elements with size $h = 1.25$ mm ($l_0  = 2h$) for 2D and $h =2.5$ mm for 3D. We choose constant displacement increment $\Delta u  = 5\times10^{-6}$ mm for each time step. In this work, we test different critical energy release rates, i.e. $G_c  =$ 25, 50, 75 and 100 J/m$^2$, respectively.

\begin{figure}[htbp]
	\centering
	\subfigure[$G_c= 25$ J/m$^2$]{\includegraphics[width = 5cm]{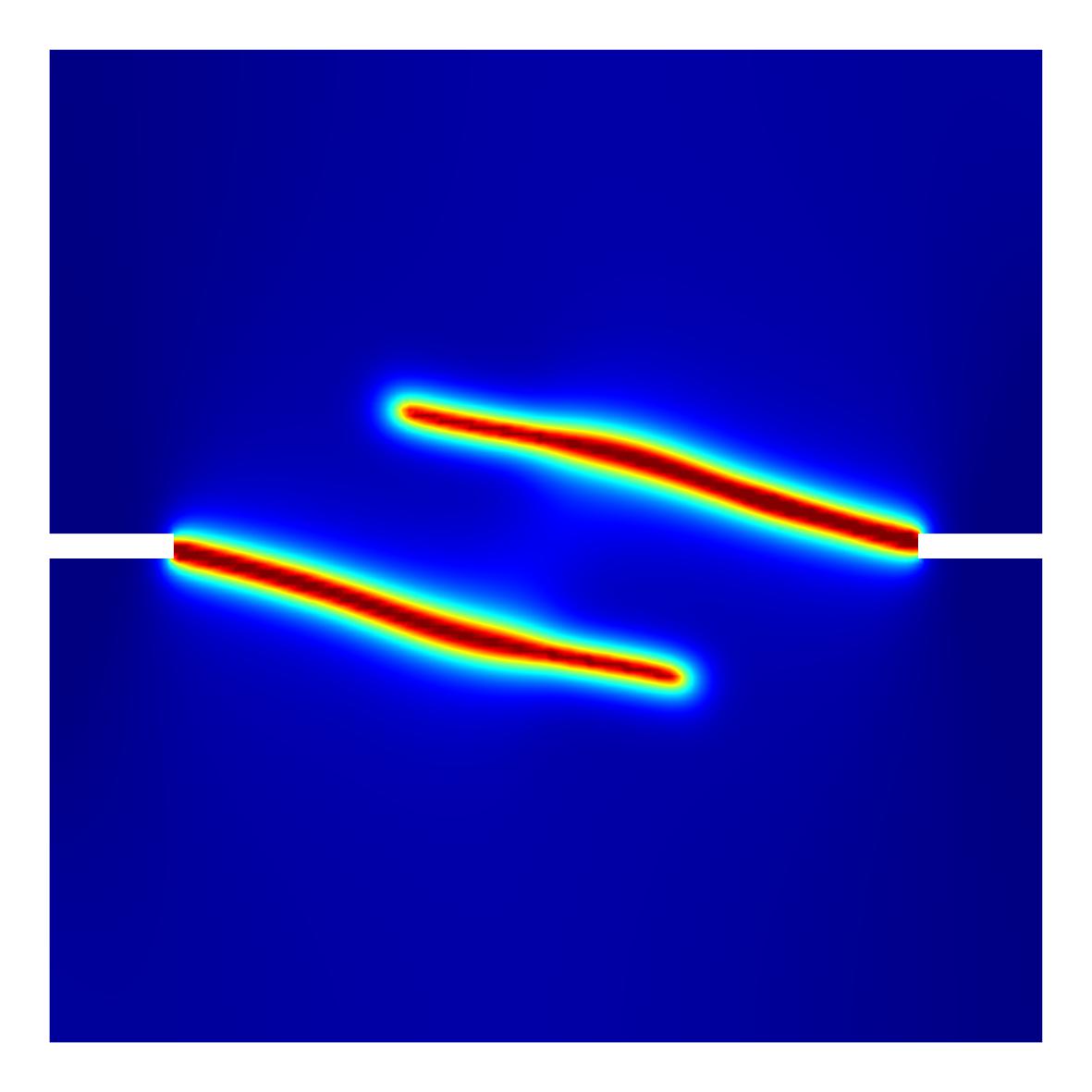}}
	\subfigure[$G_c= 50$ J/m$^2$]{\includegraphics[width = 5cm]{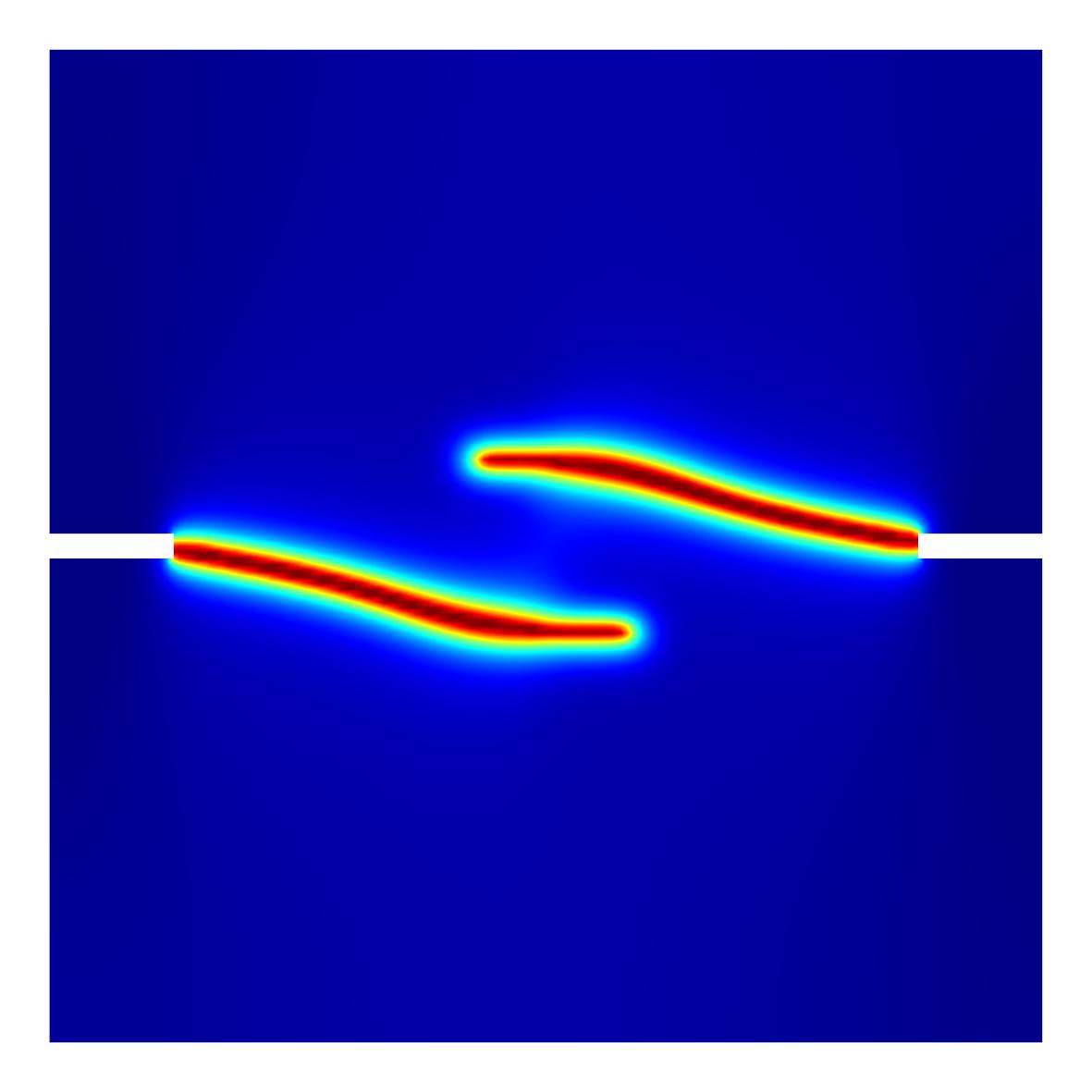}}
	
	\subfigure[$G_c= 75$ J/m$^2$]{\includegraphics[width = 5cm]{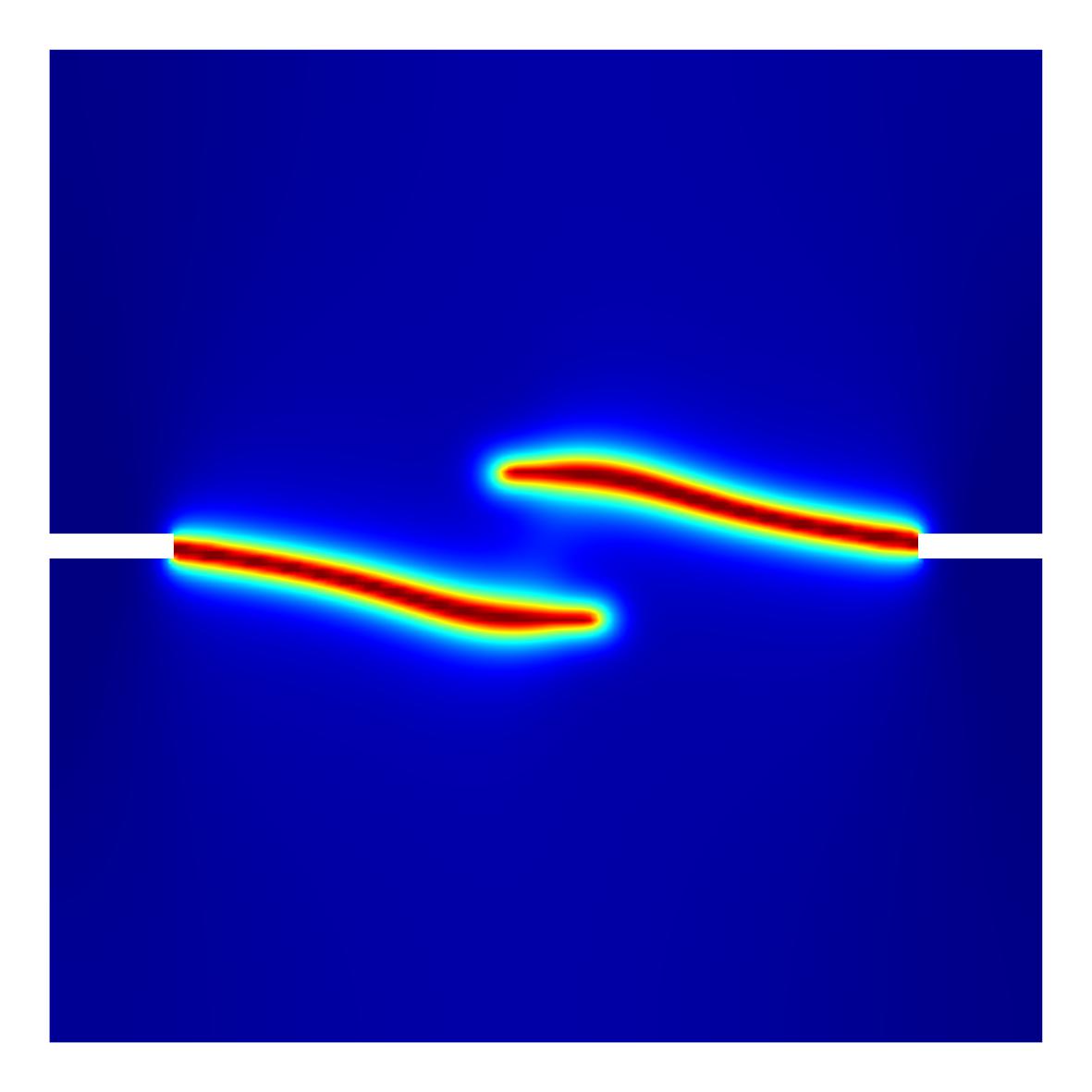}}
	\subfigure[$G_c= 100$ J/m$^2$]{\includegraphics[width = 5cm]{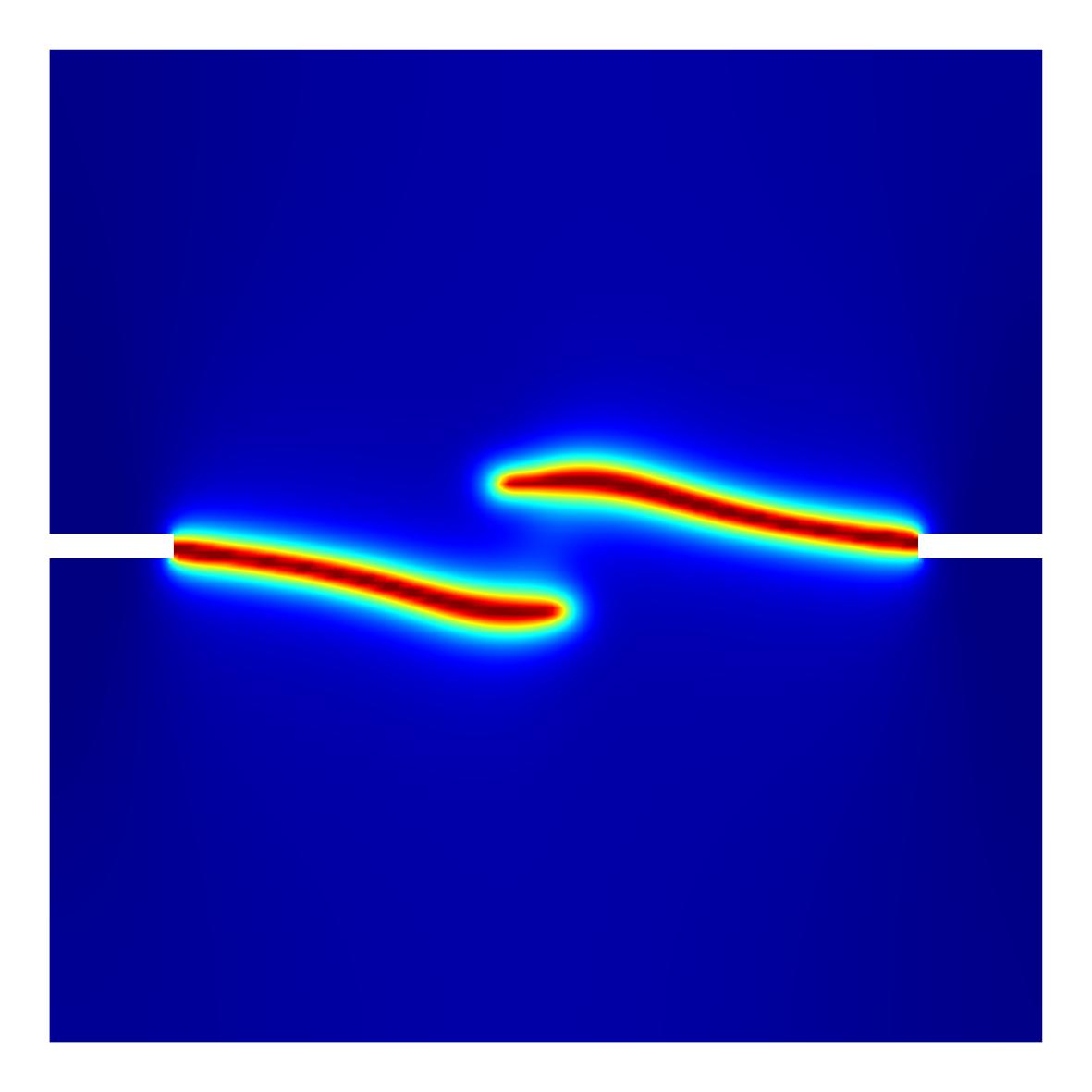}}
	\caption{Crack patterns of the 2D notched plate under shear and tension at the placement $\delta = 0.026$ mm}
	\label{Crack patterns of the 2D notched plate under shear and tension at the placement}
	\end{figure}

\begin{figure}[htbp]
	\centering
	\includegraphics[width = 8cm]{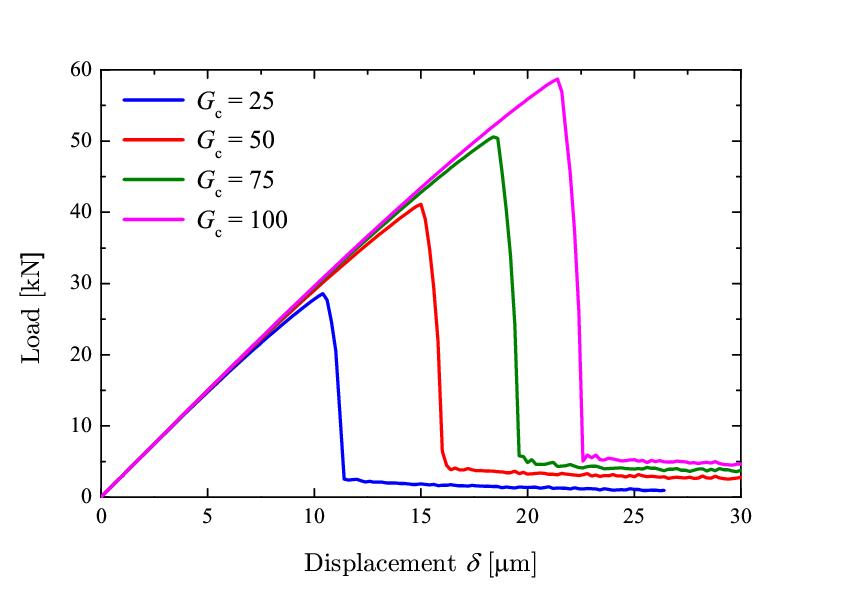}
	\caption{Load-displacement curves for the 2D notched plate under shear and tension}
	\label{Load-displacement curves for the 2D notched plate under shear and tension}
	\end{figure}

\begin{figure}[htbp]
	\centering
	\subfigure[]{\includegraphics[width = 5cm]{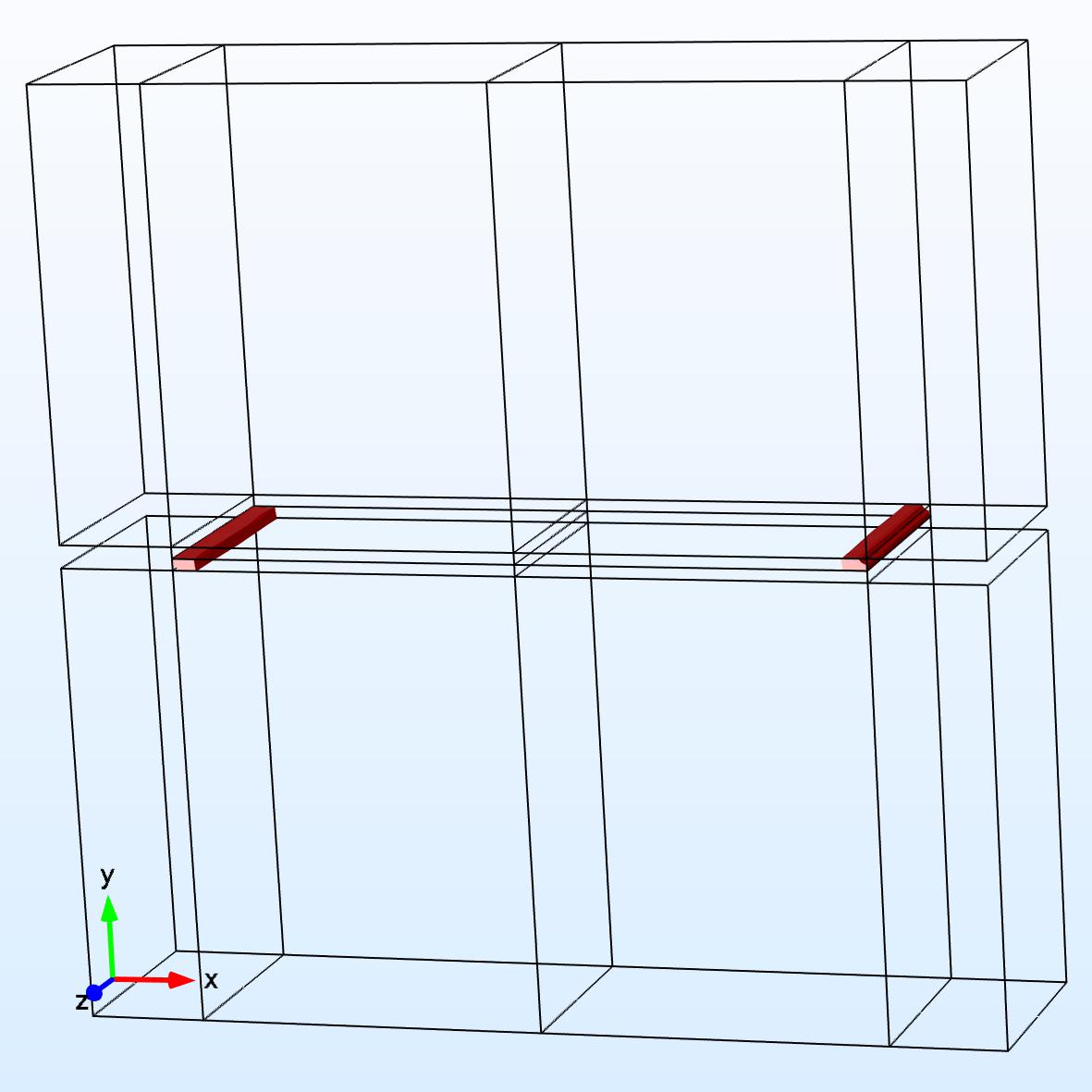}}
	\subfigure[]{\includegraphics[width = 5cm]{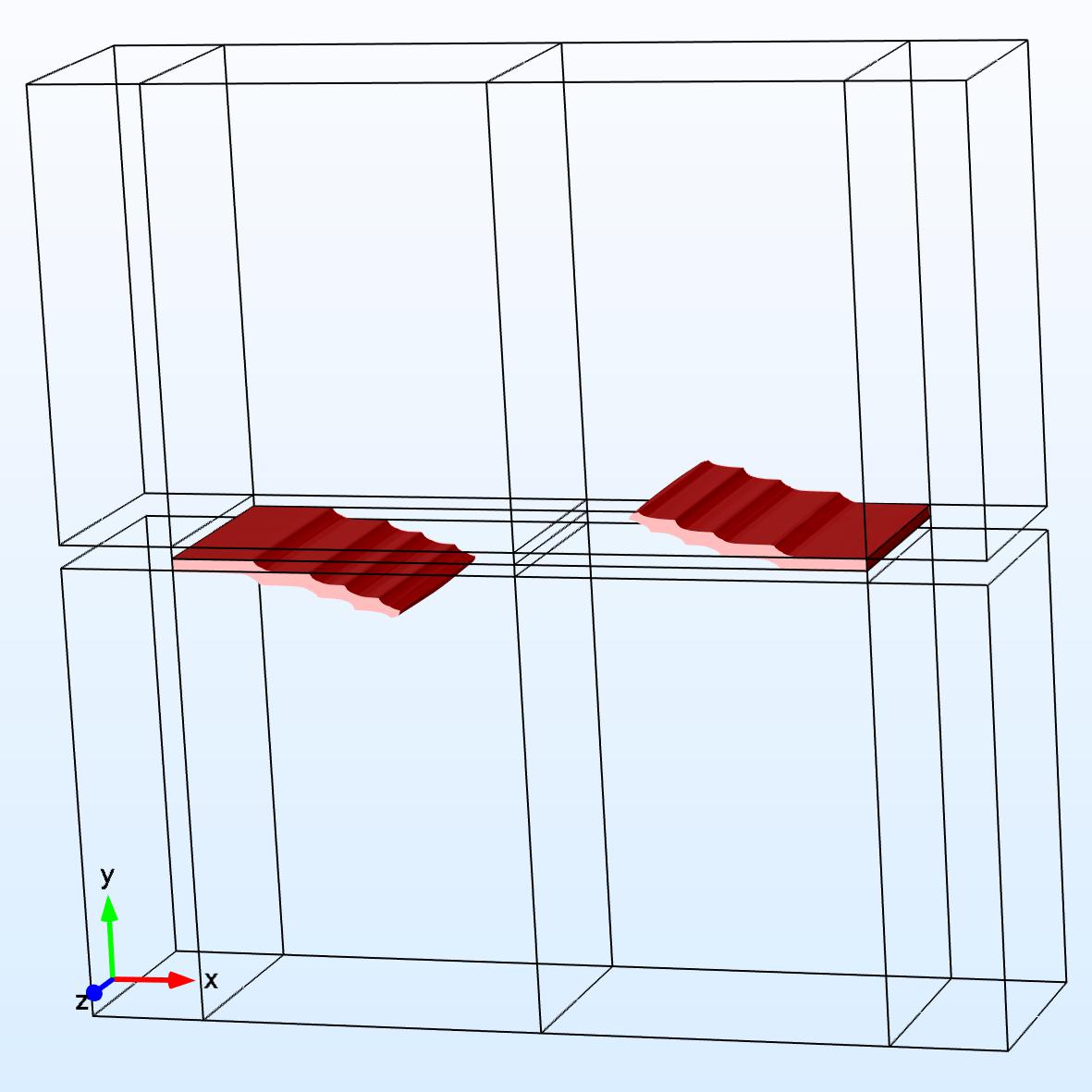}}
	\subfigure[]{\includegraphics[width = 5cm]{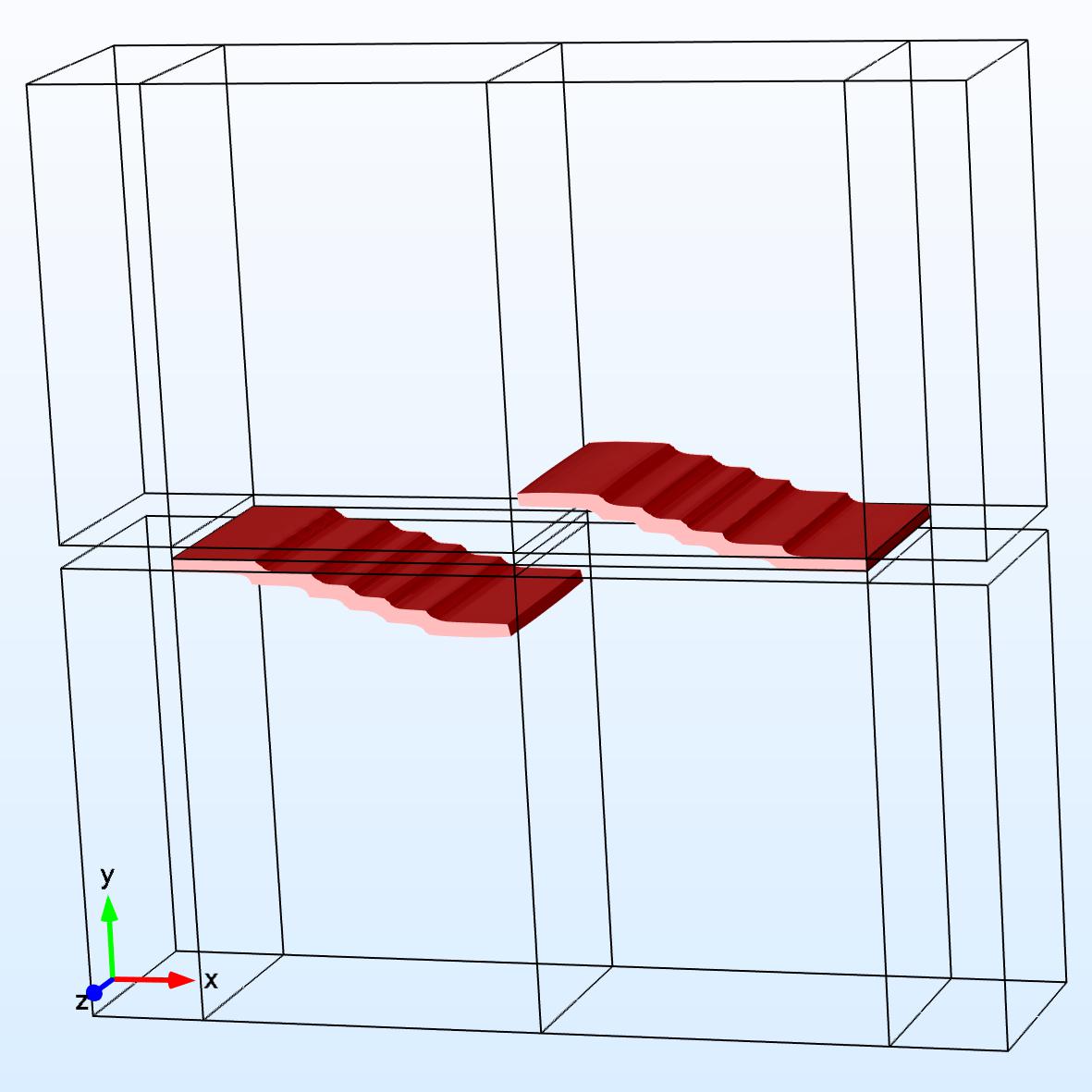}}
	
	\subfigure[]{\includegraphics[width = 5cm]{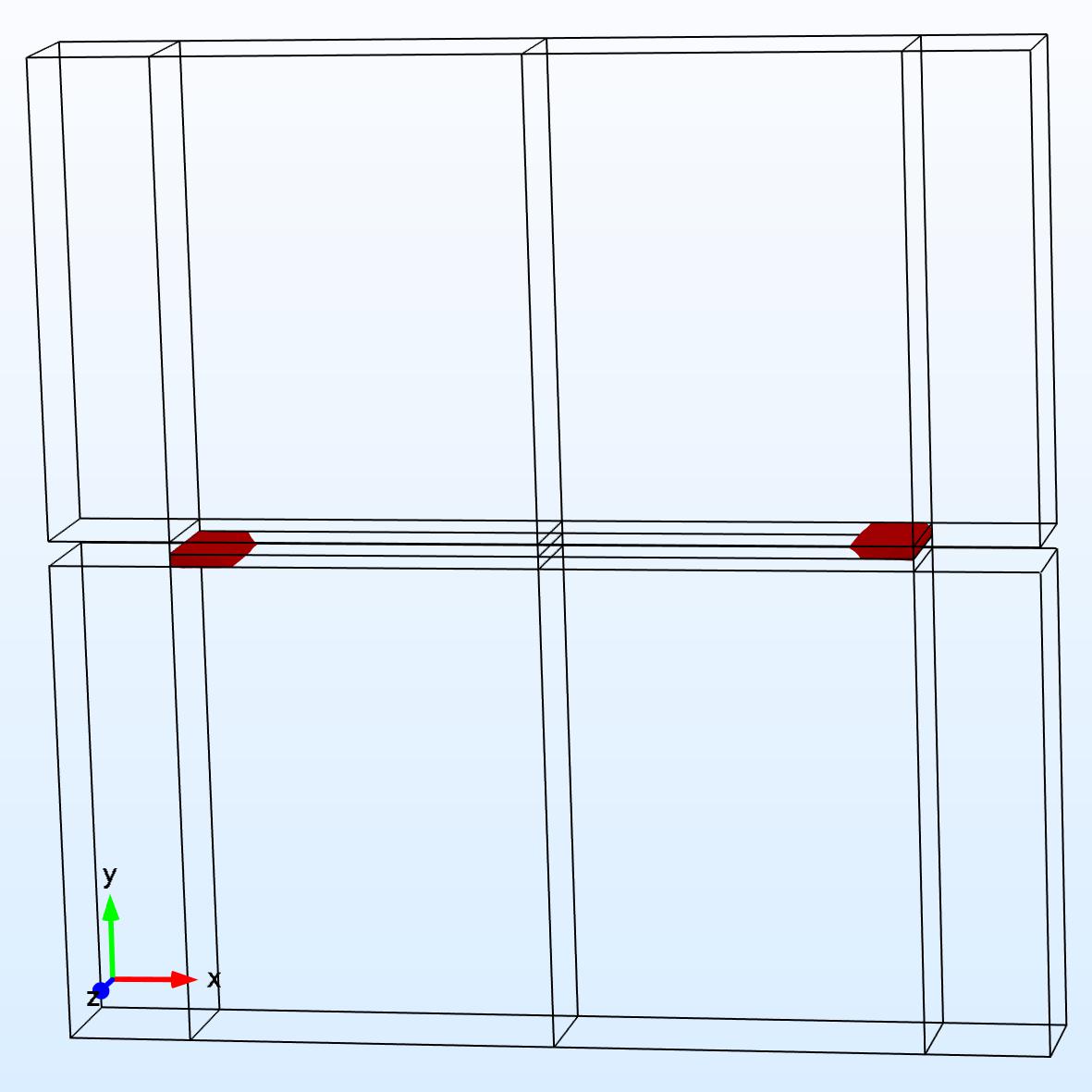}}
	\subfigure[]{\includegraphics[width = 5cm]{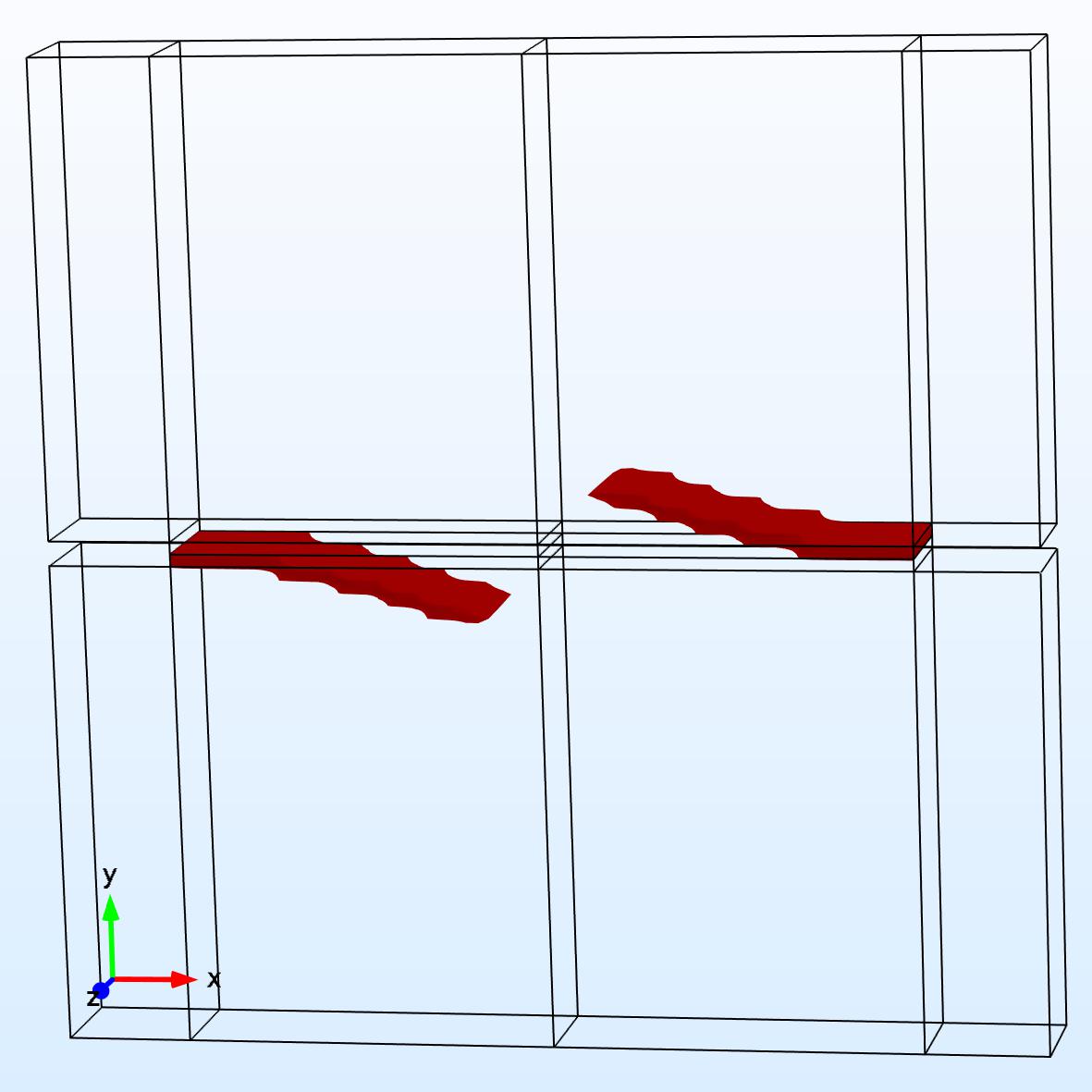}}
	\subfigure[]{\includegraphics[width = 5cm]{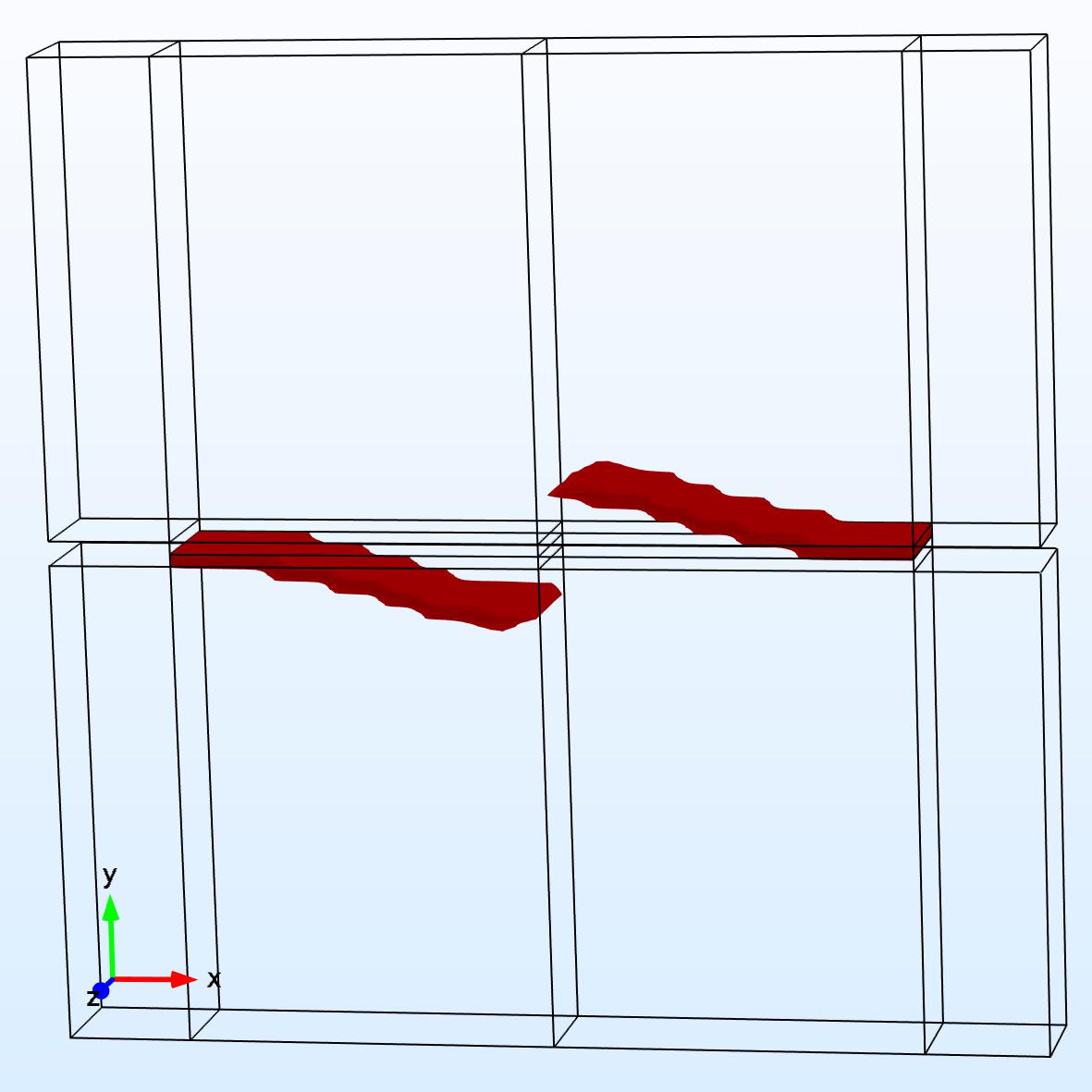}}
	\caption{3D notched plate under shear and tension. Crack pattern at a displacement of (a) $\delta = 0.025$ mm, (b) $\delta = 0.0265$ mm, (c) $\delta = 0.028$ mm for $G_c=75$ J/m$^2$ and (d) $\delta = 0.0285$ mm, (e) $\delta=0.03$ mm, and (f)$\delta=0.0315$ mm for $G_c=100$ J/m$^2$.}
	\label{3D notched plate under shear and tension. Crack pattern}
	\end{figure}

\begin{figure}[htbp]
	\centering
	\includegraphics[width = 8cm]{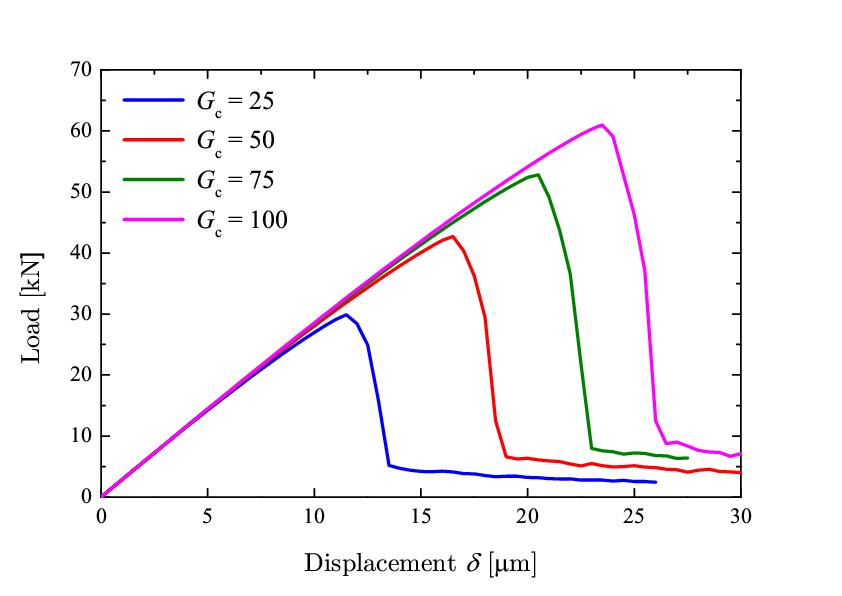}
	\caption{Load-displacement curves for the 3D notched plate under shear and tension}
	\label{Load-displacement curves for the 3D notched plate under shear and tension}
	\end{figure}

Figure \ref{Crack patterns of the 2D notched plate under shear and tension at the placement} shows the crack patterns of the 2D notched plate when the displacement is $\delta = 0.026$ mm. The crack for lager $G_c$ has a smaller inclination angle to the horizontal direction and is more curved. This results in a smaller distance between the two parallel cracks. Figure \ref{Load-displacement curves for the 2D notched plate under shear and tension} presents the load-displacement curves of the 2D notched plated under shear and tension for different critical energy release rates. A quick drop of the load after a nearly linear increase is observed. The peak and residual values of the load increases with the increase in the critical energy release rate. Figure \ref{3D notched plate under shear and tension. Crack pattern} shows the crack propagation in the 3D notched square plate for $G_c=75$ and 100 J/m$^2$. The crack patterns of the 2D and 3D simulations are in good agreement.  The load-displacement curves of the 3D plate are depicted in Fig. \ref{Load-displacement curves for the 3D notched plate under shear and tension}, which are  less steep  after the peak compared with the 2D plate. The reason is that the 2D plate section suffers from tension perpendicular to the section under the plane strain assumption, which accelerates the drop in the load bearing capacity of the plate.

\subsection{3D three-point bending test}

\begin{figure}[htbp]
	\centering
	\includegraphics[width= 7cm]{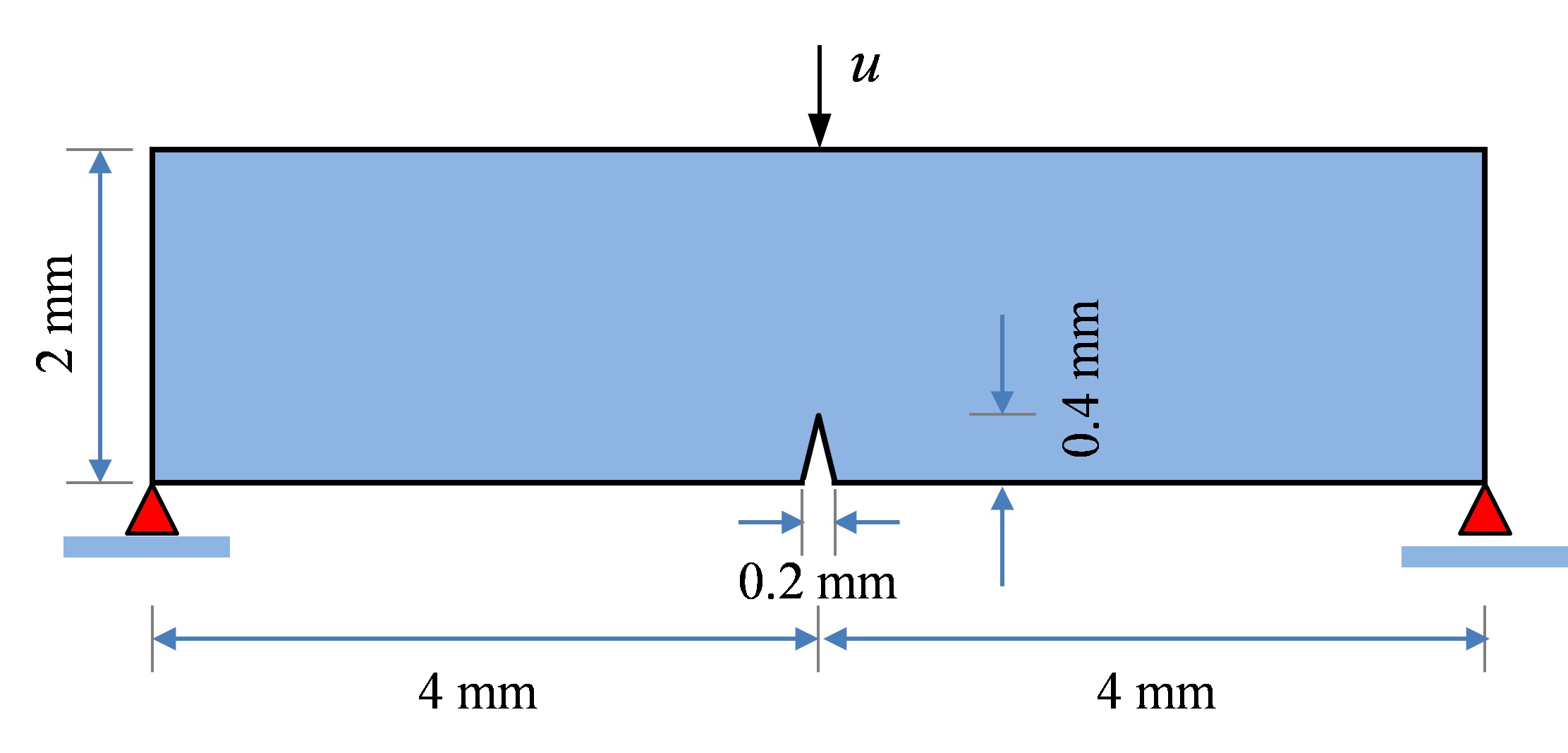}
	\caption{Geometry and boundary conditions of the three-point bending test }
	\label{Geometry and boundary conditions of the three-point bending test}
	\end{figure}

Let us consider a  3D three-point bending test shown in Fig. \ref{Geometry and boundary conditions of the three-point bending test}. 2D results were presented by \citet{miehe2010phase, miehe2010thermodynamically}. The thickness of the beam is 0.4 mm and the following  material parameters are used: $\lambda= 12$ kN/mm$^2$,  $\mu= 8$ kN/mm$^2$, and $G_c = 0.5$ N/mm. Prism elements are used to discretize the beam with a maximum element size of $h=6\times10^{-2}$ mm except $h=1.5\times10^{-2}$ mm in the region where the crack is expected to propagate. The specimen is loaded displacement-driven  with a constant displacement increment $\Delta u  = 5\times10^{-5}$ mm is applied. 

\begin{figure}[htbp]
	\centering
		\begin{tabular}{p{0.05cm}p{5.5cm}p{0.05cm}p{5.5cm}}
		(a) &\includegraphics[width = 5.5cm]{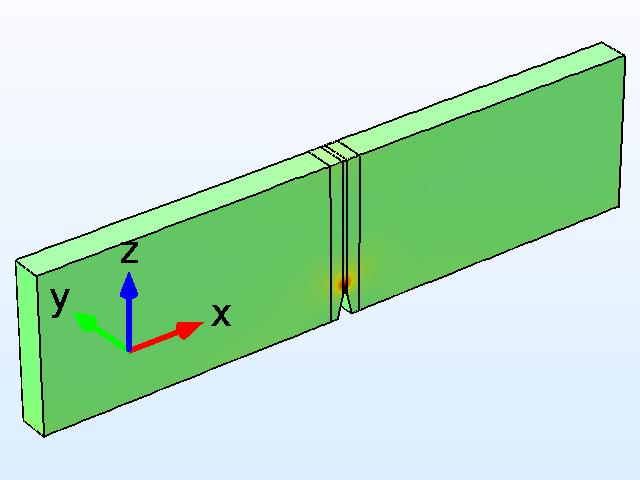} & (d) &\includegraphics[width = 5.5cm]{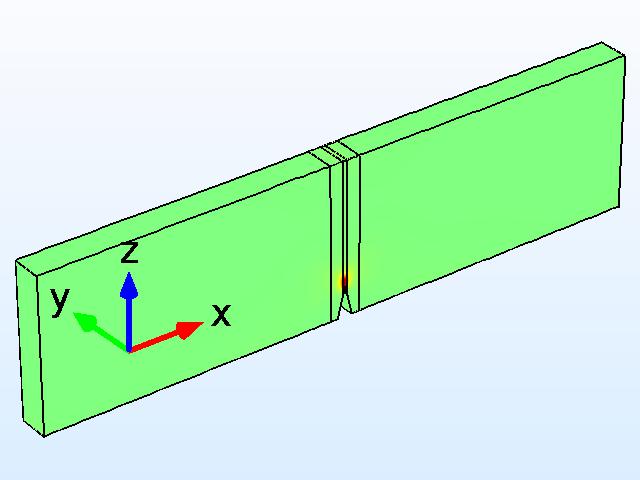}\\ 
		(b) &\includegraphics[width = 5.5cm]{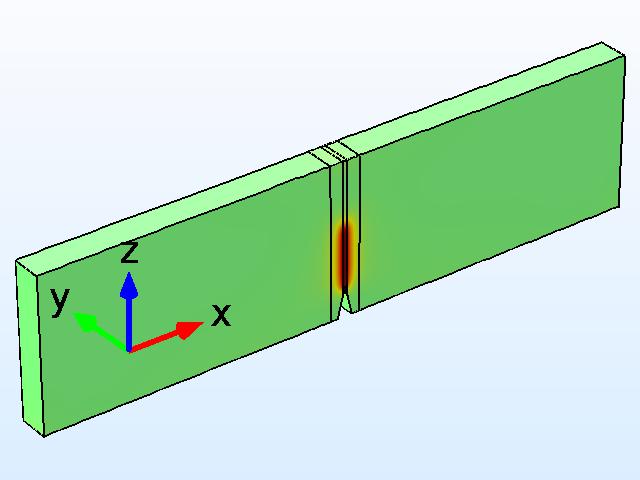} & (e) &\includegraphics[width = 5.5cm]{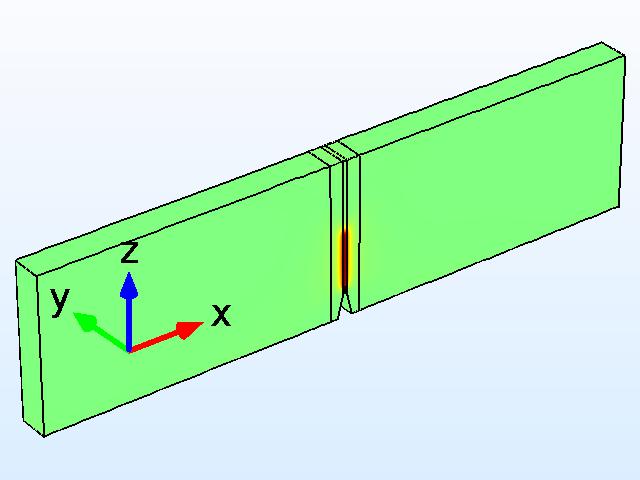}\\ 
		(c) &\includegraphics[width = 5.5cm]{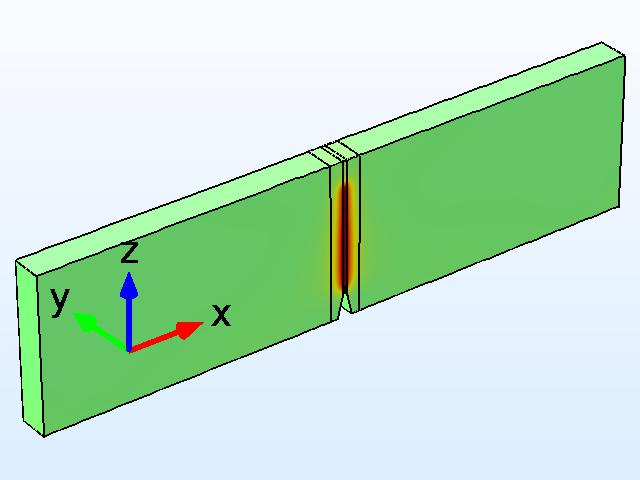} & (f) &\includegraphics[width = 5.5cm]{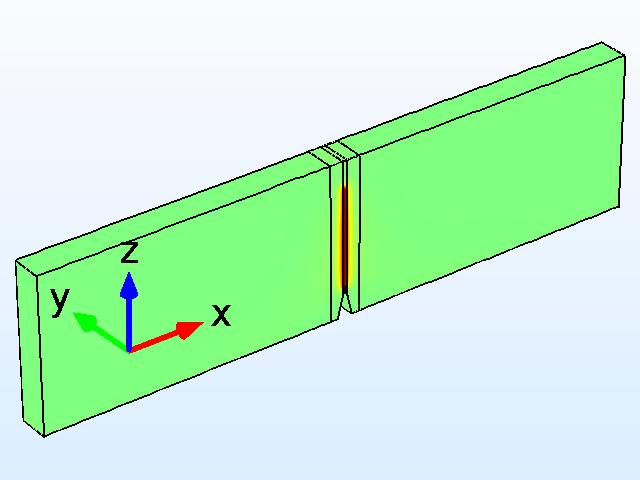}\\ 	
		\end{tabular}
	\caption{3D three-point bending test. Crack pattern at a displacement of (a) $u = 4.7\times10^{-2}$ mm, (b) $u = 5.2\times10^{-2}$ mm, (c) $u = 8\times10^{-2}$ mm for a length scale $l_0$ of $6\times10^{-2}$ mm and (d) $u = 5\times10^{-2}$ mm, (e) $u = 5.5\times10^{-2}$ mm, and (f) $u = 8\times10^{-2}$ mm for a length scale $l_0$ of $3\times10^{-2}$ mm.}
	\label{Crack patterns of the 3D three-point bending test}
	\end{figure}

We choose the length scale $l_0  = 0.06$ mm and 0.03 mm and show the crack patterns of the simply supported notched beam in Fig. \ref{Crack patterns of the 3D three-point bending test}. The simulation shows a larger crack width when $l_0  = 0.06$ mm.  Figure \ref{Load-displacement curves of  the 3D three point bending test} presents the reaction force at the top of the beam versus the applied displacement. The presented results in 3D are then compared with those 2D results proposed by \citet{miehe2010phase}. As observed, the 3D results by the present results are in good agreement with the 2D simulations by \citet{miehe2010phase}.

	\begin{figure}[htbp]
	\centering
	\subfigure[]{\includegraphics[width = 6cm]{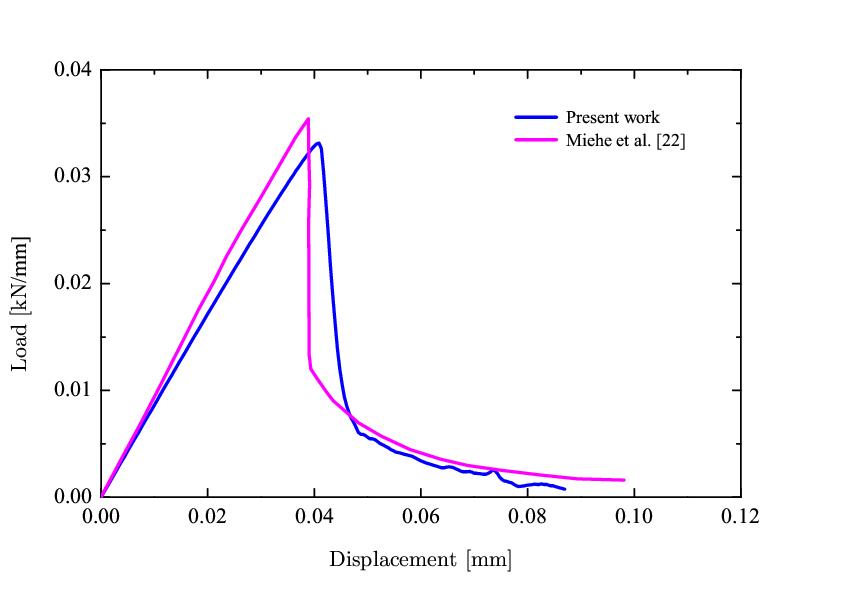}}
	\subfigure[]{\includegraphics[width = 6cm]{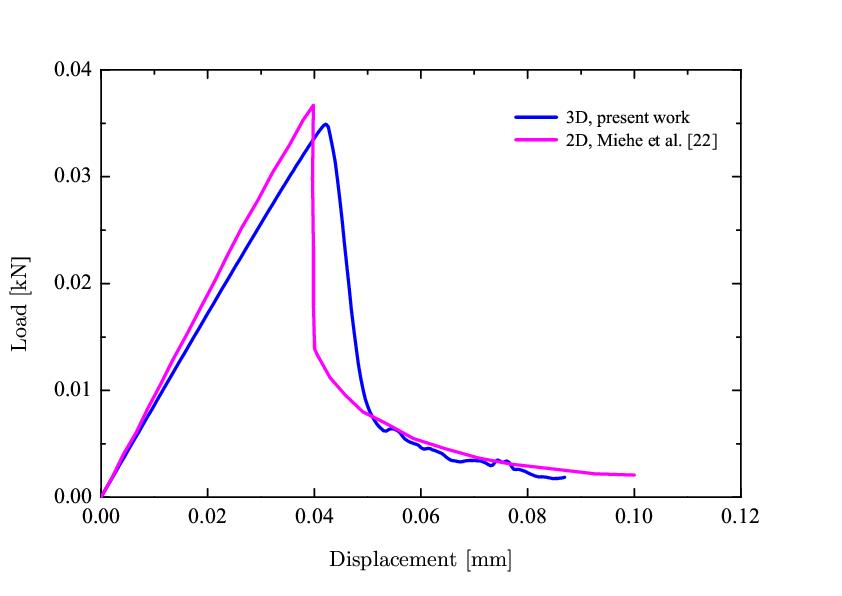}}\\
	\caption{Load-displacement curves of the 3D three point bending test for a length scale (a) $l_0 = 6\times10^{-2}$ mm and (b) $l_0 = 3\times10^{-2}$ mm.}
	\label{Load-displacement curves of the 3D three point bending test}
	\end{figure}

\subsection{2D dynamic shear loading of Kalthoff experiment}

\begin{figure}[htbp]
	\centering
	\includegraphics[width = 8cm]{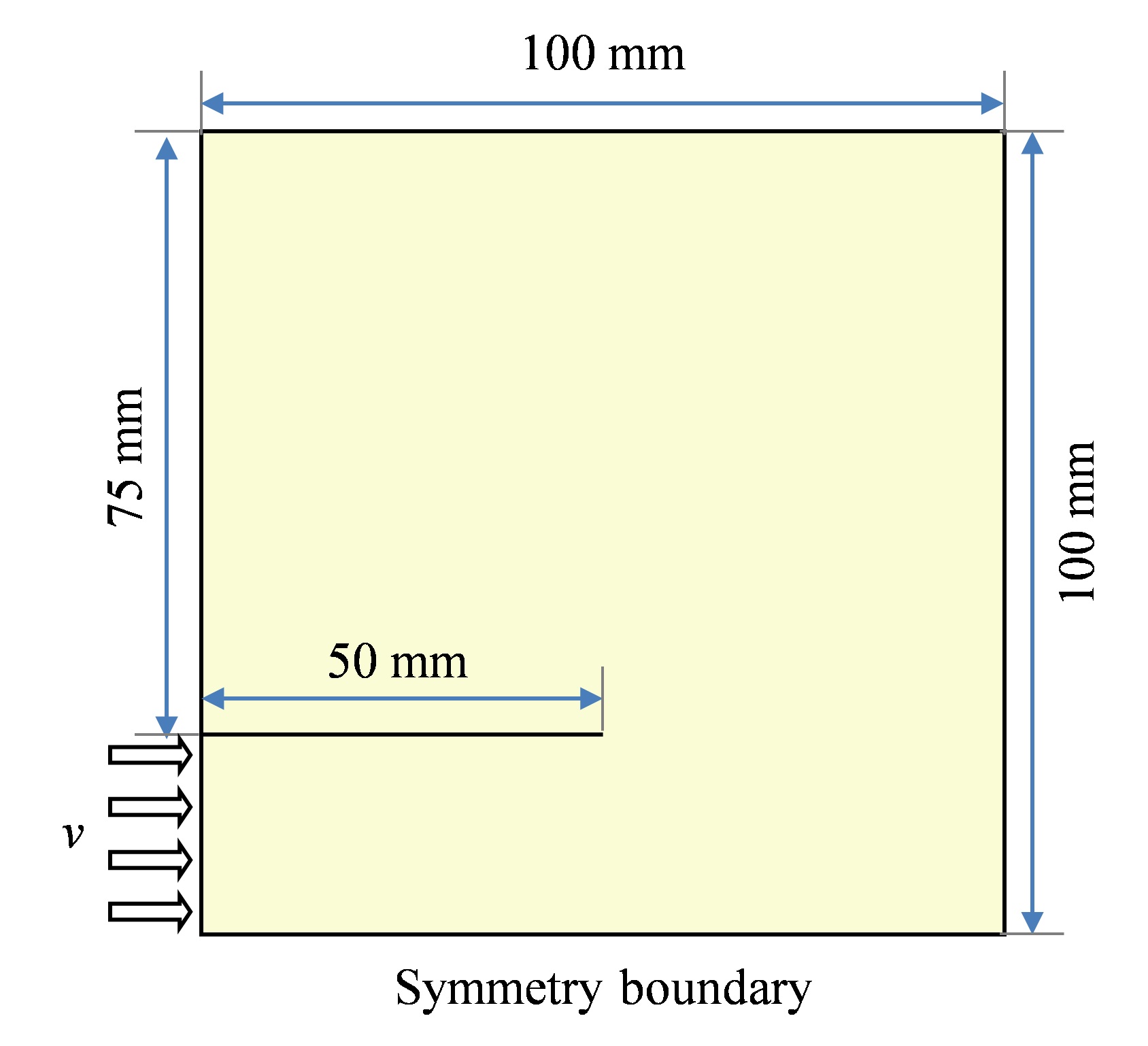}
	\caption{Geometry and boundary conditions of Kalthoff experiment}
	\label{Geometry and boundary conditions of Kalthoff experiment}
	\end{figure}

We next test our method for dynamic fracture by taking advantage of the Kalthoff-Winkler experiments \citep{kalthoff2000modes} which has been studied by several other researchers \citep{rabczuk2008new,rabczuk2007meshfree,rabczuk2007three,ren2017dual,ren2016dual,rabczuk2008discontinuous,rabczuk2007simplified,rabczuk2010simple}. We adopt the symmetry condition to reduce the computational cost and the dimensions and loading conditions are shown in Fig. \ref{Geometry and boundary conditions of Kalthoff experiment}.

The impactor is modelled by applying the following velocity $v$:

	\begin{equation}
	v=\left\{
	\begin{aligned}
	&\frac t {t_0}v_0\qquad &t\le t_0
	\\&v_0 \quad &t>t_0
	\end{aligned}
	\right.
	\end{equation}

\noindent with $v_0  = 16.5$ m/s and  $t_0 = 1$  $\mu$s. Moreover, the initial crack is assumed to be traction free.

The material parameters are from \citep{borden2012phase}: $\rho = 8000$ kg/m$^3$, $E  = 190$ GPa, $ \nu  = 0.3$, $G_c  = 2.213\times10^4$ J/m$^2$, and $k = 1\times10^{-9}$. The Rayleigh wave speed of the plate is $v_R  = 2803$ m/s. The length scale parameter $l_0$ is fixed as $3.9\times10^{-4}$ m. The initial crack is modeled as a notch with a width of $l_0$. Here, we use Q4 elements to discretize the plate and we simulate the crack patterns for two mesh levels: Mesh 1 with element size $h  = 3.9\times10^{-4}$ m ( $l_0 = h$) and Mesh 2 with element size $h  = 1.95\times10^{-4}$ m ($l_0  = 2h$). The time steps are set as $\Delta t = 0.04$ $\mu$s (for Mesh 1, $\Delta t  = 0.288h/v_R$  and for Mesh 2, $\Delta t  = 0.576h/v_R$) and  $\Delta t= 0.01$ $\mu$s (for Mesh 1, $\Delta t = 0.072h/v_R$   and for Mesh 2, $\Delta t  = 0.144h/v_R$), respectively.

\begin{figure}[htbp]
	\centering
	\subfigure[Mesh 1,  $\Delta t = 0.04$ $\mu$s]{\includegraphics[width = 5cm]{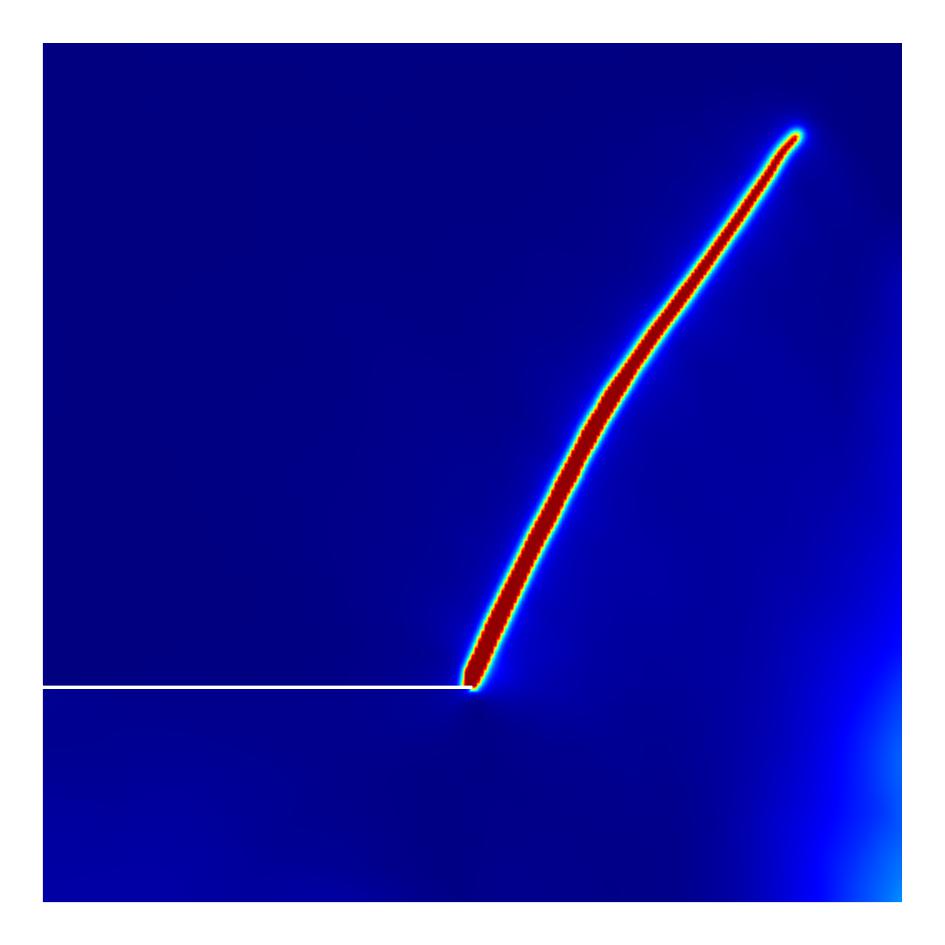}}
	\subfigure[Mesh 1,  $\Delta t = 0.01$ $\mu$s]{\includegraphics[width = 5cm]{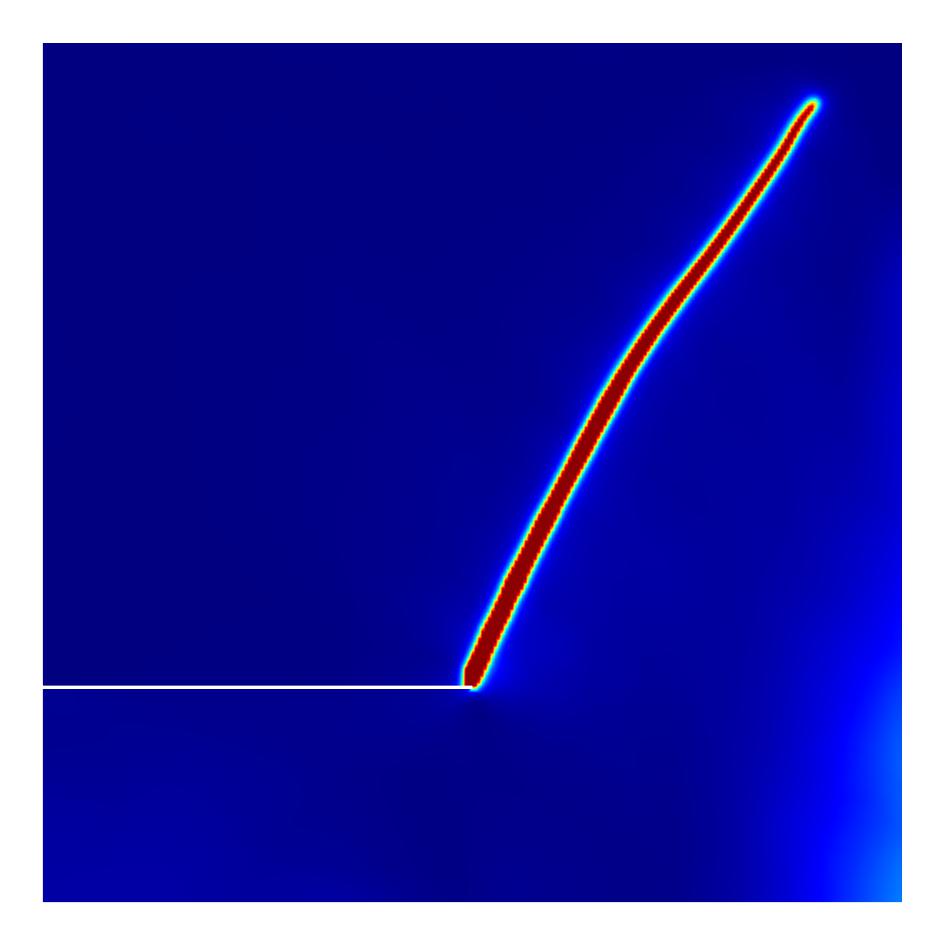}}
	
	\subfigure[Mesh 2,  $\Delta t = 0.04$ $\mu$s]{\includegraphics[width = 5cm]{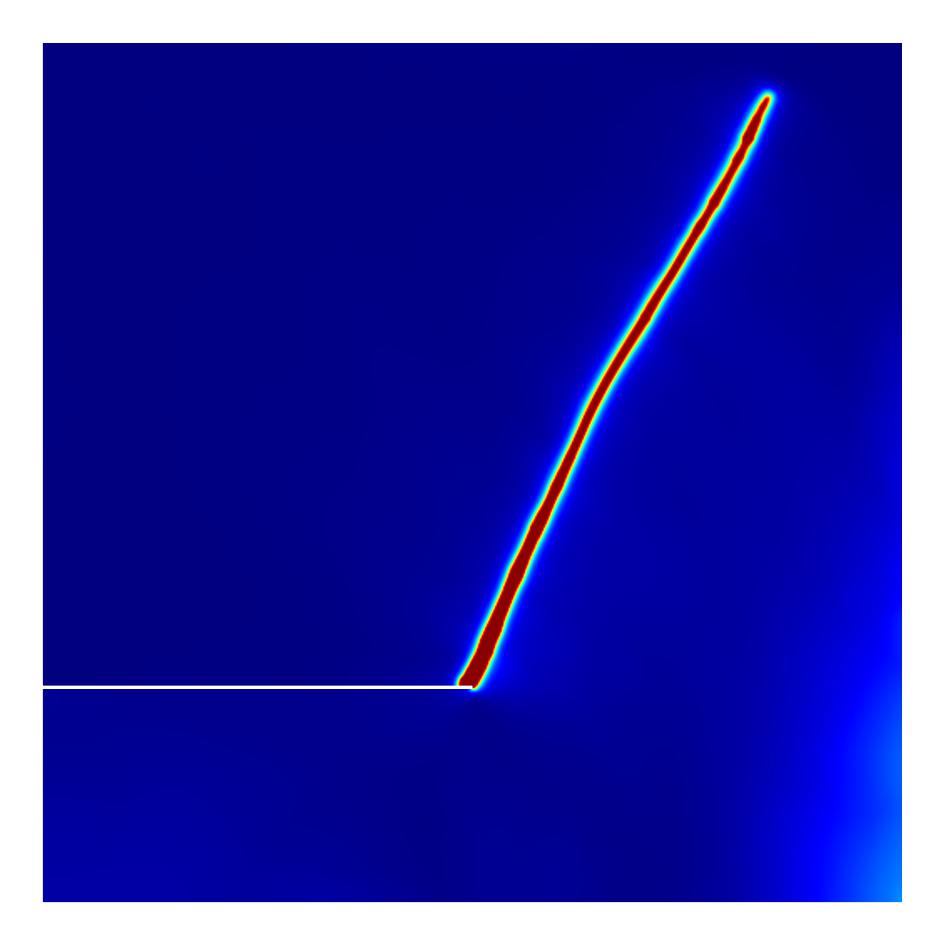}}
	\subfigure[Mesh 2,  $\Delta t = 0.01$ $\mu$s]{\includegraphics[width = 5cm]{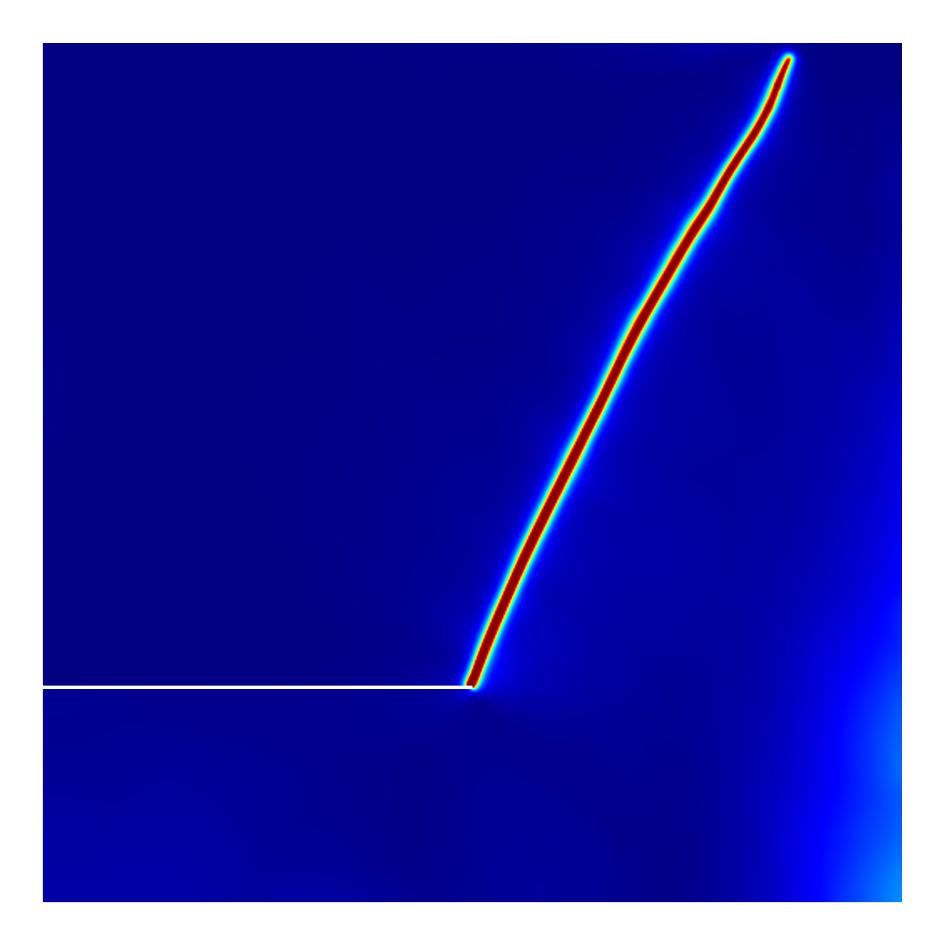}}
	\caption{Phase field at 90 $\mu$s for dynamic shear loading tests by using different meshes and time steps}
	\label{Phase field at 90 s for dynamic shear loading tests by using different meshes and time steps}
	\end{figure}

The phase-field of our simulation at 90 $\mu$s is shown in Fig. \ref{Phase field at 90 s for dynamic shear loading tests by using different meshes and time steps} for different mesh levels and time steps. The crack starts to propagate at 26 $\mu$s. The crack angle versus the horizontal axis varies from $63^\circ$ to $67^\circ$ which matches well the experimental results and other numerical results \citep{borden2012phase}. As shown in Fig. \ref{Phase field at 90 s for dynamic shear loading tests by using different meshes and time steps}, the crack tip by the coarser mesh (Mesh 1) has larger distance from the upper boundary than that by the finer mesh (Mesh 2). The crack has a larger angle and goes more close to the upper boundary for the smaller time step $\Delta t = 0.01$ $\mu$s.

\begin{figure}[htbp]
	\centering
	\subfigure[$\Delta t = 0.04$ $\mu$s]{\includegraphics[width = 6cm]{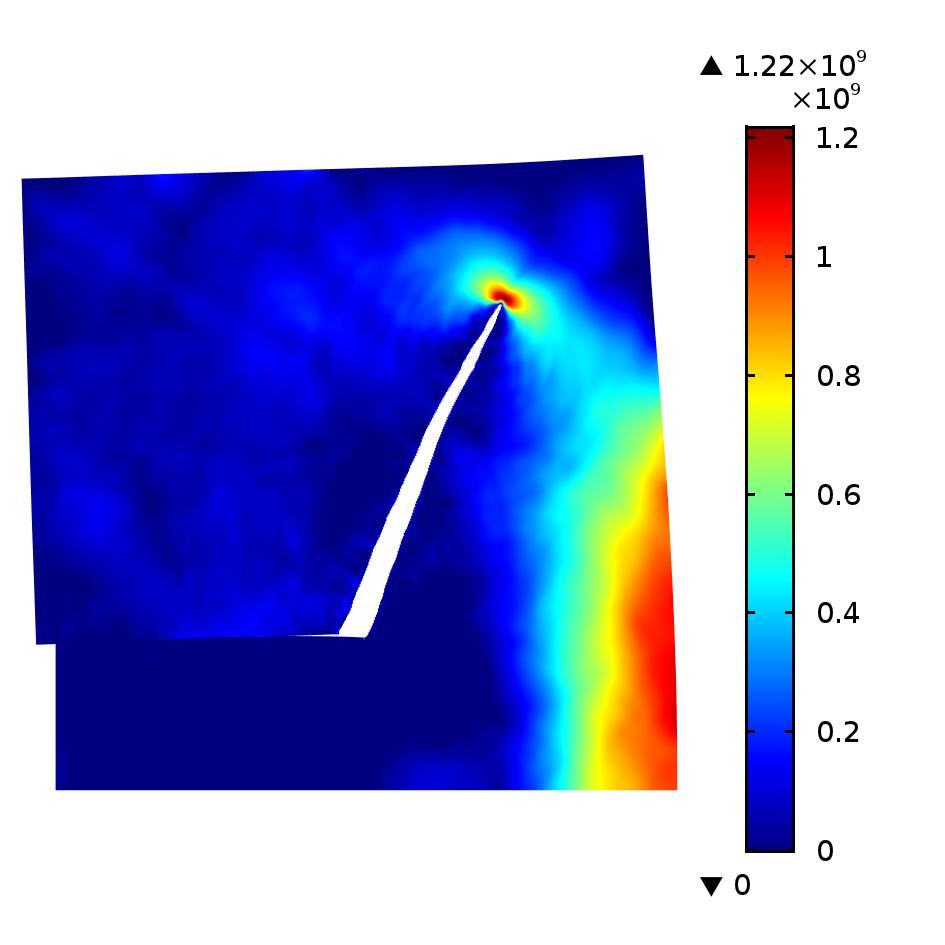}}
	\subfigure[$\Delta t = 0.01$ $\mu$s]{\includegraphics[width = 6cm]{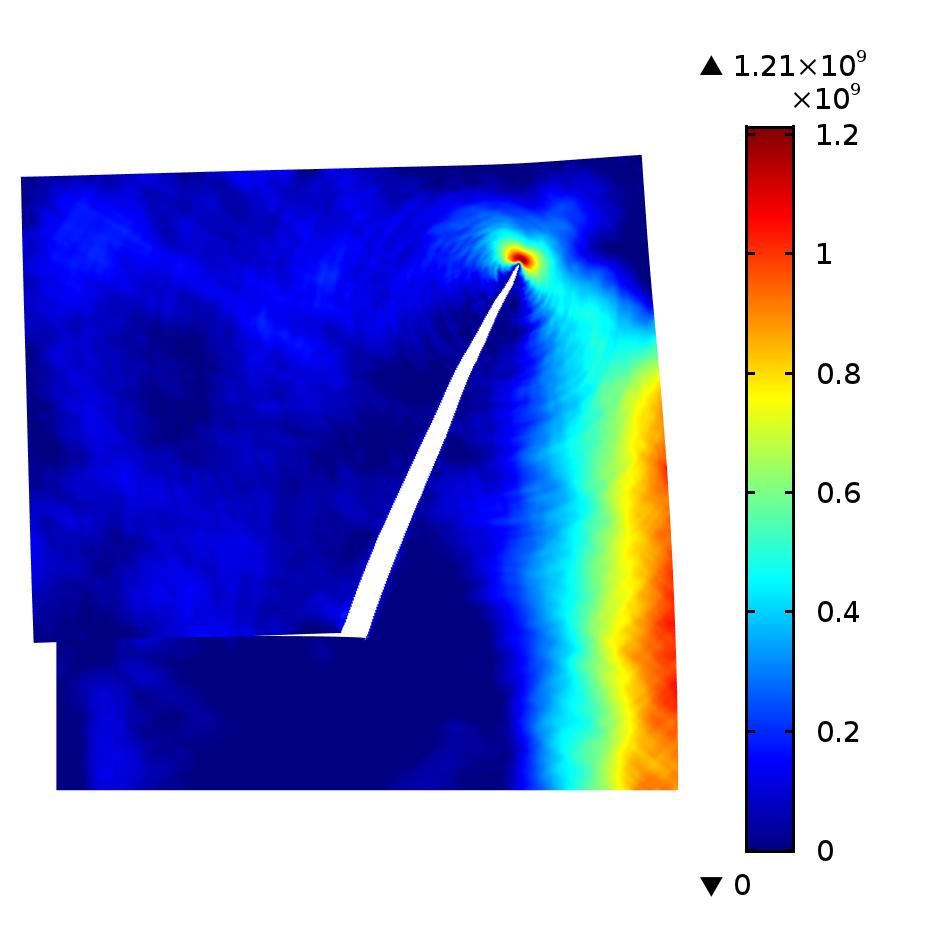}}
	\caption{Maximum tensile stress of the dynamic shear loading example at  $t = 75$ $\mu$s for Mesh 2. The stress is measured in Pa}
	\label{Maximum tensile stress of the dynamic shear loading example at   = 75 µs for Mesh 2. The stress is measured in Pa}
	\end{figure}

Figure \ref{Maximum tensile stress of the dynamic shear loading example at   = 75 µs for Mesh 2. The stress is measured in Pa} shows the contour plot of the maximum principal tensile stress for Mesh 2 at $t  = 75$ $\mu$s under different time steps. Note that the deformation is scaled by a factor of 5. The region with  $\phi>0.95$ is also removed from the figure to see the broken geometry of the plate. 

\begin{figure}[htbp]
	\centering
	\includegraphics[width = 8cm]{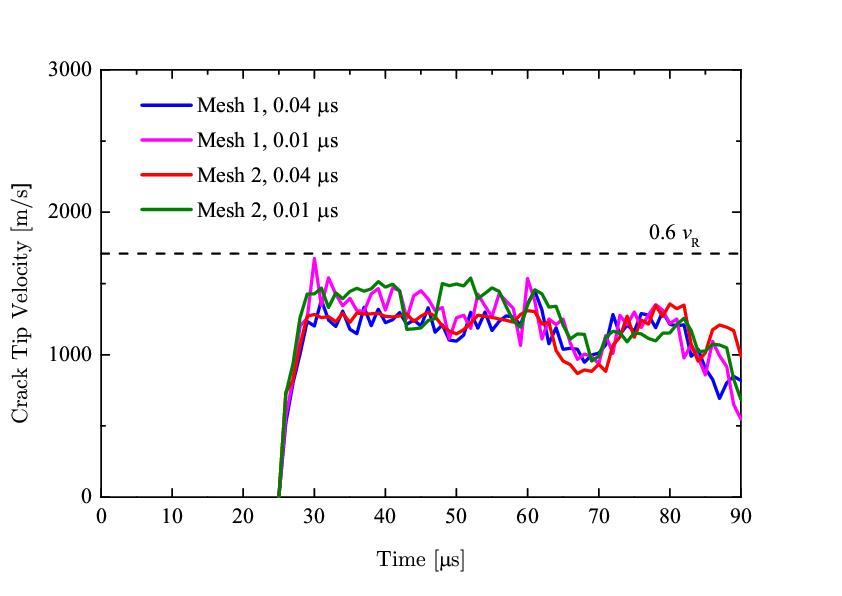}
	\caption{The crack-tip velocity curves for the dynamic shear loading example}
	\label{The crack-tip velocity curves for the dynamic shear loading example}
	\end{figure}

We also obtain the velocity of crack tip as shown in Fig. \ref{The crack-tip velocity curves for the dynamic shear loading example}. The velocity $\bm v_n$  at the crack tip is calculated as follows:

	\begin{equation}
	\bm v_n = (\bm x_n-\bm x_{n-1})/\Delta t
	\end{equation}

\noindent with $\bm x_n$  the position of current crack tip at time $t_n$. The position of crack tip is determined from the iso-curve of the phase-field $\phi  = 0.75$ according to \citet{borden2012phase}. The crack speed increases to a velocity of  $0.6v_R$ and remains nearly constant until the end of the simulation. Figure \ref{The crack-tip velocity curves for the dynamic shear loading example} also shows that the results are independent for two different time steps.

\begin{figure}[htbp]
	\centering
	\includegraphics[width = 8cm]{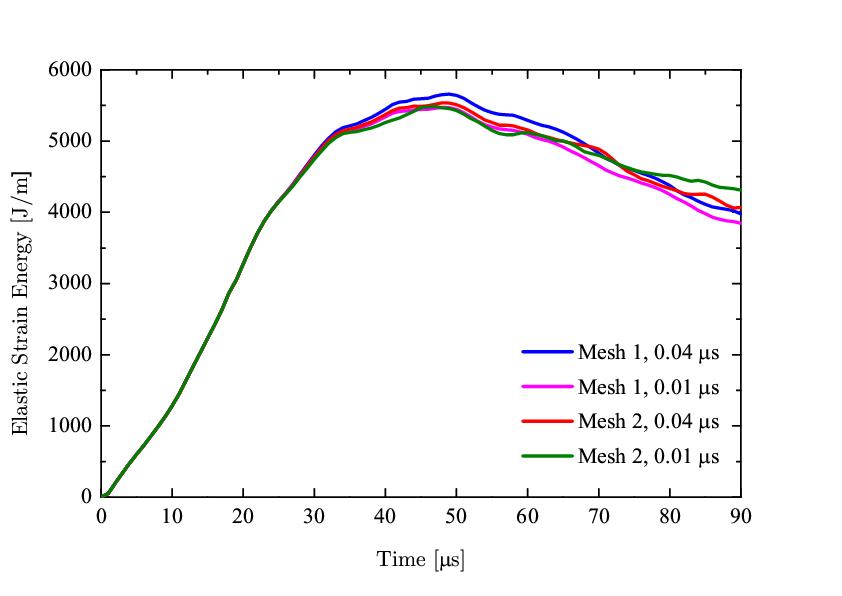}
	\caption{Elastic strain energy curves for the dynamic shear loading example}
	\label{The elastic strain energy curves for the dynamic shear loading example}
	\end{figure}

\begin{figure}[htbp]
	\centering
	\includegraphics[width = 8cm]{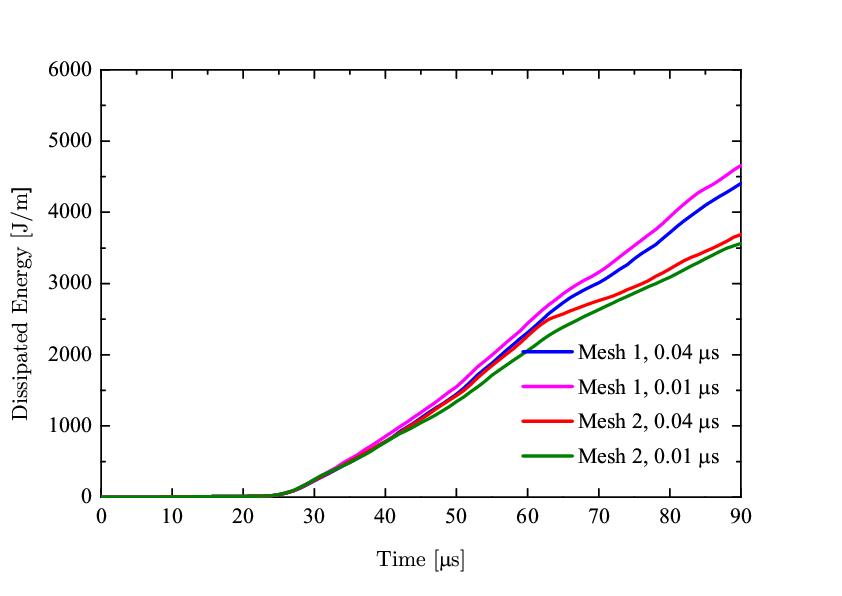}
	\caption{Dissipated energy curves for the dynamic shear loading example}
	\label{The dissipated energy curves for the dynamic shear loading example}
	\end{figure}

Figures \ref{The elastic strain energy curves for the dynamic shear loading example} and \ref{The dissipated energy curves for the dynamic shear loading example} present the elastic strain energy and dissipated energy curves, respectively. The elastic strain energy is calculated as

	\begin{equation}
	E_{\varepsilon}=\int_{\Omega} \left\{[(1-k)(1-\phi)^2+k]\psi_{\varepsilon}^+ +\psi_{\varepsilon}^- \right\}d\Omega
	\end{equation}

\noindent while the dissipated energy is obtained by

	\begin{equation}
	E_{d}=\int_{\Omega} G_c \left [\frac {\phi^2} {2l_0}+\frac {l_0} 2 \frac {\partial \phi}{\partial x_i} \frac{\partial \phi}{\partial x_i}\right] d{\Omega}
	\end{equation}

As shown in Fig. \ref{The elastic strain energy curves for the dynamic shear loading example}, the elastic strain energy curves for Mesh 1 and Mesh 2 are in good agreement. The elastic strain energy increases quickly before reaching a maximum. After that, the elastic strain energy starts to decrease slowly. In Fig. \ref{The dissipated energy curves for the dynamic shear loading example}, the dissipated energy increases as the time increases. . 

\begin{figure}[htbp]
	\centering
	\subfigure[$G_c=5\times10^3$ J/m$^2$]{\includegraphics[width = 4cm]{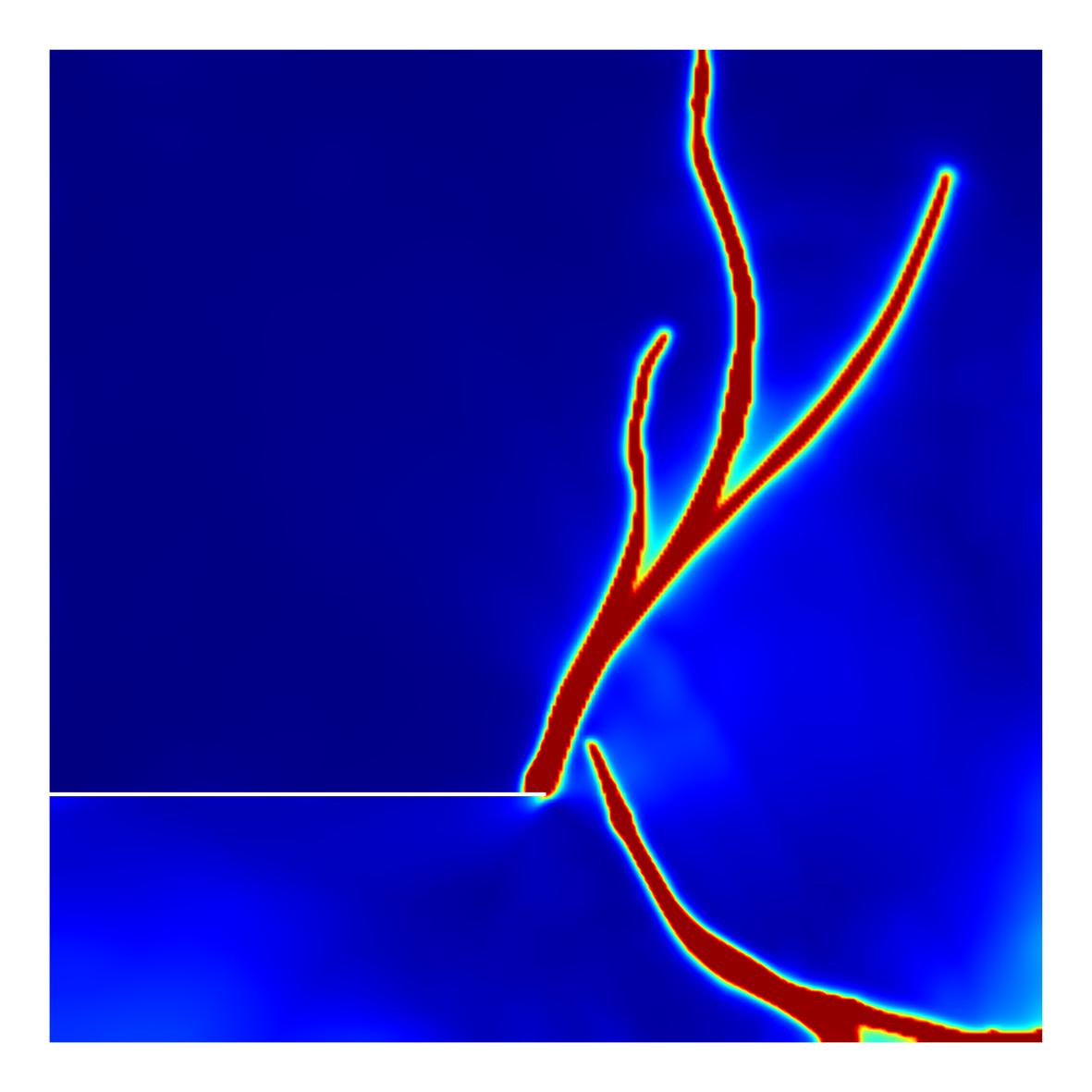}}
	\subfigure[$G_c=1\times10^4$ J/m$^2$]{\includegraphics[width = 4cm]{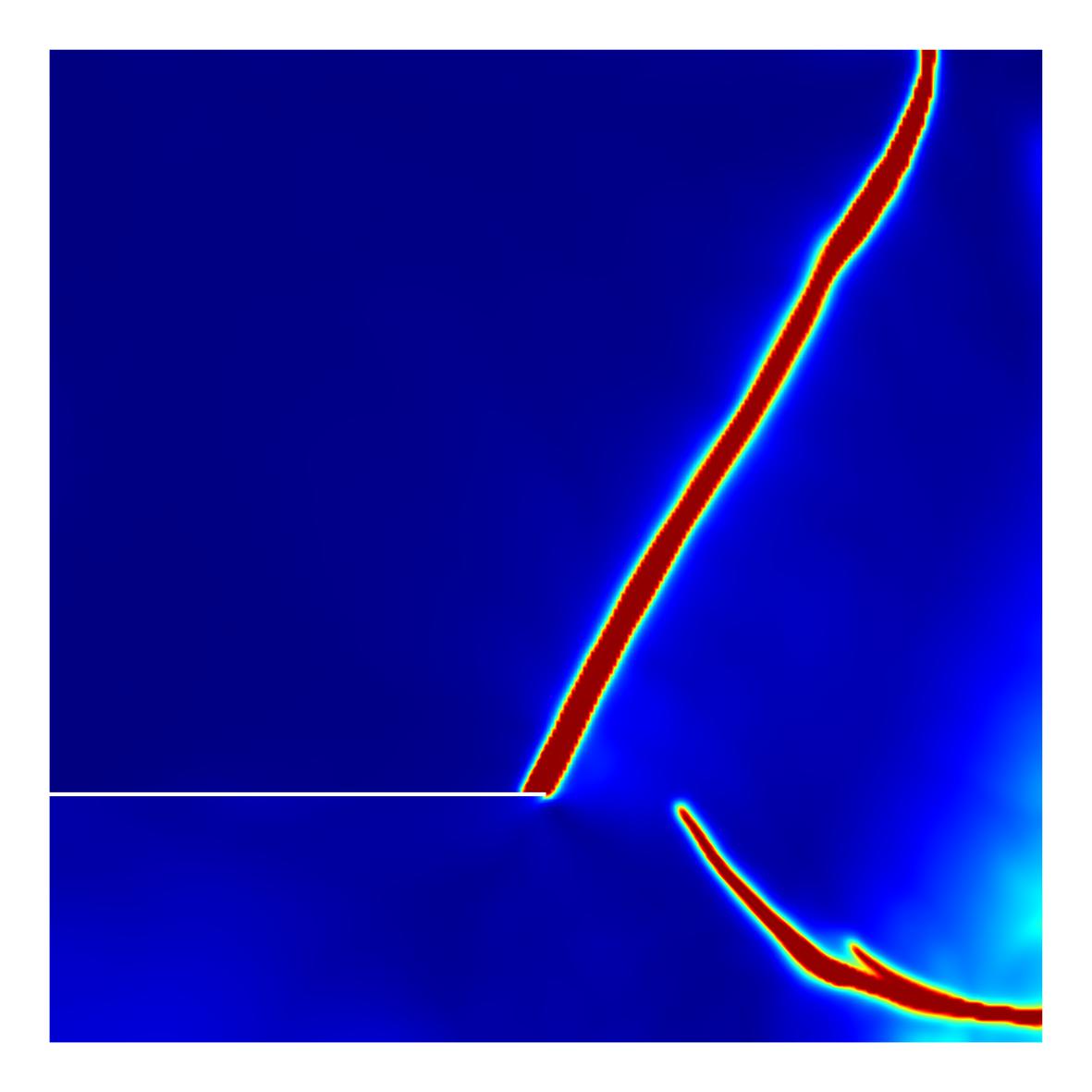}}
	\subfigure[$G_c=2.213\times10^4$ J/m$^2$]{\includegraphics[width = 4cm]{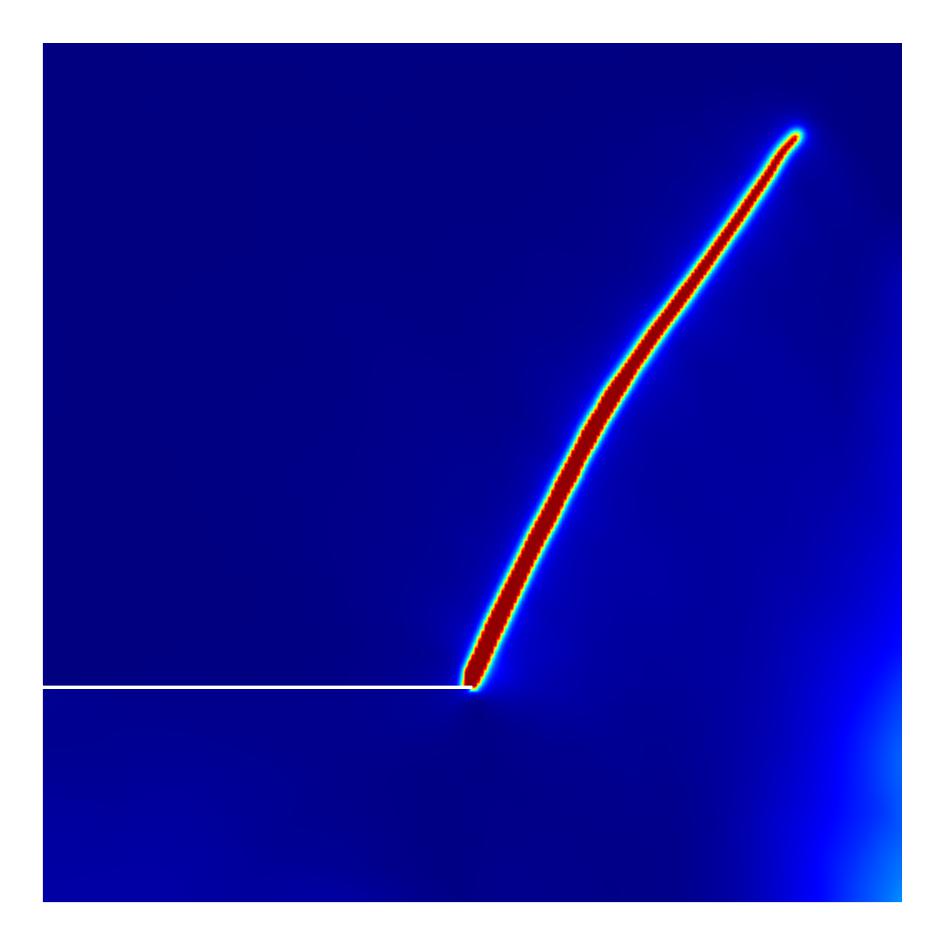}}
	\subfigure[$G_c=3\times10^4$ J/m$^2$]{\includegraphics[width = 4cm]{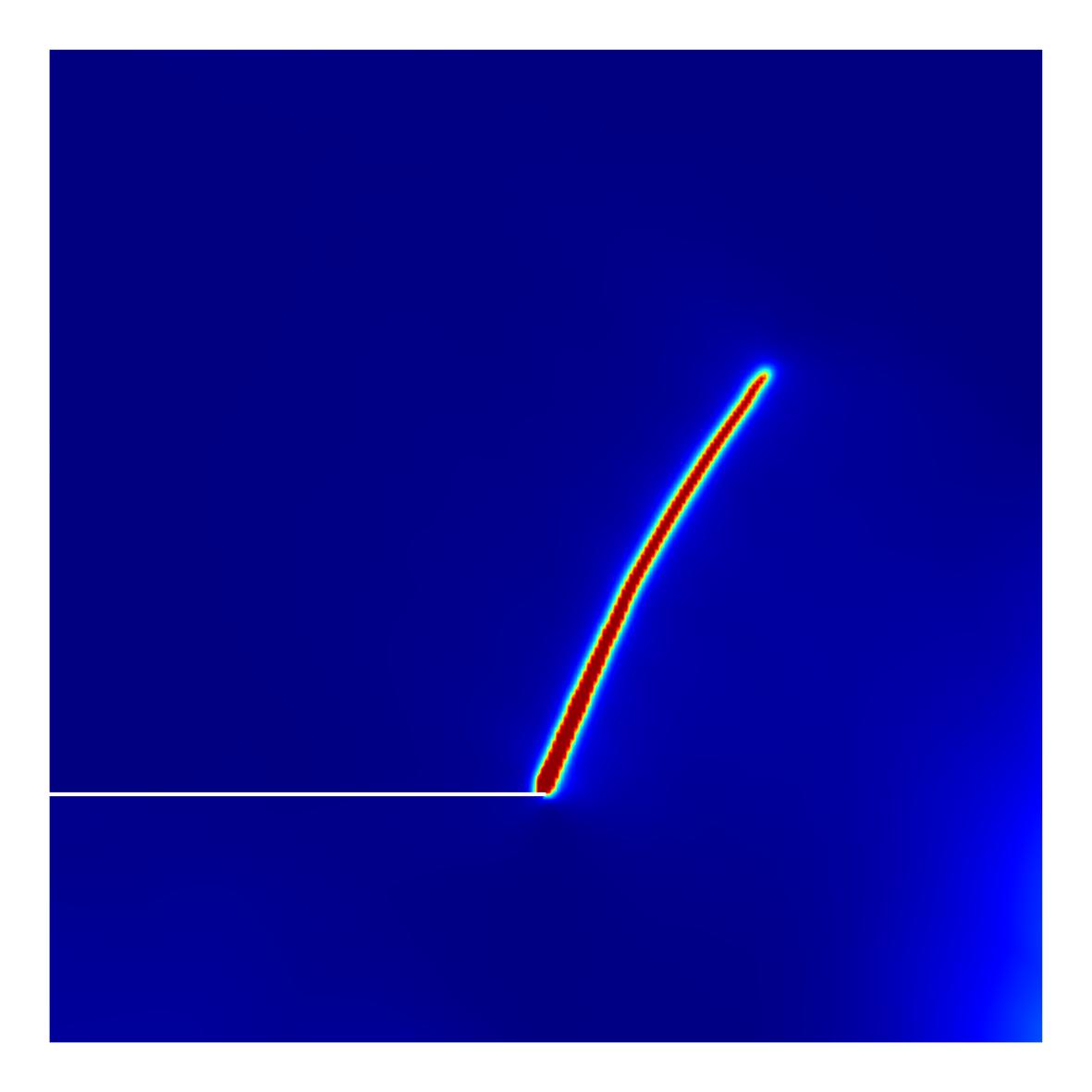}}
	\caption{Phase field  of dynamic shear loading tests at 90 $\mu$s for different $G_c$}
	\label{Phase field  of dynamic shear loading tests at 90 for different G_c}
	\end{figure}

In the end of this example, we test the influence of the critical energy release rate $G_c$. We present the results of $G_c=5\times10^3$, $1\times10^4$, $2.213\times10^4$, and $3\times10^4$ J/m$^2$ for Mesh 1 and $\Delta t=0.04$ $\mu$s. Figure \ref{Phase field  of dynamic shear loading tests at 90 for different G_c} gives the phase field at 90 $\mu$s for different $G_c$. More complex crack patterns are observed for smaller $G_c$. Crack branching occurs when $G_c=5\times10^3$ and $1\times10^4$ J/m$^2$. Secondary cracks occur in the bottom right corner of the model because of wave reflections \citep{song2008comparative}. However, for a larger $G_c$, only a single crack is seen. The crack is more hard to reach the upper boundary of the plate when $G_c$ becomes larger. The variation in the critical energy release rate $G_c$ does not change the main crack pattern. The main cracks initiate from the tip of the pre-existing crack. 

\begin{figure}[htbp]
	\centering
	\includegraphics[width = 8cm]{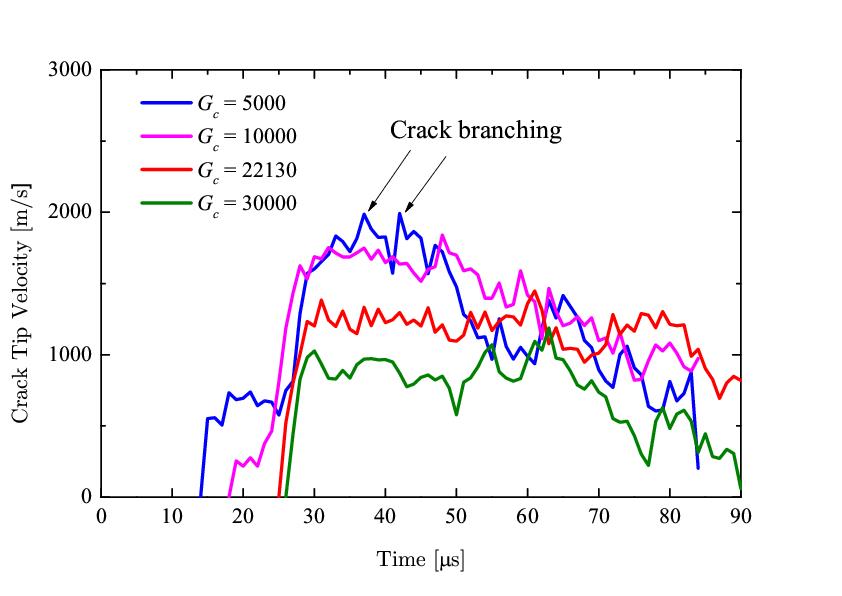}
	\caption{Crack-tip velocity of the dynamic shear loading example for different $G_c$}
	\label{Crack-tip velocity of the dynamic shear loading example for different G_c}
	\end{figure}

Figure \ref{Crack-tip velocity of the dynamic shear loading example for different G_c} presents the crack-tip velocity for different $G_c$. The velocity is calculated on the longest cracks in Fig. \ref{Phase field  of dynamic shear loading tests at 90 for different G_c}. Figure \ref{Crack-tip velocity of the dynamic shear loading example for different G_c} shows that the maximum crack-tip velocity decreases and the time for crack initiation increases with the increase in $G_c$. Particularly, for $G_c=5\times10^3$ J/m$^2$, two crack branching successively occur when the crack-tip velocity reaches the maximum. 

\subsection{2D dynamic crack branching under tension}

\begin{figure}[htbp]
	\centering
	\includegraphics[width = 8cm]{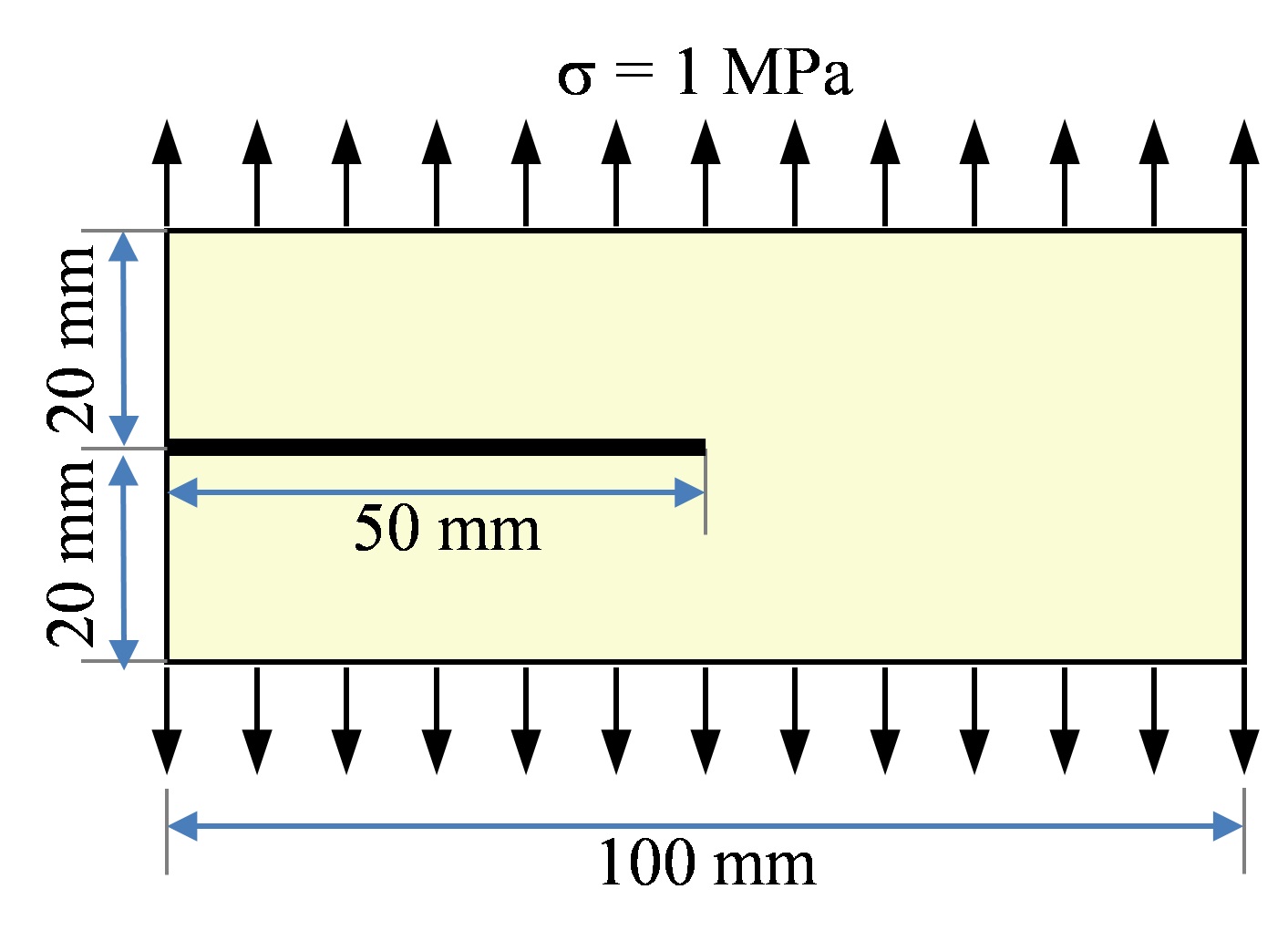}
	\caption{Geometry and boundary conditions for the case of dynamic crack branching}
	\label{Geometry and boundary conditions for the case of dynamic crack branching}
	\end{figure}

The last example is another classical benchmark problem for dynamic fracture: a pre-notched rectangular plate subjected to uni-axial traction. Figure \ref{Geometry and boundary conditions for the case of dynamic crack branching} gives the geometry of the plate along with the boundary conditions. This benchmark test has been calculated by \citet{song2008comparative}, \citet{liu2016abaqus} and \citet{rabczuk2007three2,rabczuk2004cracking}.

The material parameters are adopted from \citep{borden2012phase} as $\rho  = 2450$ kg/m$^3$, $E  = 32$ GPa,  $\nu = 0.2$ and $G_c  = 3$ J/m$^2$.  Plane strain conditions are assumed. The length scale parameter $l_0$  is fixed as $5.0\times10^{-4}$ m. The pre-existing notch is modeled by introducing an initial history strain field as explained in Section 3.2. Q4 elements are used to discretize the plate with a uniform mesh; $k=1\times10^{-9}$  is also picked to avoid the singularity during calculation. We conduct the simulation using two different mesh levels: Mesh 1 with size $h  = 2.5\times10^{-4}$ m ($l_0=2h$) and Mesh 2 with $h  = 1.25\times10^{-4}$ m ($l_0=4h$), respectively. For each mesh, the time step is chosen as 0.1 $\mu$s, 0.05 $\mu$s and 0.025 $\mu$s.

\begin{figure}[htbp]
	\centering
	\subfigure[Mesh 1,  $\Delta t = 0.1$ $\mu$s]{\includegraphics[width = 8cm]{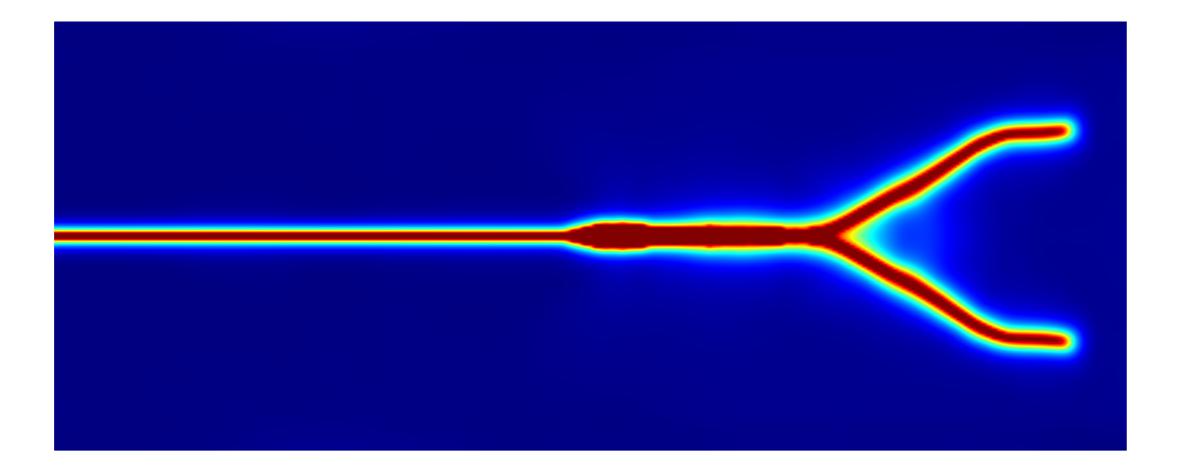}}
	\subfigure[Mesh 2,  $\Delta t = 0.1$ $\mu$s]{\includegraphics[width = 8cm]{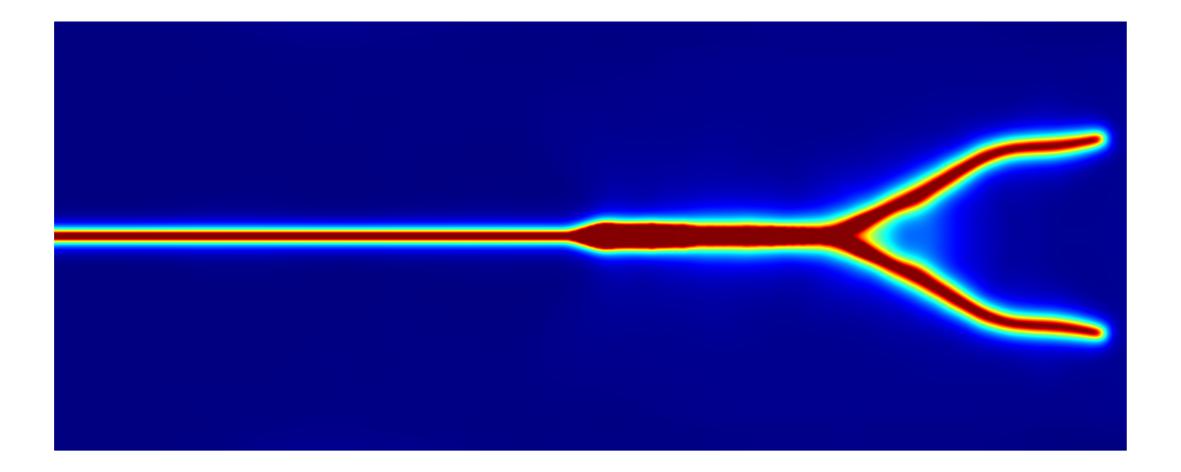}}
	
	\subfigure[Mesh 1,  $\Delta t = 0.05$ $\mu$s]{\includegraphics[width = 8cm]{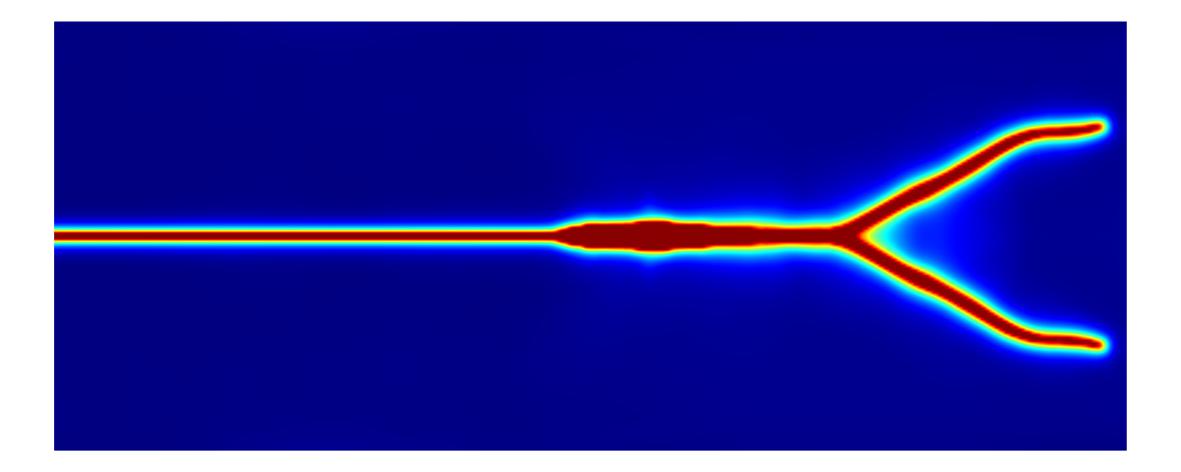}}
	\subfigure[Mesh 2,  $\Delta t = 0.05$ $\mu$s]{\includegraphics[width = 8cm]{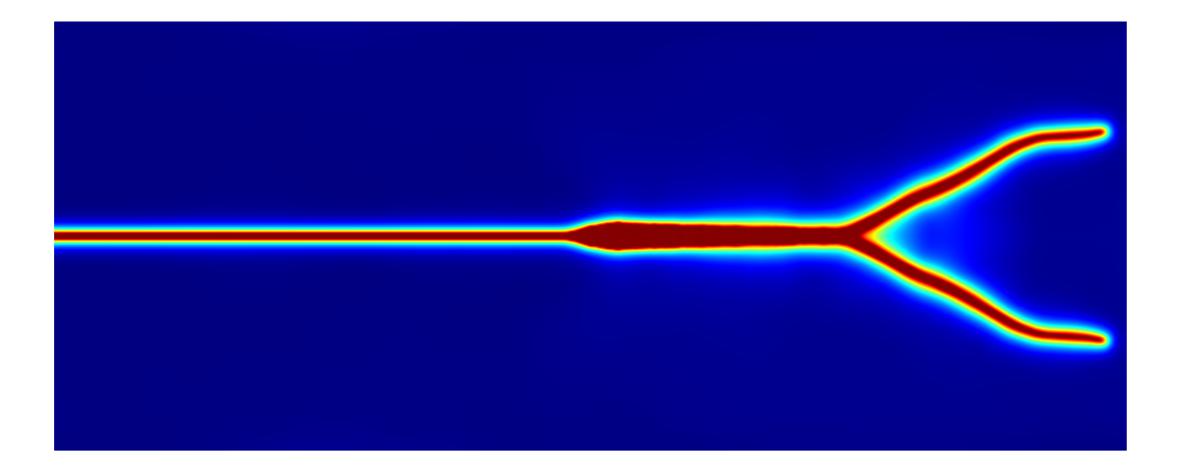}}

	\subfigure[Mesh 1,  $\Delta t = 0.025$ $\mu$s]{\includegraphics[width = 8cm]{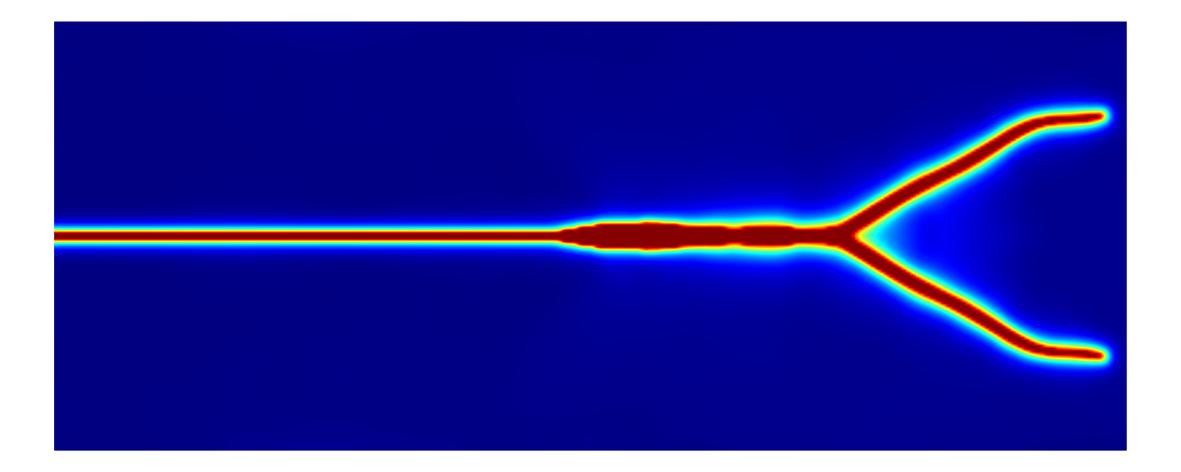}}
	\subfigure[Mesh 2,  $\Delta t = 0.025$ $\mu$s]{\includegraphics[width = 8cm]{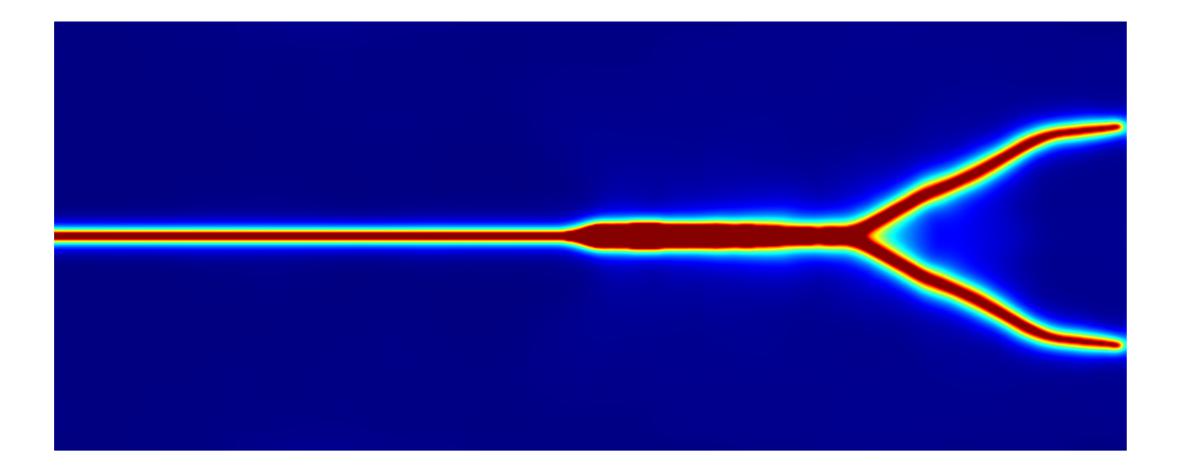}}
	\caption{2D crack-branching example. Phase field at $t= 80$ $\mu$s for $G_c=3$ J/m$^2$.}
	\label{2D crack-branching example. Phase field at 80}
	\end{figure}

Figure \ref{2D crack-branching example. Phase field at 80} shows the results of the phase field at 80 $\mu$s. Mesh 1 and Mesh 2 have similar crack patterns for different time steps and the crack fail to reach the boundary because the staggered scheme is adopted. As shown in \citep{liu2016abaqus} and \citep{borden2012phase}, the monolithic scheme is easy to reach the boundary. Thus, the staggered scheme in this work seems to delay the crack branching. In addition, Fig. \ref{2D crack-branching example. Phase field at 80} shows that a coarser mesh and a larger time step have wider cracks and the cracks are much more difficult to reach the boundary than a finer mesh and a smaller time step.

\begin{figure}[htbp]
	\centering
	\subfigure[$\Delta t = 0.05$ $\mu$s]{\includegraphics[width = 9cm]{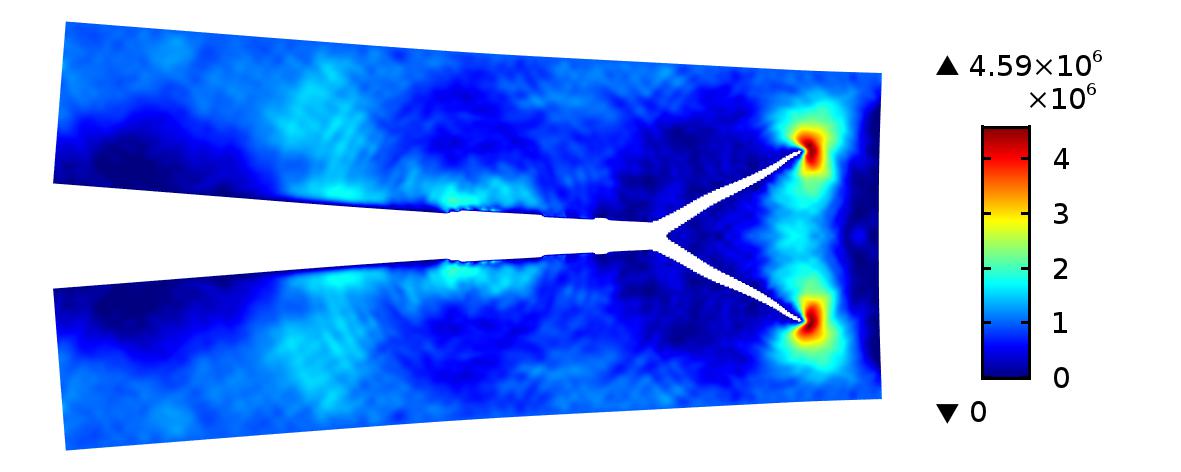}}

	\subfigure[$\Delta t = 0.025$ $\mu$s]{\includegraphics[width = 9cm]{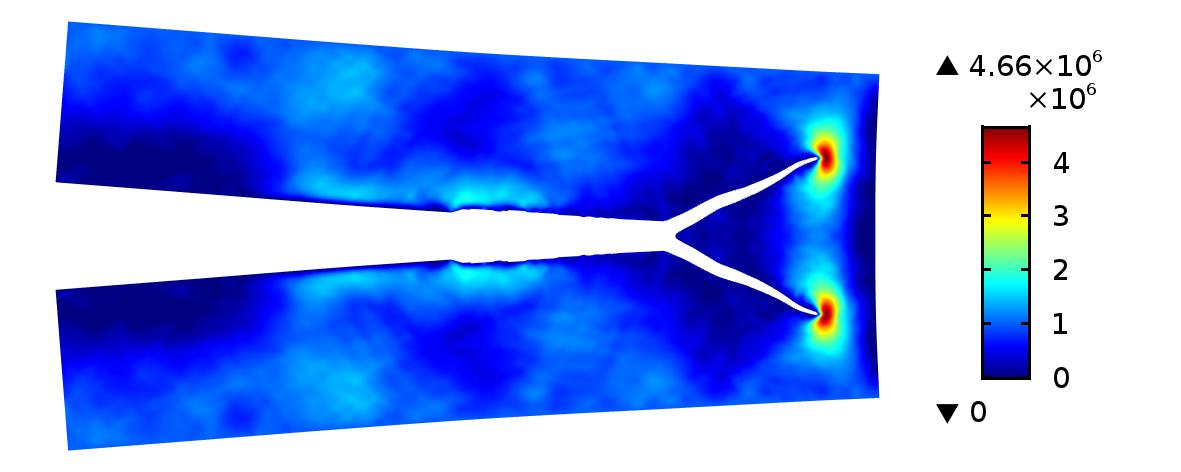}}
	\caption{Maximum tensile stress of the 2D crack branching example at $t  = 70$ $\mu$s for Mesh 2. The stress is measured in Pa.}
	\label{Maximum tensile stress of the 2D crack branching example}
	\end{figure}

Figure \ref{Maximum tensile stress of the 2D crack branching example} gives the maximum tensile stress for Mesh 1 and Mesh 2 at $t  = 70$ $\mu$s. In the figure, we scale the displacement field by a factor of 100 and the regions where the phase field is more than 0.95 are also removed from the figure to display the broken geometry of the plate. Figure \ref{Maximum tensile stress of the 2D crack branching example} shows a tensile stress concentration at the crack tip and the results of both meshes are similar and in good agreement with the results of \citet{liu2016abaqus}.

\begin{figure}[htbp]
	\centering
	\includegraphics[width = 8cm]{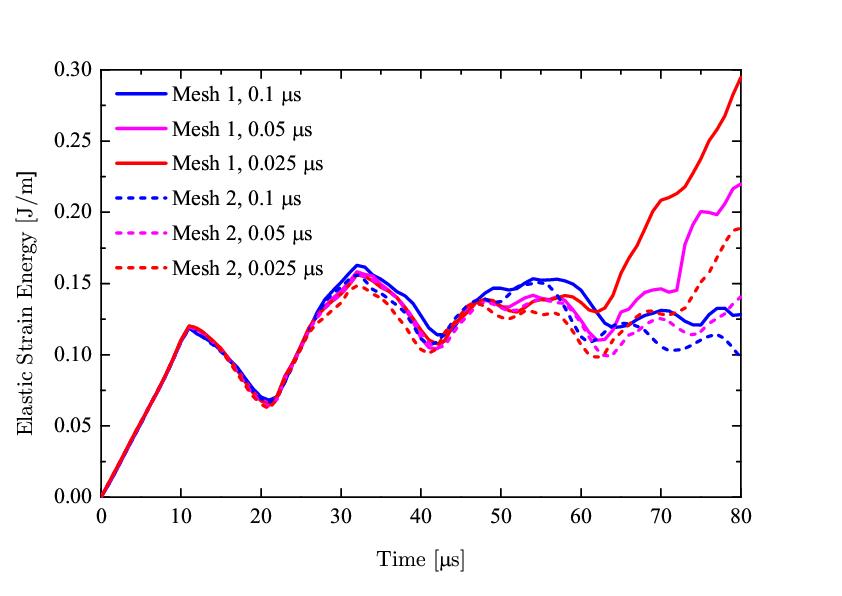}
	\caption{Elastic strain energy curves for the 2D crack branching example}
	\label{Elastic strain energy curves for the 2D crack branching example}
	\end{figure}

\begin{figure}[htbp]
	\centering
	\includegraphics[width = 8cm]{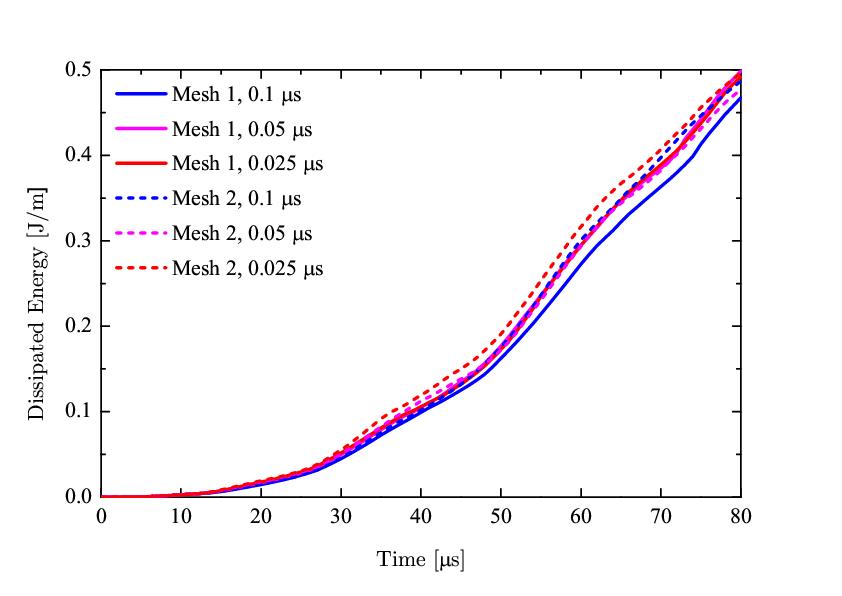}
	\caption{Dissipated energy curves for the 2D crack branching example}
	\label{Dissipated energy curves for the 2D crack branching example}
	\end{figure}

Figure \ref{Elastic strain energy curves for the 2D crack branching example} presents the elastic strain energy curves for all the meshes and time steps. All the curves are in good agreement before 60 $\mu$s. However, some discrepancies exist after 60 $\mu$s. For a smaller time step, the elastic strain energy approximately decreases with the increase in time. But for a larger time step, the elastic strain energy increases as the time goes. Additionally, the elastic strain energy for the coarser mesh (Mesh 1) is larger than that for the finer mesh (Mesh 2). Figure \ref{Dissipated energy curves for the 2D crack branching example} represent the dissipated energy curves for the 2D crack branching example. In Fig. \ref{Dissipated energy curves for the 2D crack branching example}, the dissipated energy increases as the time increases. Meanwhile, the results for both meshes and all the time steps are in quite good agreement.

\begin{figure}[htbp]
	\centering
	\includegraphics[width = 8cm]{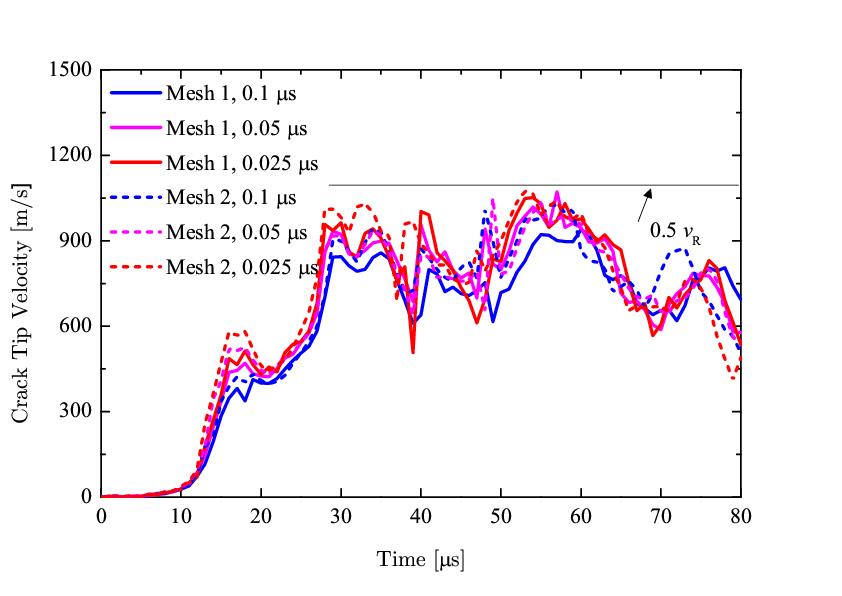}
	\caption{Crack tip velocity curves for the 2D crack branching example}
	\label{Crack tip velocity curves for the 2D crack branching example}
	\end{figure}

Figure \ref{Crack tip velocity curves for the 2D crack branching example} presents the curves of crack tip velocity achieved from the post-processing. As depicted in previous example, the crack tip is found in the iso-curve of the phase-field $\phi  = 0.75$. All the curves in Fig. \ref{Crack tip velocity curves for the 2D crack branching example} have the similar trend for the crack branching case. As we notice in the simulation, all the velocities are smaller than $0.5v_R$. This finding is in good agreement with the results of \citet{borden2012phase}. In addition, the crack widening occurs when the time  $t = 28$ $\mu$s $\sim$ 30 $\mu$s. The crack starts to branch when the time goes to 48 $\mu$s $\sim$ 51 $\mu$s, which is later than the results of \citet{borden2012phase}. This also proves that the staggered scheme can delay the time for the crack branching.

\begin{figure}[htbp]
	\centering
	\subfigure[$G_c=0.5$ J/m$^2$, $t=56$ $\mu$s]{\includegraphics[width = 8cm]{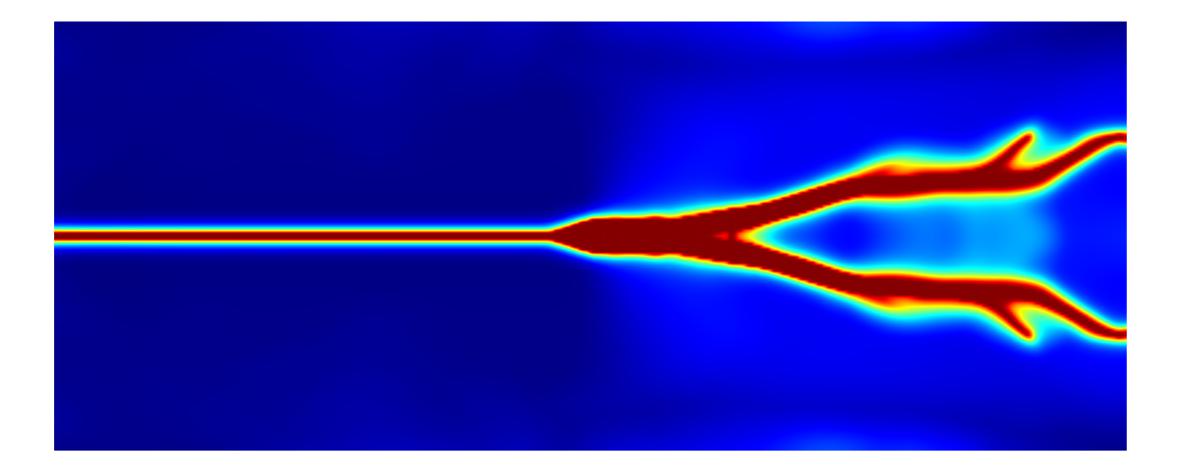}}
	\subfigure[$G_c=1$ J/m$^2$, $t=66$ $\mu$s]{\includegraphics[width = 8cm]{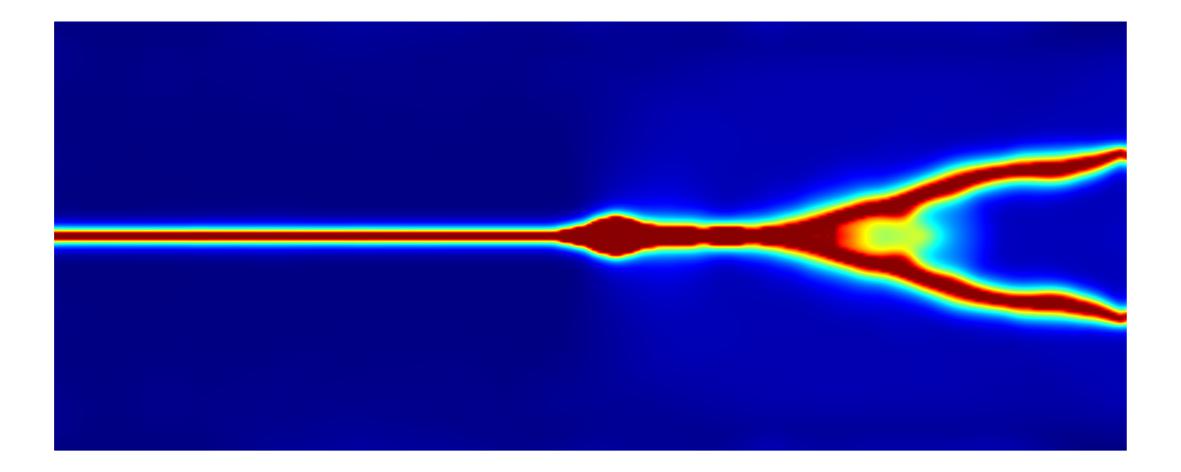}}
	\subfigure[$G_c=5$ J/m$^2$, $t=113$ $\mu$s]{\includegraphics[width = 8cm]{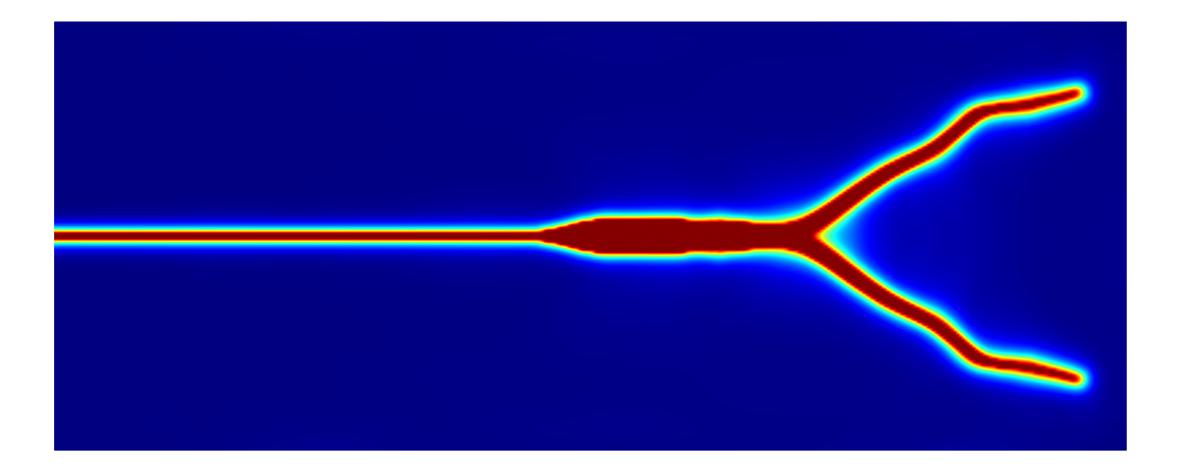}}
	\subfigure[$G_c=10$ J/m$^2$, $t=142$ $\mu$s]{\includegraphics[width = 8cm]{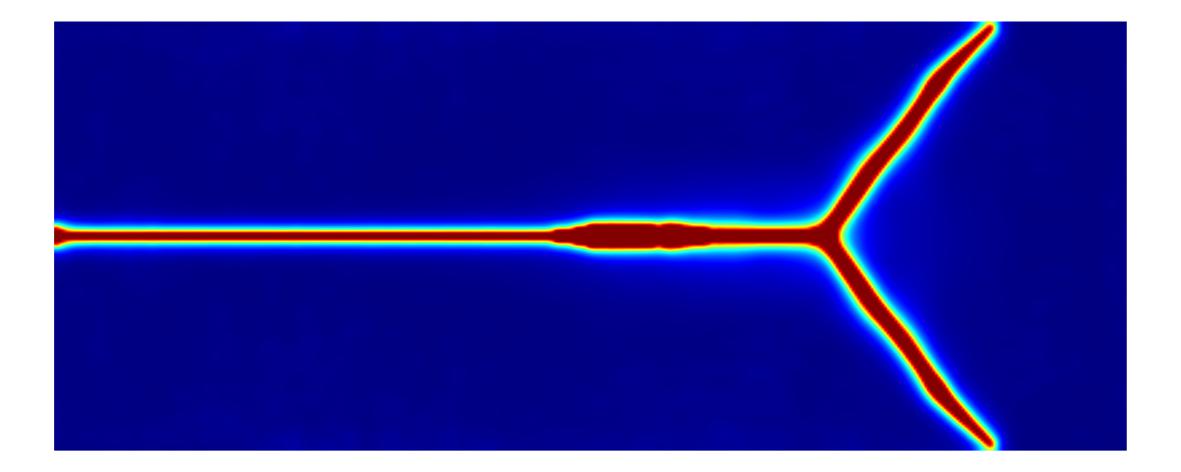}}
	\caption{Crack patterns of the 2D crack branching example for different $G_c$}
	\label{Crack patterns of the 2D crack branching example for different G_c}
	\end{figure}

\begin{figure}[htbp]
	\centering
	\includegraphics[width = 8cm]{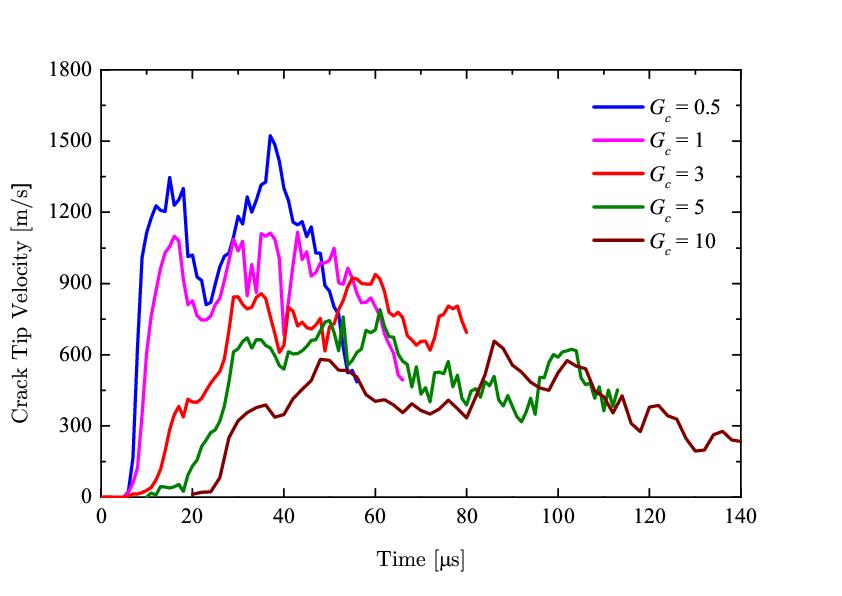}
	\caption{Crack-tip velocity of the 2D crack branching example for different $G_c$}
	\label{Crack-tip velocity of the 2D crack branching example for different G_c}
	\end{figure}

To end the example of 2D crack branching, we test the influence of the critical energy release rate $G_c$ on the crack pattern and  the crack-tip velocity as shown in Figs. \ref{Crack patterns of the 2D crack branching example for different G_c} and \ref{Crack-tip velocity of the 2D crack branching example for different G_c}.  The results are now presented only using Mesh 1 and $\Delta t=0.1$ $\mu$s. When $G_c=0.5$ J/m$^2$, multiple crack branching can be seen in Fig. \ref{Crack patterns of the 2D crack branching example for different G_c}. Figure \ref{Crack patterns of the 2D crack branching example for different G_c} also shows that,  for a larger $G_c$, the crack will propagate at a larger angle with the horizontal after the first branching. The crack is more hard to propagate as well. Figure \ref{Crack-tip velocity of the 2D crack branching example for different G_c} depicts the crack-tip velocity for different $G_c$. The maximum crack-tip velocity decreases with the increase in $G_c$. Larger $G_c$ will cause crack propagation at a larger velocity, which is in good agreement with the 2D dynamic shear example.

\section {Conclusions}

In this work, we present the implementation of the numerical phase-field modeling for crack propagation in commercial finite element software COMSOL. The previous existing phase-field models for quasi-static and dynamic crack propagation problems are reviewed and implemented in proper forms in COMSOL. The elastic strain energy density is decomposed into two individual parts resulting from compression and tension, respectively. Thus, only the tension induced cracks are obtained. In COMSOL, the simulation is facilitated by establishing four modules and placing the coupling terms appropriately. The coupled system is solved using a staggered scheme. Here, an implicit time integration scheme is used and employ the Newton-Raphson iteration is adopted to compute the individual fields.


A number of 2D and 3D benchmark examples for quasi-static and dynamic crack propagations are tested. We also check the effects of length scale parameter, critical energy release rate, mesh size, and time step size on the results. All the simulations have the correct crack patterns and satisfactory accuracy, showing the feasibility of implementing phase-field model for crack propagation by COMSOL, even in 3D spaces. For future work, COMSOL will be more helpful and effective for implementing and extending the phase-field model to problems with more fields.

\section*{Acknowledgement}
The financial support provided by the Sino-German (CSC-DAAD) Postdoc Scholarship Program 2016, the Natural Science Foundation of China (51474157), and RISE-project BESTOFRAC (734370) is gratefully acknowledged.

\bibliography{references}

\begin{thebibliography}{61}
\providecommand{\natexlab}[1]{#1}
\providecommand{\url}[1]{\texttt{#1}}
\expandafter\ifx\csname urlstyle\endcsname\relax
  \providecommand{\doi}[1]{doi: #1}\else
  \providecommand{\doi}{doi: \begingroup \urlstyle{rm}\Url}\fi

\bibitem[Anderson(2005)]{anderson2005fracture}
TL~Anderson.
\newblock \emph{Fracture mechanics: fundamentals and applications}.
\newblock CRC press, 2005.

\bibitem[Liu et~al.(2016)Liu, Li, Msekh, and Zuo]{liu2016abaqus}
Guowei Liu, Qingbin Li, Mohammed~A Msekh, and Zheng Zuo.
\newblock Abaqus implementation of monolithic and staggered schemes for
  quasi-static and dynamic fracture phase-field model.
\newblock \emph{Computational Materials Science}, 121:\penalty0 35--47, 2016.

\bibitem[Klinsmann et~al.(2015)Klinsmann, Rosato, Kamlah, and
  McMeeking]{klinsmann2015assessment}
Markus Klinsmann, Daniele Rosato, Marc Kamlah, and Robert~M McMeeking.
\newblock An assessment of the phase field formulation for crack growth.
\newblock \emph{Computer Methods in Applied Mechanics and Engineering},
  294:\penalty0 313--330, 2015.

\bibitem[Budarapu and Rabczuk(2017)]{budarapu2017multiscale}
PR~Budarapu and T~Rabczuk.
\newblock Multiscale methods for fracture: A review.
\newblock \emph{Journal of the Indian Institute of Science}, 97\penalty0
  (3):\penalty0 339--376, 2017.

\bibitem[Budarapu et~al.(2017)Budarapu, Reinoso, and
  Paggi]{budarapu2017concurrently}
Pattabhi~R Budarapu, Jos{\'e} Reinoso, and Marco Paggi.
\newblock Concurrently coupled solid shell-based adaptive multiscale method for
  fracture.
\newblock \emph{Computer Methods in Applied Mechanics and Engineering},
  319:\penalty0 338--365, 2017.

\bibitem[Yang et~al.(2015)Yang, Budarapu, Mahapatra, Bordas, Zi, and
  Rabczuk]{yang2015meshless}
Shih-Wei Yang, Pattabhi~R Budarapu, D~Roy Mahapatra, St{\'e}phane~PA Bordas,
  Goangseup Zi, and Timon Rabczuk.
\newblock A meshless adaptive multiscale method for fracture.
\newblock \emph{Computational Materials Science}, 96:\penalty0 382--395, 2015.

\bibitem[Budarapu et~al.(2014{\natexlab{a}})Budarapu, Gracie, Yang, Zhuang, and
  Rabczuk]{budarapu2014efficient}
Pattabhi~R Budarapu, Robert Gracie, Shih-Wei Yang, Xiaoying Zhuang, and Timon
  Rabczuk.
\newblock Efficient coarse graining in multiscale modeling of fracture.
\newblock \emph{Theoretical and Applied Fracture Mechanics}, 69:\penalty0
  126--143, 2014{\natexlab{a}}.

\bibitem[Budarapu et~al.(2014{\natexlab{b}})Budarapu, Gracie, Bordas, and
  Rabczuk]{budarapu2014adaptive}
Pattabhi~R Budarapu, Robert Gracie, St{\'e}phane~PA Bordas, and Timon Rabczuk.
\newblock An adaptive multiscale method for quasi-static crack growth.
\newblock \emph{Computational Mechanics}, 53\penalty0 (6):\penalty0 1129--1148,
  2014{\natexlab{b}}.

\bibitem[Ingraffea and Saouma(1985)]{ingraffea1985numerical}
AR~Ingraffea and V~Saouma.
\newblock Numerical modeling of discrete crack propagation in reinforced and
  plain concrete.
\newblock In \emph{Fracture mechanics of concrete: structural application and
  numerical calculation}, pages 171--225. Springer, 1985.

\bibitem[Mo{\"e}s and Belytschko(2002)]{moes2002extended}
Nicolas Mo{\"e}s and Ted Belytschko.
\newblock Extended finite element method for cohesive crack growth.
\newblock \emph{Engineering fracture mechanics}, 69\penalty0 (7):\penalty0
  813--833, 2002.

\bibitem[Chen et~al.(2012)Chen, Rabczuk, Bordas, Liu, Zeng, and
  Kerfriden]{chen2012extended}
L~Chen, T~Rabczuk, Stephane Pierre~Alain Bordas, GR~Liu, KY~Zeng, and Pierre
  Kerfriden.
\newblock Extended finite element method with edge-based strain smoothing
  (esm-xfem) for linear elastic crack growth.
\newblock \emph{Computer Methods in Applied Mechanics and Engineering},
  209:\penalty0 250--265, 2012.

\bibitem[Fries and Belytschko(2010)]{fries2010extended}
Thomas-Peter Fries and Ted Belytschko.
\newblock The extended/generalized finite element method: an overview of the
  method and its applications.
\newblock \emph{International Journal for Numerical Methods in Engineering},
  84\penalty0 (3):\penalty0 253--304, 2010.

\bibitem[Chau-Dinh et~al.(2012)Chau-Dinh, Zi, Lee, Rabczuk, and
  Song]{chau2012phantom}
Thanh Chau-Dinh, Goangseup Zi, Phill-Seung Lee, Timon Rabczuk, and Jeong-Hoon
  Song.
\newblock Phantom-node method for shell models with arbitrary cracks.
\newblock \emph{Computers \& Structures}, 92:\penalty0 242--256, 2012.

\bibitem[Rabczuk et~al.(2008{\natexlab{a}})Rabczuk, Zi, Gerstenberger, and
  Wall]{rabczuk2008new}
Timon Rabczuk, Goangseup Zi, Axel Gerstenberger, and Wolfgang~A Wall.
\newblock A new crack tip element for the phantom-node method with arbitrary
  cohesive cracks.
\newblock \emph{International Journal for Numerical Methods in Engineering},
  75\penalty0 (5):\penalty0 577--599, 2008{\natexlab{a}}.

\bibitem[Zhou and Molinari(2004)]{zhou2004dynamic}
Fenghua Zhou and Jean-Francois Molinari.
\newblock Dynamic crack propagation with cohesive elements: a methodology to
  address mesh dependency.
\newblock \emph{International Journal for Numerical Methods in Engineering},
  59\penalty0 (1):\penalty0 1--24, 2004.

\bibitem[Nguyen et~al.(2001)Nguyen, Repetto, Ortiz, and
  Radovitzky]{nguyen2001cohesive}
O~Nguyen, EA~Repetto, Michael Ortiz, and RA~Radovitzky.
\newblock A cohesive model of fatigue crack growth.
\newblock \emph{International Journal of Fracture}, 110\penalty0 (4):\penalty0
  351--369, 2001.

\bibitem[Rabczuk et~al.(2008{\natexlab{b}})Rabczuk, Zi, Bordas, and
  Nguyen-Xuan]{rabczuk2008geometrically}
Timon Rabczuk, Goangseup Zi, St{\'e}phane Bordas, and Hung Nguyen-Xuan.
\newblock A geometrically non-linear three-dimensional cohesive crack method
  for reinforced concrete structures.
\newblock \emph{Engineering Fracture Mechanics}, 75\penalty0 (16):\penalty0
  4740--4758, 2008{\natexlab{b}}.

\bibitem[Belytschko and Lin(1987)]{belytschko1987three}
Ted Belytschko and Jerry~I Lin.
\newblock A three-dimensional impact-penetration algorithm with erosion.
\newblock \emph{International Journal of Impact Engineering}, 5\penalty0
  (1-4):\penalty0 111--127, 1987.

\bibitem[Johnson and Stryk(1987)]{johnson1987eroding}
Gordon~R Johnson and Robert~A Stryk.
\newblock Eroding interface and improved tetrahedral element algorithms for
  high-velocity impact computations in three dimensions.
\newblock \emph{International Journal of Impact Engineering}, 5\penalty0
  (1-4):\penalty0 411--421, 1987.

\bibitem[Liu et~al.(2014)Liu, Filonova, Hu, Yuan, Fish, Yuan, and
  Belytschko]{liu2014regularized}
Yang Liu, Vasilina Filonova, Nan Hu, Zifeng Yuan, Jacob Fish, Zheng Yuan, and
  Ted Belytschko.
\newblock A regularized phenomenological multiscale damage model.
\newblock \emph{International Journal for Numerical Methods in Engineering},
  99\penalty0 (12):\penalty0 867--887, 2014.

\bibitem[Song et~al.(2008)Song, Wang, and Belytschko]{song2008comparative}
Jeong-Hoon Song, Hongwu Wang, and Ted Belytschko.
\newblock A comparative study on finite element methods for dynamic fracture.
\newblock \emph{Computational Mechanics}, 42\penalty0 (2):\penalty0 239--250,
  2008.

\bibitem[Miehe et~al.(2010{\natexlab{a}})Miehe, Hofacker, and
  Welschinger]{miehe2010phase}
Christian Miehe, Martina Hofacker, and Fabian Welschinger.
\newblock A phase field model for rate-independent crack propagation: Robust
  algorithmic implementation based on operator splits.
\newblock \emph{Computer Methods in Applied Mechanics and Engineering},
  199\penalty0 (45):\penalty0 2765--2778, 2010{\natexlab{a}}.

\bibitem[Miehe et~al.(2010{\natexlab{b}})Miehe, Welschinger, and
  Hofacker]{miehe2010thermodynamically}
C~Miehe, F~Welschinger, and M~Hofacker.
\newblock Thermodynamically consistent phase-field models of fracture:
  Variational principles and multi-field fe implementations.
\newblock \emph{International Journal for Numerical Methods in Engineering},
  83\penalty0 (10):\penalty0 1273--1311, 2010{\natexlab{b}}.

\bibitem[Borden et~al.(2012)Borden, Verhoosel, Scott, Hughes, and
  Landis]{borden2012phase}
Michael~J Borden, Clemens~V Verhoosel, Michael~A Scott, Thomas~JR Hughes, and
  Chad~M Landis.
\newblock A phase-field description of dynamic brittle fracture.
\newblock \emph{Computer Methods in Applied Mechanics and Engineering},
  217:\penalty0 77--95, 2012.

\bibitem[Hesch and Weinberg(2014)]{hesch2014thermodynamically}
C~Hesch and K~Weinberg.
\newblock Thermodynamically consistent algorithms for a finite-deformation
  phase-field approach to fracture.
\newblock \emph{International Journal for Numerical Methods in Engineering},
  99\penalty0 (12):\penalty0 906--924, 2014.

\bibitem[Rabczuk(2013)]{rabczuk2013computational}
Timon Rabczuk.
\newblock Computational methods for fracture in brittle and quasi-brittle
  solids: state-of-the-art review and future perspectives.
\newblock \emph{ISRN Applied Mathematics}, 2013, 2013.

\bibitem[Bourdin et~al.(2008)Bourdin, Francfort, and
  Marigo]{bourdin2008variational}
Blaise Bourdin, Gilles~A Francfort, and Jean-Jacques Marigo.
\newblock The variational approach to fracture.
\newblock \emph{Journal of elasticity}, 91\penalty0 (1):\penalty0 5--148, 2008.

\bibitem[Karma et~al.(2001)Karma, Kessler, and Levine]{karma2001phase}
Alain Karma, David~A Kessler, and Herbert Levine.
\newblock Phase-field model of mode iii dynamic fracture.
\newblock \emph{Physical Review Letters}, 87\penalty0 (4):\penalty0 045501,
  2001.

\bibitem[Verhoosel and Borst(2013)]{verhoosel2013phase}
Clemens~V Verhoosel and Ren{\'e} Borst.
\newblock A phase-field model for cohesive fracture.
\newblock \emph{International Journal for numerical methods in Engineering},
  96\penalty0 (1):\penalty0 43--62, 2013.

\bibitem[Ulmer et~al.(2013)Ulmer, Hofacker, and Miehe]{ulmer2013phase}
Heike Ulmer, Martina Hofacker, and Christian Miehe.
\newblock Phase field modeling of brittle and ductile fracture.
\newblock \emph{PAMM}, 13\penalty0 (1):\penalty0 533--536, 2013.

\bibitem[Badnava et~al.(2017)Badnava, Etemadi, and Msekh]{badnava2017phase}
Hojjat Badnava, Elahe Etemadi, and Mohammed~A Msekh.
\newblock A phase field model for rate-dependent ductile fracture.
\newblock \emph{Metals}, 7\penalty0 (5):\penalty0 180, 2017.

\bibitem[Lee et~al.(2016)Lee, Wheeler, and Wick]{lee2016pressure}
Sanghyun Lee, Mary~F Wheeler, and Thomas Wick.
\newblock Pressure and fluid-driven fracture propagation in porous media using
  an adaptive finite element phase field model.
\newblock \emph{Computer Methods in Applied Mechanics and Engineering},
  305:\penalty0 111--132, 2016.

\bibitem[Miehe et~al.(2015)Miehe, Hofacker, Schaenzel, and
  Aldakheel]{miehe2015phase}
Christian Miehe, M~Hofacker, L-M Schaenzel, and F~Aldakheel.
\newblock Phase field modeling of fracture in multi-physics problems. part ii.
  coupled brittle-to-ductile failure criteria and crack propagation in
  thermo-elastic--plastic solids.
\newblock \emph{Computer Methods in Applied Mechanics and Engineering},
  294:\penalty0 486--522, 2015.

\bibitem[Badnava et~al.(2018)Badnava, Msekh, Etemadi, and
  Rabczuk]{badnava2018h}
Hojjat Badnava, Mohammed~A Msekh, Elahe Etemadi, and Timon Rabczuk.
\newblock An h-adaptive thermo-mechanical phase field model for fracture.
\newblock \emph{Finite Elements in Analysis and Design}, 138:\penalty0 31--47,
  2018.

\bibitem[Miehe et~al.(2010{\natexlab{c}})Miehe, Welschinger, and
  Hofacker]{miehe2010phase2}
C~Miehe, F~Welschinger, and M~Hofacker.
\newblock A phase field model of electromechanical fracture.
\newblock \emph{Journal of the Mechanics and Physics of Solids}, 58\penalty0
  (10):\penalty0 1716--1740, 2010{\natexlab{c}}.

\bibitem[Amiri et~al.(2014)Amiri, Mill{\'a}n, Shen, Rabczuk, and
  Arroyo]{amiri2014phase}
Fatemeh Amiri, Daniel Mill{\'a}n, Yongxing Shen, Timon Rabczuk, and M~Arroyo.
\newblock Phase-field modeling of fracture in linear thin shells.
\newblock \emph{Theoretical and Applied Fracture Mechanics}, 69:\penalty0
  102--109, 2014.

\bibitem[Jamshidian and Rabczuk(2014)]{jamshidian2014phase1}
Mostafa Jamshidian and Timon Rabczuk.
\newblock Phase field modelling of stressed grain growth: Analytical study and
  the effect of microstructural length scale.
\newblock \emph{Journal of Computational Physics}, 261:\penalty0 23--35, 2014.

\bibitem[Jamshidian et~al.(2014)Jamshidian, Zi, and
  Rabczuk]{jamshidian2014phase2}
M~Jamshidian, G~Zi, and T~Rabczuk.
\newblock Phase field modeling of ideal grain growth in a distorted
  microstructure.
\newblock \emph{Computational Materials Science}, 95:\penalty0 663--671, 2014.

\bibitem[Jamshidian et~al.(2016)Jamshidian, Thamburaja, and
  Rabczuk]{jamshidian2016multiscale}
Mostafa Jamshidian, P~Thamburaja, and Timon Rabczuk.
\newblock A multiscale coupled finite-element and phase-field framework to
  modeling stressed grain growth in polycrystalline thin films.
\newblock \emph{Journal of Computational Physics}, 327:\penalty0 779--798,
  2016.

\bibitem[Msekh et~al.(2015)Msekh, Sargado, Jamshidian, Areias, and
  Rabczuk]{msekh2015abaqus}
Mohammed~A Msekh, Juan~Michael Sargado, Mostafa Jamshidian, Pedro~Miguel
  Areias, and Timon Rabczuk.
\newblock Abaqus implementation of phase-field model for brittle fracture.
\newblock \emph{Computational Materials Science}, 96:\penalty0 472--484, 2015.

\bibitem[Zhou et~al.(2015{\natexlab{a}})Zhou, Xia, Du, Zhang, and
  Zhou]{zhou2015analytical}
Shu-Wei Zhou, Cai-Chu Xia, Shi-Gui Du, Ping-Yang Zhang, and Yu~Zhou.
\newblock An analytical solution for mechanical responses induced by
  temperature and air pressure in a lined rock cavern for underground
  compressed air energy storage.
\newblock \emph{Rock Mechanics and Rock Engineering}, 48\penalty0 (2):\penalty0
  749--770, 2015{\natexlab{a}}.

\bibitem[Zhou et~al.(2017{\natexlab{a}})Zhou, Xia, Zhao, Mei, and
  Zhou]{zhou2017numerical}
Shu-Wei Zhou, Cai-Chu Xia, Hai-Bin Zhao, Song-Hua Mei, and Yu~Zhou.
\newblock Numerical simulation for the coupled thermo-mechanical performance of
  a lined rock cavern for underground compressed air energy storage.
\newblock \emph{Journal of Geophysics and Engineering}, 14\penalty0
  (6):\penalty0 1382, 2017{\natexlab{a}}.

\bibitem[Xia et~al.(2015)Xia, Zhou, Zhang, Hu, and Zhou]{xia2015strength}
Cai-Chu Xia, Shu-Wei Zhou, Ping-Yang Zhang, Yong-Sheng Hu, and Yu~Zhou.
\newblock Strength criterion for rocks subjected to cyclic stress and
  temperature variations.
\newblock \emph{Journal of Geophysics and Engineering}, 12\penalty0
  (5):\penalty0 753, 2015.

\bibitem[Zhou et~al.(2015{\natexlab{b}})Zhou, Xia, Hu, Zhou, and
  Zhang]{zhou2015damage}
SW~Zhou, CC~Xia, YS~Hu, Y~Zhou, and PY~Zhang.
\newblock Damage modeling of basaltic rock subjected to cyclic temperature and
  uniaxial stress.
\newblock \emph{International Journal of Rock Mechanics and Mining Sciences},
  77:\penalty0 163--173, 2015{\natexlab{b}}.

\bibitem[Zhou et~al.(2017{\natexlab{b}})Zhou, Xia, Zhao, Mei, and
  Zhou]{zhou2017statistical}
Shu-Wei Zhou, Cai-Chu Xia, Hai-Bin Zhao, Song-Hua Mei, and Yu~Zhou.
\newblock Statistical damage constitutive model for rocks subjected to cyclic
  stress and cyclic temperature.
\newblock \emph{Acta Geophysica}, 65\penalty0 (5):\penalty0 893--906,
  2017{\natexlab{b}}.

\bibitem[Zhou et~al.(2018)Zhou, Xia, and Zhou]{zhou2018analyicaltheory}
Shuwei Zhou, Caichu Xia, and Yu~Zhou.
\newblock A theoretical approach to quantify the effect of random cracks on
  rock deformation in uniaxial compression.
\newblock \emph{Journal of Geophysics and Engineering}, 15\penalty0
  (3):\penalty0 627, 2018.
\newblock URL \url{http://stacks.iop.org/1742-2140/15/i=3/a=627}.

\bibitem[Francfort and Marigo(1998)]{francfort1998revisiting}
Gilles~A Francfort and J-J Marigo.
\newblock Revisiting brittle fracture as an energy minimization problem.
\newblock \emph{Journal of the Mechanics and Physics of Solids}, 46\penalty0
  (8):\penalty0 1319--1342, 1998.

\bibitem[Miehe(1998)]{miehe1998comparison}
Ch~Miehe.
\newblock Comparison of two algorithms for the computation of fourth-order
  isotropic tensor functions.
\newblock \emph{Computers \& structures}, 66\penalty0 (1):\penalty0 37--43,
  1998.

\bibitem[Miehe(1993)]{miehe1993computation}
C~Miehe.
\newblock Computation of isotropic tensor functions.
\newblock \emph{International Journal for Numerical Methods in Biomedical
  Engineering}, 9\penalty0 (11):\penalty0 889--896, 1993.

\bibitem[Comsol(2005)]{comsol2005comsol}
AB~Comsol.
\newblock Comsol multiphysics user¡¯s guide.
\newblock \emph{Version: September}, 10:\penalty0 333, 2005.

\bibitem[Bobi{\'n}ski and Tejchman(2011)]{bobinski2011simulations}
Jerzy Bobi{\'n}ski and Jacek Tejchman.
\newblock Simulations of fracture in concrete elements using continuous and
  discontinuous models.
\newblock \emph{Mechanics and Control}, 30\penalty0 (4), 2011.

\bibitem[Kalthoff(2000)]{kalthoff2000modes}
Joerg~F Kalthoff.
\newblock Modes of dynamic shear failure in solids.
\newblock \emph{International Journal of Fracture}, 101\penalty0 (1):\penalty0
  1--31, 2000.

\bibitem[Rabczuk and Zi(2007)]{rabczuk2007meshfree}
Timon Rabczuk and Goangseup Zi.
\newblock A meshfree method based on the local partition of unity for cohesive
  cracks.
\newblock \emph{Computational Mechanics}, 39\penalty0 (6):\penalty0 743--760,
  2007.

\bibitem[Rabczuk et~al.(2007{\natexlab{a}})Rabczuk, Bordas, and
  Zi]{rabczuk2007three}
Timon Rabczuk, St{\'e}phane Bordas, and Goangseup Zi.
\newblock A three-dimensional meshfree method for continuous multiple-crack
  initiation, propagation and junction in statics and dynamics.
\newblock \emph{Computational Mechanics}, 40\penalty0 (3):\penalty0 473--495,
  2007{\natexlab{a}}.

\bibitem[Ren et~al.(2017)Ren, Zhuang, and Rabczuk]{ren2017dual}
Huilong Ren, Xiaoying Zhuang, and Timon Rabczuk.
\newblock Dual-horizon peridynamics: A stable solution to varying horizons.
\newblock \emph{Computer Methods in Applied Mechanics and Engineering},
  318:\penalty0 762--782, 2017.

\bibitem[Ren et~al.(2016)Ren, Zhuang, Cai, and Rabczuk]{ren2016dual}
Huilong Ren, Xiaoying Zhuang, Yongchang Cai, and Timon Rabczuk.
\newblock Dual-horizon peridynamics.
\newblock \emph{International Journal for Numerical Methods in Engineering},
  108\penalty0 (12):\penalty0 1451--1476, 2016.

\bibitem[Rabczuk and Samaniego(2008)]{rabczuk2008discontinuous}
T~Rabczuk and E~Samaniego.
\newblock Discontinuous modelling of shear bands using adaptive meshfree
  methods.
\newblock \emph{Computer Methods in Applied Mechanics and Engineering},
  197\penalty0 (6):\penalty0 641--658, 2008.

\bibitem[Rabczuk et~al.(2007{\natexlab{b}})Rabczuk, Areias, and
  Belytschko]{rabczuk2007simplified}
Timon Rabczuk, PMA Areias, and Ted Belytschko.
\newblock A simplified mesh-free method for shear bands with cohesive surfaces.
\newblock \emph{International Journal for Numerical Methods in Engineering},
  69\penalty0 (5):\penalty0 993--1021, 2007{\natexlab{b}}.

\bibitem[Rabczuk et~al.(2010)Rabczuk, Zi, Bordas, and
  Nguyen-Xuan]{rabczuk2010simple}
Timon Rabczuk, Goangseup Zi, Stephane Bordas, and Hung Nguyen-Xuan.
\newblock A simple and robust three-dimensional cracking-particle method
  without enrichment.
\newblock \emph{Computer Methods in Applied Mechanics and Engineering},
  199\penalty0 (37):\penalty0 2437--2455, 2010.

\bibitem[Rabczuk and Belytschko(2007)]{rabczuk2007three2}
T~Rabczuk and T~Belytschko.
\newblock A three-dimensional large deformation meshfree method for arbitrary
  evolving cracks.
\newblock \emph{Computer Methods in Applied Mechanics and Engineering},
  196\penalty0 (29):\penalty0 2777--2799, 2007.

\bibitem[Rabczuk and Belytschko(2004)]{rabczuk2004cracking}
T~Rabczuk and T~Belytschko.
\newblock Cracking particles: a simplified meshfree method for arbitrary
  evolving cracks.
\newblock \emph{International Journal for Numerical Methods in Engineering},
  61\penalty0 (13):\penalty0 2316--2343, 2004.

\end{thebibliography}

\end{document}